\documentclass[letterpaper, traditabstract]{aa}

\usepackage{txfonts}
\usepackage{natbib}
\bibpunct{(}{)}{;}{a}{}{,} 

\usepackage{graphicx}

\newcommand{\sbpp}{{\scshape{Superbox}\footnotesize\raise.4ex\hbox{+\kern-.2em+}}}
\newcommand{\cpp}{C{\footnotesize\raise.4ex\hbox{+\kern-.2em+}}}
\newcommand{\subo}{{\scshape Superbox}}

\begin{document}

\title{Making Counter-Orbiting Tidal Debris}
\subtitle{The Origin of the Milky Way Disc of Satellites}

\author{M.~S.~Pawlowski
\and P.~Kroupa
\and K.~S.~de~Boer}

\institute{Argelander Institute for Astronomy, University of Bonn, Auf dem H\"{u}gel 71, D-53121 Bonn, Germany\\
\email{mpawlow@astro.uni-bonn.de,pavel@astro.uni-bonn.de,deboer@astro.uni-bonn.de}
}

\date{Received -- ; accepted --}

\abstract{ Using stellar-dynamical calculations it is shown for the first time that counter-orbiting material emerges naturally in tidal interactions of disc galaxies. Model particles on both pro- and retrograde orbits can be formed as tidal debris in single encounters with disc galaxies of 1-to-1 and 4-to-1 mass ratios. A total of 74 model calculations are performed for a range of different initial parameters. Interactions include fly-by and merger cases. The fraction of counter-orbiting material produced varies over a wide range (from a few up to 50 percent). All fly-by models show a similar two-phase behaviour, with retrograde material forming first. Properties of the prograde and retrograde populations are extracted to make an observational discrimination possible.\\
During such encounters the tidal debris occupies a certain region in
phase space. In this material, tidal-dwarf galaxies may form. The modelling
therefore can explain why galaxies may have dwarf galaxies orbiting counter
to the bulk of their dwarf galaxies. An example is the Sculptor dwarf of
the Milky Way, which orbits counter to the bulk of the disc of satellites.
The modelling thus supports the scenario of the MW satellites being
ancient tidal-dwarf galaxies formed from gaseous material stripped from another galaxy during an encounter with the young MW.\\
A possible candidate for this galaxy is identified as the Magellanic
Cloud progenitor galaxy. Its angular motion fits the angular motion of the
MW disc of satellites objects. This scenario is in agreement with
Lynden-Bell's original suggestion for the origin of the dSph satellites and the near-unbound orbit of the LMC.}


\titlerunning{Making Counter-Orbiting Tidal Debris}
\authorrunning{Pawlowski, Kroupa \& de Boer}

\maketitle

\section{Introduction}
\label{introsect}
The early-recognised anisotropic distribution of Milky Way (MW) satellite galaxies naturally lead to the proposition that they are phase-space correlated tidal dwarf galaxies \citep{Lynden-Bell76, kunkel79}. With the advent of cold-dark-matter cosmology the satellites were however viewed as the unmerged remnants of the bottom-up hierarchical structure formation process.
In (standard) cold dark matter (CDM) cosmology, galaxies form through accretion and merging of smaller systems. This suggests a simple origin for the dwarf spheroidal (dSph) satellite galaxies of the Milky Way: they are believed to be dark matter dominated subhaloes, with a luminous matter content, that are gravitationally trapped in the Milky Way halo and have not yet merged completely with their host \citep{white78}. The details of this origin, in contrast, are still disputed and not even the nature of the dominant dark matter component in dSphs is well defined \citep{gilmore07}. However, Zwicky's (1956) proposition that new dwarf galaxies form when galaxies interact and his suggestion \citep{Zwicky37} that galaxies contain dark matter today lead to the Fritz Zwicky Paradox \citep{kroupa10} since the implied standard cosmological model leads to a large population of TDGs that ought to have the properties of dE and dSph galaxies \citep{Okazaki00}. This would leave little room for the existence of cold-dark-matter dominated satellite dwarf galaxies, clashing with the expected existence of large numbers of such objects.

In addition to this, the number of observed satellites was reported to be about an order of magnitude lower than expected from CDM simulations, resulting in what is termed the `missing satellite problem' \citep{moore99,klypin99,diemand08}. There are several proposed scenarios to solve this issue, like reducing the number of expected satellites as only the most massive progenitors \citep{libeskind05}, the earliest formed haloes \citep{strigari07} or even only the low-mass systems \citep{sales07} are to be expected to produce satellite galaxies surviving until today. Other possibilities are to assume that only the most massive subhaloes were able to form stars \citep{stoehr02}. All of these proposals are rather 'ad hoc'. The alternative proposed solution is that very faint satellites, that were until now overlooked because of their low stellar densities \citep{tollerud08}, are still to be discovered. None of these proposals are satisfactory.

Still, not only the observed number of Milky Way satellite galaxies is in conflict with the expectation, they also do not match the dark-matter mass -- luminosity relation nor the form of the mass function of luminous dark matter halos \citep{kroupa10}. The distribution is exceedingly anisotropic, too \citep[e.g.:][]{Lynden-Bell76,majewski94,hartwick00}. Especially the most luminous `classical' satellites pose a problem for hierarchical structure formation. They make up a significantly pronounced disc of satellites (DoS), a structure highly inclined to the plane of the Milky Way disc \citep{kroupa05,metz07}. The inclusion of more recently found, very faint satellite galaxies supports this spatial feature \citep{metz09a}. In fact, treating the 13 new ultra-faint satellites independently leads to virtually the same DoS solution as given by the 11 bright classical satellites \citep{kroupa10}. 
As \citet{kroupa10} show, the standard CDM model \citep[e.g.][]{Libeskind09} can at best reproduce 0.4 percent of all MW-type dark matter halos having a host galaxy of similar luminosity as the MW with 11 satellites in a DoS. An infall of the satellites as a group of dwarf galaxies suggested by \citet{li08} and \citet{donghia09} has been ruled out by \citet{metz09b} because of the thin structure of the DoS in comparison to the extension of observed dwarf galaxy groups.

\begin{figure}
\centering
 \includegraphics[width=88mm]{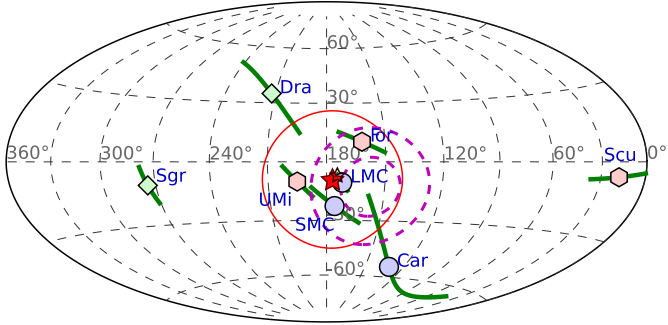}
 \caption{Orbital poles for the eight MW satellite galaxies for which proper motions are measured. The plot is an Aitoff projection of galactocentric coordinates. The green lines give the uncertainties in direction of the orbital poles. The red star marks the mean direction of the orbital poles of the six satellites concentrated in the centre, excluding Sagittarius and Sculptor, the red loop marks the spherical standard deviation. The dashed circles are the regions $15^{\circ}$\ and $30^{\circ}$\ degrees from the normal of the DoS fitted to the 11 classical MW satellites. The figure is adapted from \citet{metz}. See the on-line edition of the article for a colour version of this figure.}
 \label{orbitalpoles}
\end{figure}

Investigating the available proper motions of the satellite galaxies \citep[e.g.:][who also include globular clusters in their studies]{lynden-bell95,palma02}, the situation becomes even more puzzling. Out of eight satellite proper motions analysed by \citet{metz}, six are found to be in agreement with being co-orbiting within the DoS, as visible in their figure 1 (reproduced in Fig. \ref{orbitalpoles} here). The authors found that such a situation is extremely unlikely even if the satellites were drawn from CDM simulations specifically tailored to solve the Milky Way DoS problem. Only two satellites have orbital poles not directly associated with other poles. Of these, one (Sagittarius) is orbiting perpendicularly to the DoS, while Sculptor is on a counter-rotating orbit within the DoS.

An alternative origin of the Milky Way satellite galaxies would be them forming as tidal dwarf galaxies (TDGs) perhaps about 10 Gyr ago in an interaction of the young MW with another galaxy. \citet{sawa05} proposed that the MW had an off-centre interaction with the primordial Andromeda galaxy M31 about 10 Gyr ago, whereby the M31 and MW satellite population was born. Another possibility would be a merger of two progenitor galaxies forming part of today's MW. The tidal expulsion of material during such an interaction might have led, through gravitational instabilities followed by clumping of matter within tidal tails, to the creation of TDGs. This concept was first proposed by \citet{zwicky56} and it was first observed to happen in the Antennae by \citet{mirabel92}. TDGs were subsequently found in many observations of interacting galaxies \citep[e.g.:][]{hunsberger96,weilbacher03,walter06} as well as in N-body calculations \citep{wetzstein07, bournaud06}. They would naturally lie in the plane of the interaction. \citet{Lynden-Bell76} first suggested that the Milky Way satellites might be of tidal origin.

The common direction of orbit around the Milky Way, found for most satellite galaxies with measured proper motions, can be considered to be an indication for this scenario. The only exceptions are Sagittarius and Sculptor. While Sagittarius is near to the Milky Way disc and thus may have suffered precession or even a scattering encounter with another satellite \citep{zhao98} and consequently can be expected to be far from the direction of the initial orbit, the Sculptor dwarf spheroidal galaxy is another case. The direction of Sculptor's orbitel angular momentum vector indicates that it indeed orbits within the DoS, but on a counter-rotating orbit. While this might, on the first glance, seem to pose a problem for the common origin of the DoS galaxies as TDGs, the opposing direction could in fact be another indication for this origin. If tidal material in galaxy encounters were to have only one orbital direction in most interactions, this would make a tidal origin of the MW satellite system unlikely.

The aim of this contribution is to investigate whether counter-orbiting material can be formed in tidal interactions. Not the creation of actual TDGs is studied, but only the motions of particles in N-Body models which represent stellar streams. Modelling of galaxy interactions and studies of the formation of TDGs in high-resolution calculations of gas-rich galaxy-encounters have previously been done \citep{bournaud06,wetzstein07,bournaud08}. But until now they have not focused on the orbital angular momenta of the created dwarf galaxies and the authors of this contribution know of no published work examining the possible creation of counter-rotating orbits. In the following, angular momenta of particles in interacting galaxies are analysed for the first time, searching for the creation of counter-orbiting material.

In total 96 N-Body calculations including gas dynamics have been performed by \citet{bournaud06}. They identified almost 600 substructures, but due to computing-time requirements the resolution was only sufficient to resolve objects more massive than $10^8~\rm{M}_{\sun}$, while all the satellite Galaxies of the MW except the Large and Small Magellanic Clouds have stellar masses in the range of $10^4$\ to $10^7~\rm{M}_{\sun}$\ \citep{strigari08}. \citet{bournaud06} state that only a small fraction of their TDG candidates survive, a statement which has to be understood in the light of their detection criterion: their TDG candidates have to have a mass of at least $10^8~\rm{M}_{\sun}$. TDGs that loose mass and drop below this threshold are counted as non-surviving, whereas TDGs still exist (and can be formed) at much lower masses. Indeed, explicit high-resolution calculations of early \citep{recchi07} and long-term \citep{kroupa97, Klessen98} evolution of $10^7 - 10^8~\rm{M}_{\sun}$ TDGs demonstrate that they readily survive.
Thus the mass and spatial-resolution of the \citet{bournaud06} models influences the deduced survival of TDGs and also their formation. In the highest-resolution calculation to date that includes gas, more than 100 young stellar objects, star clusters and TDGs, with masses between $10^5$\ and $10^9~\rm{M}_{\sun}$, have been identified in one single galaxy collision \citep{bournaud08}\footnote{The more massive super star clusters identified in their work, with masses below $10^8~\rm{M}_{\sun}$, can be interpreted to be compact spheroidal galaxies in contrast to massive and rotating TDGs \citep{Bournaud2008}.}. Furthermore, TDGs might also form from star cluster complexes which are observed to exist in tidal tails \citep{Fellhauer2002}. Finally, \citet{bournaud06} performed merger-only calculations, whereas this contribution also includes fly-by interactions. This is why their results about the dependence of TDG survival-chance on the place of formation along the tidal tails can not simply be adopted here\footnote{They find that TDGs formed at the tip of the tidal tail have the best chances to survive. However, in contrast to a merger where the tip will remain relatively undisturbed for a long time, that tip in a fly-by hits the second (target) galaxy first, which might lead to different conclusions.}.

A full analysis of the orbital motions of formed TDGs would therefore demand calculations with very high resolutions that include gas dynamics. Performing a parameter scan with these is currently not feasible. Gas dynamics is therefore neglected in the present work, to allow for a higher number of calculations scanning over initial velocities and directions between the interacting galaxies. This results in not looking for the formation of TDGs, but only in a study of the phase-space properties of the formed tails, testing if the concept of counter-orbiting material originating in a single encounter is feasible.

This approach is applicable as angular momentum conservation implies that a condensed object (a TDG) will have the composite orbital angular momentum of the material (the particles) accreted into it. While the effect of dynamical friction is stronger for a dwarf-galaxy-sized object than for a single star, it is still negligible for TDGs \citep{kroupa97}. According to equation 7-26 in \citep{BT}, the dynamical friction timescale, $t_{\rm{fric}}$, for a $M = 10^{8}~\rm{M}_{\sun}$ galaxy orbiting at a galactocentric distance $d = 60 \rm{kpc}$\ with a velocity of $v = 125~\rm{km s}^{-1}$ is of the order of $10^2$\ Gyr, much longer than a Hubble time. Note that $t_{\rm{fric}} \propto \frac{d^2 v}{M}$, thus the parameters chosen here are conservative, as the velocity $v$ is low, $d$\ is small and $t_{\rm{fric}}$ will be even longer for TDGs forming with a variety of masses $M < 10^{8}~\rm{M}_{\sun}$. Therefore, the dynamics of TDGs and the stellar streams will not separate and the results can be interpreted in the light of a tidal origin for the MW satellites. 

Using a new version of the original \citep{madejsky93, superbox} \subo\ code developed by M. Metz, \sbpp, it is shown that particles on both pro- and retrograde orbits are indeed a \textit{natural} outcome of an encounter of two disc galaxies. Calculations are performed for perpendicularly oriented galaxies of 1:1 and 1:4 mass-ratios, scanning a range of different initial parameters and interactions including fly-by and merger cases. In almost all cases counter-orbiting material is found.

Section \ref{nomenclatesect} defines the nomenclature used in the following. In Sect. \ref{modelsect}, the galaxy model, initial conditions and the thoughts leading to it are introduced. The results of two exemplary computations are described in Sect. \ref{resultssect}. Section \ref{parascansect} gives an overview of the results from the other models calculated, scanning a range of initial parameters. A discussion including extraction of possible observational constraints is performed in Sect. \ref{discusssect}, where also the possibility of a tidal origin of the MW satellites system is discussed. Note that Sect. \ref{resultssect} concentrates on equal-mass encounters, while Sects. \ref{parascansect}-\ref{discusssect} generalize the encounters to non-equal galaxies. Finally there are concluding remarks in Sect. \ref{concludesect}. There is an appendix which gives details on how the scaling of the MW was performed. It also compiles the results of the parameter scans in 12 tables.

\section{Definitions}
\label{nomenclatesect}
The galaxies in the models will be referred to as:
\begin{description}
 \item \textit{Target galaxy}: around this disc galaxy the particles ripped out of the infalling galaxy orbit and are later analysed.
 \item \textit{Infalling galaxy}: the disc galaxy approaching the target. This galaxy will develop tidal tails from which particles will stream to the target and form retrograde and prograde orbiting material.
\end{description}
Thus, in the following the terms \textit{prograde} and \textit{retrograde} will be used frequently:
\begin{description}
 \item \textit{Prograde particles}: particles that, after the galaxy-interaction, have an orbital angular momentum vector oriented in the same direction as the galaxies' orbital one within a circle of acceptance (COA) of a given angular radius.
 \item \textit{Retrograde particles}: particles that, after the galaxy-interaction, have an orbital angular momentum vector oriented in the opposite direction than the galaxies' orbital one within a COA.
\end{description}
Furthermore, proper motions of satellite galaxies in the vicinity of the MW are oriented preferentially in one direction, with the Sculptor dwarf galaxy moving along an opposite orbit. It is said that Sculptor is a 'retrograde' satellite, with respect to the bulk of the satellites' motions. This statement does not include any knowledge of a possible interaction with another galaxy and thus might cause some confusion. What is seen today as a 'retrograde dwarf satellite' might have been the prograde orientation of the orbit when compared to the angular momenta of the galaxies in the initial interaction. Due to this ambiguity the use of 'prograde' and 'retrograde' is refrained from for those situations where the initial interaction geometry is unknown, as is the case in the MW-satellite system. Instead we define:
\begin{description}
 \item \textit{Co-orbiting}: orbiting in the same direction as the bulk of the material (particles in the models or satellite galaxies in the MW).
 \item \textit{Counter-orbiting}: orbiting in the opposite sense than the bulk motion.
\end{description}
Thus, from the point of view of an observer today, Sculptor is counter-orbiting with respect to the other satellites. Finally the naming scheme of the models is defined:
\begin{description}
 \item \textit{Names of models}: The names of the models presented in the next sections are composed of the direction of the initial velocity vector for the galaxies, expressed in degrees from the z-axis (e.g. `7.5deg' meaning $7.5^\circ$) and the initial relative velocity in per cent of the parabolic velocity for point-masses of the same mass (e.g. `100vel' meaning 100 percent of $v_{\rm{parab}}$).
\end{description}

\section{Models}

\subsection{Scenario and galaxy model}
\label{modelsect}

\begin{table*}
 \caption{Parameters of the dwarf disc galaxy models.}
 \label{modelparams}
 \begin{center}
 \begin{tabular}{@{}llllll}
  \hline
  ~ & ~ & equal mass ratio & \multicolumn{2}{c}{4-to-1 mass ratio} & ~ \\
  ~ & Symbol & Infalling \& Target & Infalling & Target & Description\\
  \hline
  Disc & $N_{\rm{d}}$ & 500\,000 & 500\,000 & 500\,000 & number of disc particles\\
   & $M_{\rm{d}}$ & $8.0 \times 10^9~\rm{M}_{\sun}$& $4.0 \times 10^9~\rm{M}_{\sun}$ &  $16.0 \times 10^9~\rm{M}_{\sun}$ & disc mass \\
   & $R_{\rm{d}}$ & 1.60 kpc & 1.15 kpc & 2.25 kpc & disc scale length \\
   & $R_{\rm{d}}^{\rm{max}}$ & 8.00 kpc & 5.75 kpc & 11.25 kpc & maximum disc radius \\
   & $R_{\sun}$ & 6.00 kpc & 4.30 kpc & 8.40 kpc & solar radius\\
   & $Q_{\sun}$ & 1.70 & 1.70 & 1.70 & Toomre Q at $R_{\sun}$ \\
   & $z_0$ & 0.25 kpc & 0.20 kpc & 0.35 kpc & disc scale height\\
   & $z_{\rm{max}}$ & 2.50 kpc & 2.00 kpc & 3.50 kpc & maximum particle height \\
  \hline
  Halo & $N_{\rm{H}}$ & 1\,000\,000 & 1\,000\,000 & 1\,000\,000 & number of halo particles\\
   & $M_{\rm{H}}$ & $8.0 \times 10^{10}~\rm{M}_{\sun}$ & $4.0 \times 10^{10}~\rm{M}_{\sun}$ & $16.0 \times 10^{10}~\rm{M}_{\sun}$ & halo mass \\
   & $a$ & 10.0 kpc & 6.9 kpc & 12.4 kpc & halo core radius\\
   & $R_{\rm{H}}^{\rm{max}}$ & 100.0 kpc & 100.0 kpc & 150.0 kpc & maximum halo radius\\
  \hline
 \end{tabular}
 \end{center}
 \small \smallskip

\end{table*}

The models are based on the following scenario. About 10 Gyr ago, the progenitor of the MW had an interaction with another progenitor disc galaxy. During this interaction, tidal debris in the form of gas and possibly stars was formed and possibly the initial disc was destroyed in the process that might have included a merger. A merger of the two initial discs possibly formed part of the bulge of today's MW, whereas the MW disc seen today formed later through ongoing accretion.

Investigating its chemical evolution, \citet{ballero07} found that the bulge of the MW formed rapidly, with a timescale of the order of 0.1 Gyr. This is in agreement with a merger-induced origin. However, \citet{Babusiaux2010} suggests that the MW bulge consists of two parts: a spheroidal that might originate from past merger events and a bar-like structure or pseudobulge. Thus, both a merger scenario forming a bulge as well as a fly-by interaction with less severe consequences for the early MW, but leading to the formation of a bar instability in the disc, appear plausible and are investigated in this contribution.

To form a realistic bulge in the calculations would require the inclusion of gas, out of which the bulge stars can form within a short timescale. Furthermore, dissipative processes in the gas of galactic discs would lead to a change in the remnant-galaxy dynamics compared to the purely stellar case. One therefore does not learn much from a comparison of the resulting central object in the calculations with the MW bulge, as the model only represents a solution to the collisionless Boltzmann Equation. Similarly, after the interaction the model disc is not comparable with today's MW disc as it, in this scenario, forms through ongoing accretion of intergalactic gas after the interaction took place. The present contribution therefore concentrates on an analysis of the general phase-space properties of the tidal material and does not claim to be a complete model for the formation of the MW disc, bulge and satellite system. Some remarks on the morphology of the remnant galaxy in the different interactions can be found in Section \ref{parascansect}.

Assuming the MW bulge formed in an equal-mass merger gives approximately 0.5 MW bulge masses as a natural starting point for the mass of the early MW models\footnote{This might on the one hand over-estimate the mass of the classical bulge component of the MW in merger-interactions and on the other hand lead to a too-low pseudobulge-mass in case of fly-by interactions, but is considered to be a good compromise.}.
The disc galaxy parameters are then scaled to that mass, while the dark matter halo parameters are adjusted such as to model the empirically found flat rotation curves, the origin of which is a matter of ongoing debate \citep{kroupa10}. The resulting parameters are compiled in Table \ref{modelparams} for the equal-mass and 4-to-1 mass models. The models for the 4-to-1 mass ratio interactions were determined with the same method, but this time the mass was either half or twice the original value for target and infalling galaxy, respectively. More details on how these scalings were performed can be found in Appendix \ref{scalingappendix}.

Particle realisations of the models with $5 \times 10^{5}$\ particles for the disc and $10^{6}$\ for the halo component are set up using the code \textsc{MaGalie} \citep{magalie}. After that, the galaxy models are virialised by integrating them in isolation for 1.5 Gyr. In the case of the equal-mass interactions, after virialisation the model galaxy is duplicated to form the target and the infalling galaxy.

One drawback of this approach is that it does not include the mass accretion forming the MW. The integrations cover 10 Gyr and during this time the newly accreted disc should increase its mass until its present-day value. A changing mass will inevitably change the orbital parameters of the pro- and retrograde particles, like the eccentricity and apocentre distance. But as the main aim is to show that counter-rotating orbits of tidal debris are in principle possible, that is no hindrance. The direction of the angular momenta of the particles will not be affected by a slow change, like that of a slowly forming disc, and their orbits will adjust adiabatically.

\subsection{Geometry and initial conditions}
\label{setupsect}

This section focuses on describing the initial parameters of two exemplary equal-mass models of interactions between the two galaxies: one collision ending with the merging of the two disc galaxies (model 7.5deg100vel) and one off-centre collision leading to a fly-by (model 5deg200vel). In both models, the infalling disc spins prograde with respect to the orbit of the two galaxies, because long lived tidal dwarfs form more easily in this geometry, as \citet{bournaud06} have stated. Results of further models scanning a range of initial parameters, including retrograde spins of the infalling galaxy in the case of mergers, are summarised in Sect. \ref{parascansect}. In the case of retrograde fly-bys, tidal tails transferring matter are usually not as pronounced as in a prograde orientation of the infalling disc.

The initial distance $d_{\rm{ini}}$\ of the galaxies is chosen to be two times their dark matter halo maximum radii, $d_{\rm{ini}} = 2 \times R^{\rm{max}}_{\rm{H}} = 200~\rm{kpc}$. Initial relative velocities are chosen with respect to the parabolic velocity $v_{\rm{parab}} = 87~\rm{km~s^{-1}}$\ calculated from the total galaxy masses $M_{\rm{target}}$ and $M_{\rm{infalling}}$, and their initial distance $d_{\rm{ini}}$, assuming the galaxies (including haloes) are point masses.
The relative velocities of the galaxies in models 7.5deg100vel and 5deg200vel are $1.0 \cdot v_{\rm{parab}}$ and $2.0 \cdot v_{\rm{parab}}$, respectively.

The two galaxies are put to +100 (infalling) and -100 kpc (target) along the z-axis. In the case of a merger, each is given half of the relative interaction velocity. Thus the centre of mass is at the origin at all times. In the fly-by case the target galaxy is assigned a smaller velocity than the infalling galaxy to make sure the target and the orbiting particles around it stay within the model volume.

\begin{table}
 \caption{Initial conditions of the calculations}
 \label{initialconditions}
 \begin{center}
 \begin{tabular}{@{}lcclcc}
  \hline
  Model: & \multicolumn{2}{c}{7.5deg100vel} & \multicolumn{2}{c}{5deg200vel} \\
  Type: &  \multicolumn{2}{c}{merger} & \multicolumn{2}{c}{fly-by} \\
  Galaxy: & Target & Infalling & Target & Infalling \\
  \hline
  Total mass [$10^{10} M_{\sun}$] & 8.8 & 8.8 & 8.8 & 8.8 \\
  Position [kpc] & & & &\\
  ~x & 0.0 & 0.0 & 0.0 & 0.0 \\
  ~y & 0.0 & 0.0 & 0.0 & 0.0 \\
  ~z & $-100.0$ & $+100.0$ & $-100.0$ & $+100.0$ \\
  Velocity [km s$^{-1}$] &  & & &\\
  ~$v_{\rm{x}}$ & 0.0 & 0.0 & 0.0 & 0.0 \\
  ~$v_{\rm{y}}$ & $+5.68$ & $-5.68$ & $+3.03$ & $-12.13$ \\
  ~$v_{\rm{z}}$ & $+43.13$ & $-43.13$ & $+34.67$ & $-138.67$ \\
  ~$\beta$ & 7.5$^{\circ}$ & 7.5$^{\circ}$ & 5.0$^{\circ}$ & 5.0$^{\circ}$ \\
  $\mathbf{L_{\rm{disc}}}$ & $+ z$ & $+ x$ & $+ z$ & $+ x$ \\
  $r_{\rm{min}}$ [kpc] & \multicolumn{2}{c}{6.1} & \multicolumn{2}{c}{8.3} \\
  Time & \multicolumn{4}{c}{1000 time-steps = 500 Myr} \\
  \hline
 \end{tabular}
 \end{center}

 \small \smallskip

The mass of the whole galaxy (disc and halo) is given in  $10^{10} M_{\sun}$. Position states the position of the centre of mass of each galaxy in kpc with respect to the centre of the model volume. Velocity does the same for the initial velocities in $\rm{km~s^{-1}}$, $\beta$\ is the angle between $v_{\rm{ini}}$\ and the z-axis. $\mathbf{L_{\rm{disc}}}$ gives the directions of the angular momenta of the discs parallel to the axes, for example, $+ z$ means the vector points along the z-axis in positive direction. Finally, $r_{min}$~is the perigalactic distance between the two galaxies at their first passage.
\end{table}

\begin{figure}
 \centering
 \includegraphics[width=50mm]{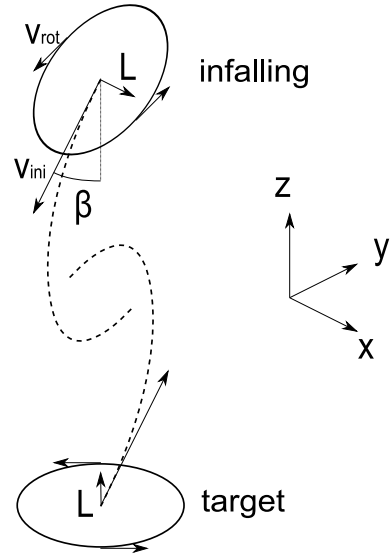}
 \caption{Schematics of the geometry of the interaction and the initial paths of the galaxies. The orientation of the axes is given at the right, the dashed lines symbolise the direction of motion of the galaxies. The vectors $\mathbf{L}$ give the direction of angular momenta of the disc spins, indicated by the small arrows on the disc edges (labelled $\rm{v}_{\rm{rot}}$).The vectors $v_{\rm{ini}}$\ sketch the direction of the initial velocity of the galaxies, the angle between it and the z-axis is $\beta$.}
 \label{sketch}
\end{figure}

The direction of the velocity vectors and the perigalactic distances during the first approach, $r_{min}$, can be found in Table \ref{initialconditions}. Also given are the direction of the internal (spin) angular momentum vectors $\mathbf{L_{\rm{disc}}}$ of the discs. The infalling galaxy's disc plane is perpendicular to the target's plane, which is motivated by the possible Local Volume source of this galaxy \citep{metz09b} and lies in the orbital plane of the encounter. The infalling galaxy is always seen edge on from the target, in other words, it approaches the target like a circular saw. The geometry of the interaction is sketched in Fig. \ref{sketch}. The whole interaction takes place in the y-z-plane of the volume, in which the retrograde and prograde particle orbits lie.

\subsection{SUPERBOX++ calculations}
\label{sbppsect}

All models in this work are integrated using a new version of the well known code \subo\ \citep{madejsky93, superbox}. \sbpp\ is a new implementation of a particle-mesh scheme with a hierarchy of high-resolution sub-grids used in \subo.
Differences in the realisation of the algorithm make the new code much more efficient than the former. \subo\ had been written in the Fortran language with a particular focus on the minimisation of usage of random access memory (RAM), which is no longer an issue. \sbpp\ is implemented in the modern \cpp\ programming language using object oriented programming techniques. The algorithm has been developed with a focus on the performance of the code, but at the same time keeping memory consumption at a low level. \sbpp\ makes optimal use of modern multi-core processor technologies. \subo\ conserves the angular momentum well, especially for the high resolution used in these models, as has been shown in tests performed by \citet{superbox}. The maximum $\Delta L / L$\ for the same grid-resolution was determined to be less than 0.05 percent after 1000 time-steps.

For the models calculated in this work, the number of grid cells per dimension is 64, and the grid sizes are 10, 30 and 400 kpc for the inner, middle and outer grid of the disc component, respectively. These values are the same as those given in \ref{modelsetupsec} for the virilisation. The models are calculated for 10 Gyr in 20000 steps. To get the best statistics to detect even small percentages of retrograde particles, positions and velocities of all disc particles of the infalling galaxy are analysed. Snapshots are produced in time-step intervals of 50, corresponding to one snapshot every 25 Myr.

\section{Results}
\label{resultssect}
At first, a general description of two exemplary calculations is given, followed by a detailed analysis of the pro- and retrograde particle populations. In Sect. \ref{parascansect} the exploration of the larger parameter space with more models analysed in an analogous manner is discussed.

\subsection{General description}

\subsubsection{Model 7.5deg100vel: merger}
In this model, the galaxies pass by each other after 1.55 Gyr at a perigalactic distance of $\approx 6.1$~kpc. After another 0.4 Gyr, the galaxies start to fall back towards each other and merge at about 2.5 Gyr after the start of the integration. The target galaxy's disc is destroyed in the collision, a spheroidal central remnant galaxy is formed. A few hundred of the target's particles are expelled to distances up to 50 kpc, but no well-defined large tidal tails form out of the target galaxy. Thus, the following deals with particles from the infalling galaxy.

The prograde particles (as defined in Sect. \ref{nomenclatesect}) form in great number because they move in the same direction as the infalling galaxy with respect to the target. Thus, when the infalling galaxy swings around and merges, many particles in the outer part will follow this direction. Furthermore, a short tidal tail between the two galaxies transports particles from the infalling to the target galaxy soon after the first passage. The majority of these is prograde.

A part of the particles becomes retrograde when the two galaxies finally merge almost head on. Even though the centres pass each other with a preferred, prograde orientation, some particles in the outer region will pass the combined centre of mass on the opposing side, leading to counter-orbiting particles. Furthermore, a significant fraction originates in the tip of the tidal tail pointing away from the target galaxy. That tail forms after the violent passing of the two galaxies and falls back towards the target together with the infalling galaxy.\footnote{The two models are also visualised in movies, available online at the Downloads Page of the AIfA: http://www.astro.uni-bonn.de/download/animations/.}

\subsubsection{Model 5deg200vel: fly-by}
In this model the galaxies approach each other until they pass by after about 1 Gyr with a minimum distance of 8.3 kpc. At 3 Gyr the distance approaches 240 kpc and the galaxies can be considered well separated. Two tidal tails form out of the infalling galaxy, one pointing towards the target galaxy, the other into the opposite direction. The latter one is of no further interest, the particles in it later fall back towards the remaining infalling (but now receding) galaxy.

The target galaxy gets disturbed, its disc heats up, thickens and a bar instability forms. However, in general it remains a rotating disc galaxy. While a pronounced spheroidal component that can be interpreted as a classical bulge is not formed, bars in disc galaxies are a typical formation mechanism of pseudobulges. Due to the interaction geometry, the spin angular momentum vector of the target disc lies within the plane of the interaction, no tidal tails are formed from the target galaxy.

\begin{figure*}
\centering
 \includegraphics[width=88mm]{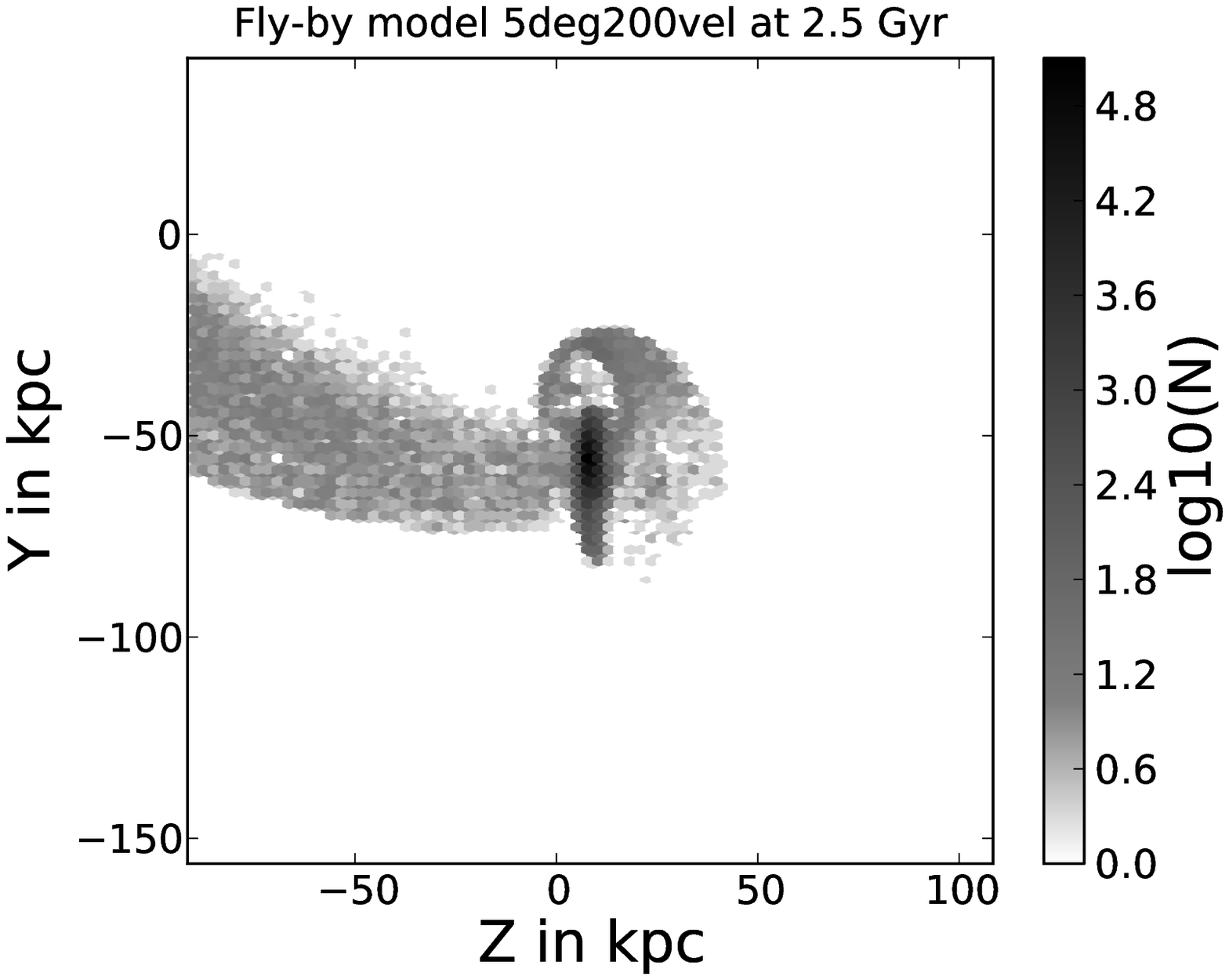}
 \includegraphics[width=88mm]{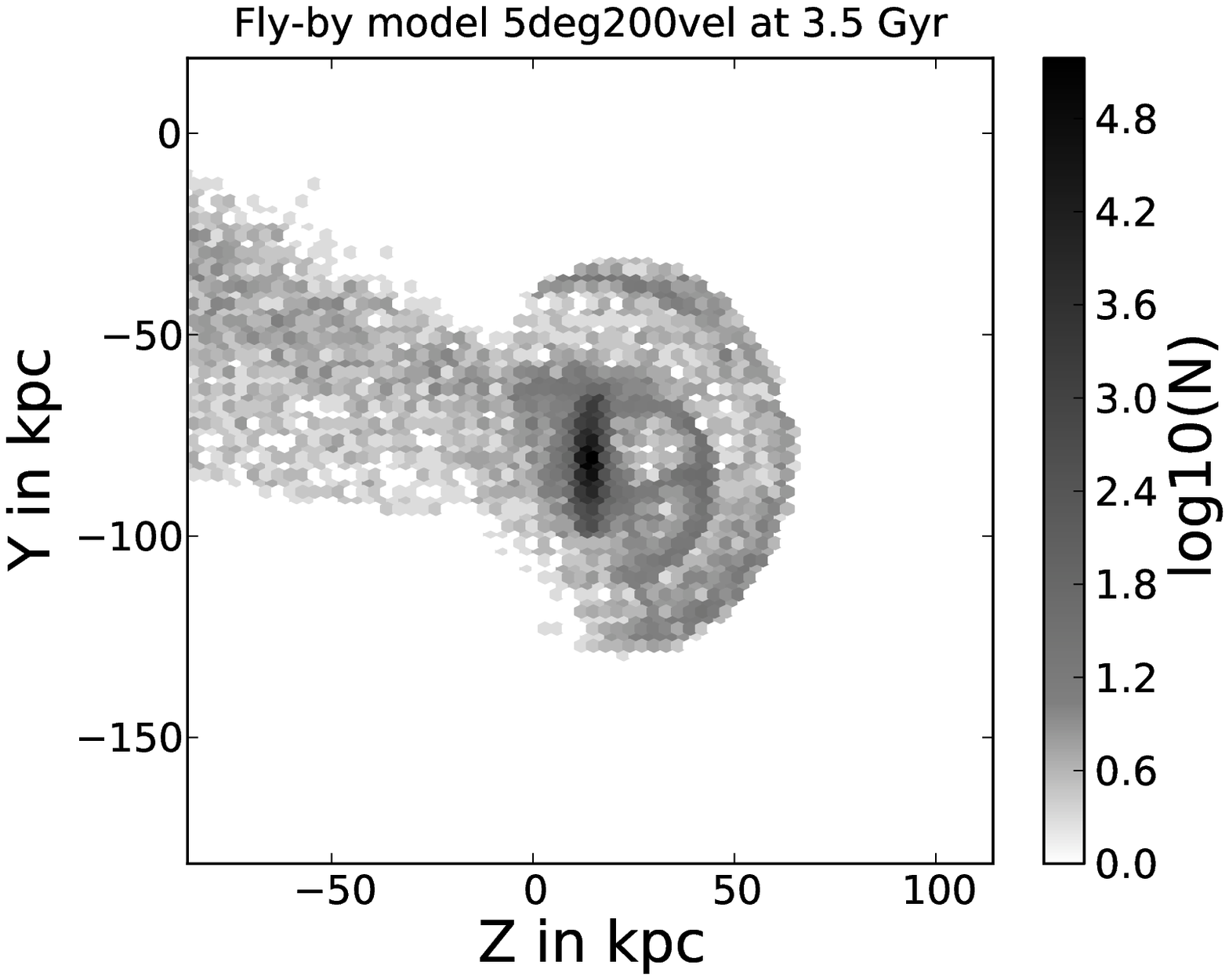}
 \caption{Time evolution of a projection onto the z-y-plane of the particle positions for model 5deg200vel, a typical fly-by. The particle numbers are logarithmically binned. Both plots are centred on the target galaxy. In the left panel the situation at 2.5 Gyr is shown, particles of the infalling galaxy (outside the upper left corner) stream towards the target galaxy (first phase). They pass the target's centre in the direction of more negative y-axis, defining their orbit to be retrograde. The right panel shows the situation in the second phase at 3.5 Gyr. Now the tidal tail has swept over the centre of the target galaxy and the particles still streaming towards the target pass it on the side of less negative y-axis. They end up on counter-rotating orbits compared to the particles of the first phase. Thus, particles on prograde orbits are formed after the retrograde ones in fly-by encounters.}
 \label{5deg200vel}
\end{figure*}

Via the tidal tail between the two galaxies, particles of the infalling galaxy's disc become stripped from their original host and stream towards the target galaxy, around whose centre they orbit from now on. The particles that are stripped in this first phase follow retrograde orbits. The second phase begins once the tidal tail, dragged along with the drifting away infalling galaxy, sweeps over the target galaxy's centre at about 2.5 Gyr. The now infalling particles pass the centre on the opposite side, resulting in a counter-rotating orbit, making these particles the prograde ones. This evolution is shown in two representative snapshots in Fig. \ref{5deg200vel}. Because they fall in from further away -- the distance between the galaxies has increased during phase one -- it can qualitatively be expected that prograde particles have higher orbital eccentricities and higher apocentre distances on average.

\subsection{Analysis: snapshots used, spherical coordinate system, choice of analysed particles}
It is the key issue of this work to analyse the movement of particles around the target galaxy. As an analysis of the orbits of thousands of particles is not feasible, a simple and easy to visualise way to trace the direction of the orbiting particles is to look at their orbital angular momentum vectors $\mathbf{L}$. Prograde and retrograde particles should show up as two separate populations in a plot of orbital angular momenta directions. That way the particles can be identified, allowing to compare other properties of the two populations, finding differences that might be observationally investigated. This makes the approach testable in nature.

The analysis of the orbits of the infalling galaxy's particles caught by the target starts at 3 Gyr into the computation. Determinations of the angular momenta are carried out for every 0.5 Gyr, until the end of the integrations at 10 Gyr. This results in 15 analysed steps per model integration. To determine the eccentricities, every snapshot from the above mentioned start is used, resulting in a temporal resolution of the particles paths of 25 Myr.

The orbital angular momenta are determined with respect to the centre of density of the target galaxy in case of fly-by interactions. For mergers, the equal-weighted mean of the infalling- and the target-galaxy centres of density is used.
To display the the orbital angular momentum vectors $\mathbf{L}$, a spherical coordinate system is set as follows: the angle $\theta$ runs from 0 (positive z-axis) to 180 degrees (negative z-axis). The angle $\phi$ gives the direction around this axis in a range from 0 to 360 degrees. $\phi = 0^{\circ}$ implies $\mathbf{L}$ has a negative y-axis component only and no x-component, whereas $\phi = 90^{\circ}$ has no y- but solely a positive x-axis component.
Thus, $\theta = 0^{\circ}$ points into the same direction as the spin angular momentum vector of the target galaxy's disc rotation. Particles exactly prograde with respect to the motion of the infalling galaxy will have $\mathbf{L}$ pointing along the positive x-axis, thus to $(\theta, \phi) = (90, 90)$. Consequently, retrograde particles have $\mathbf{L}$ pointing to $(\theta, \phi) = (90, 270)$.

\subsection{Orbital angular momenta, particle numbers and ratios}

\begin{figure*}
\centering
\includegraphics[width=88mm]{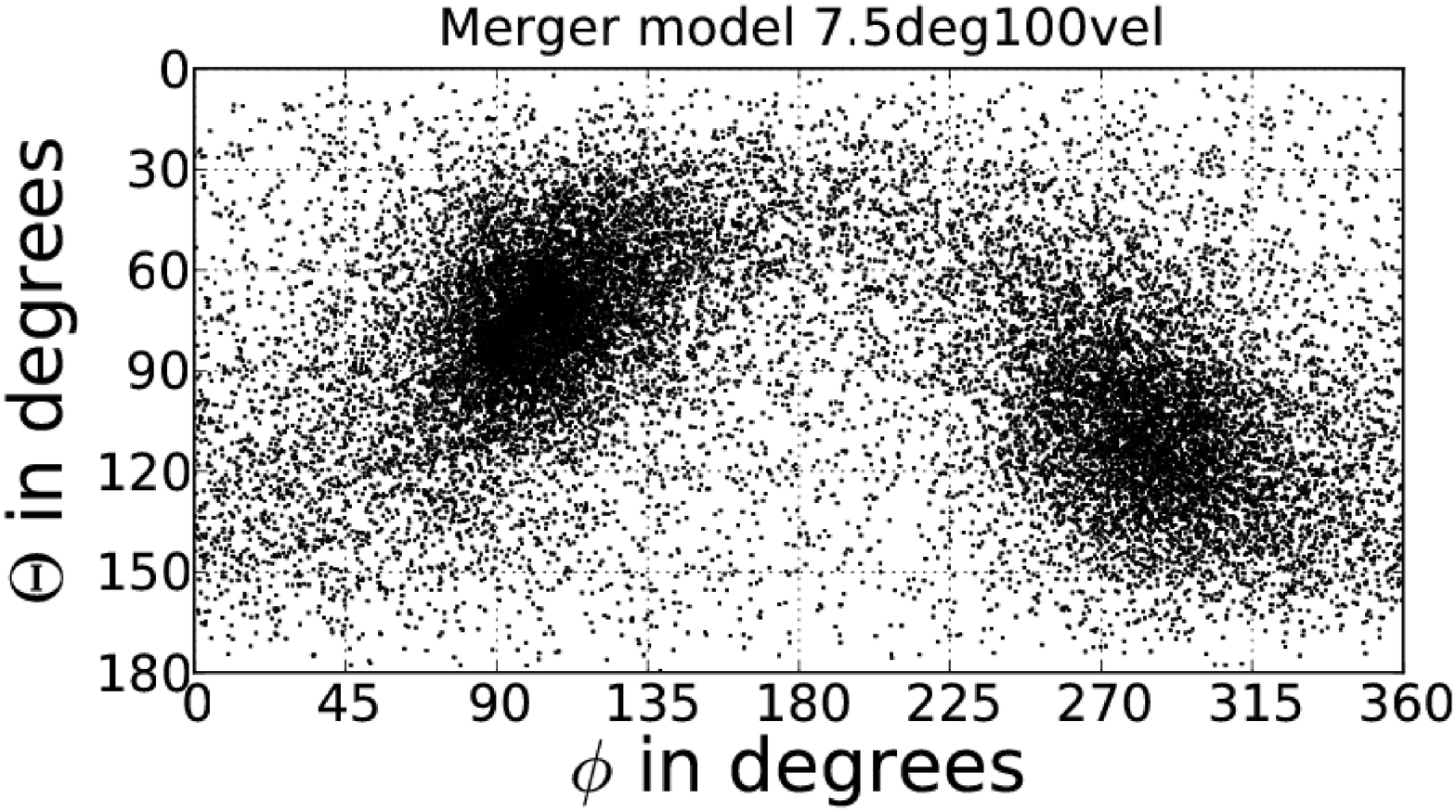}
\includegraphics[width=88mm]{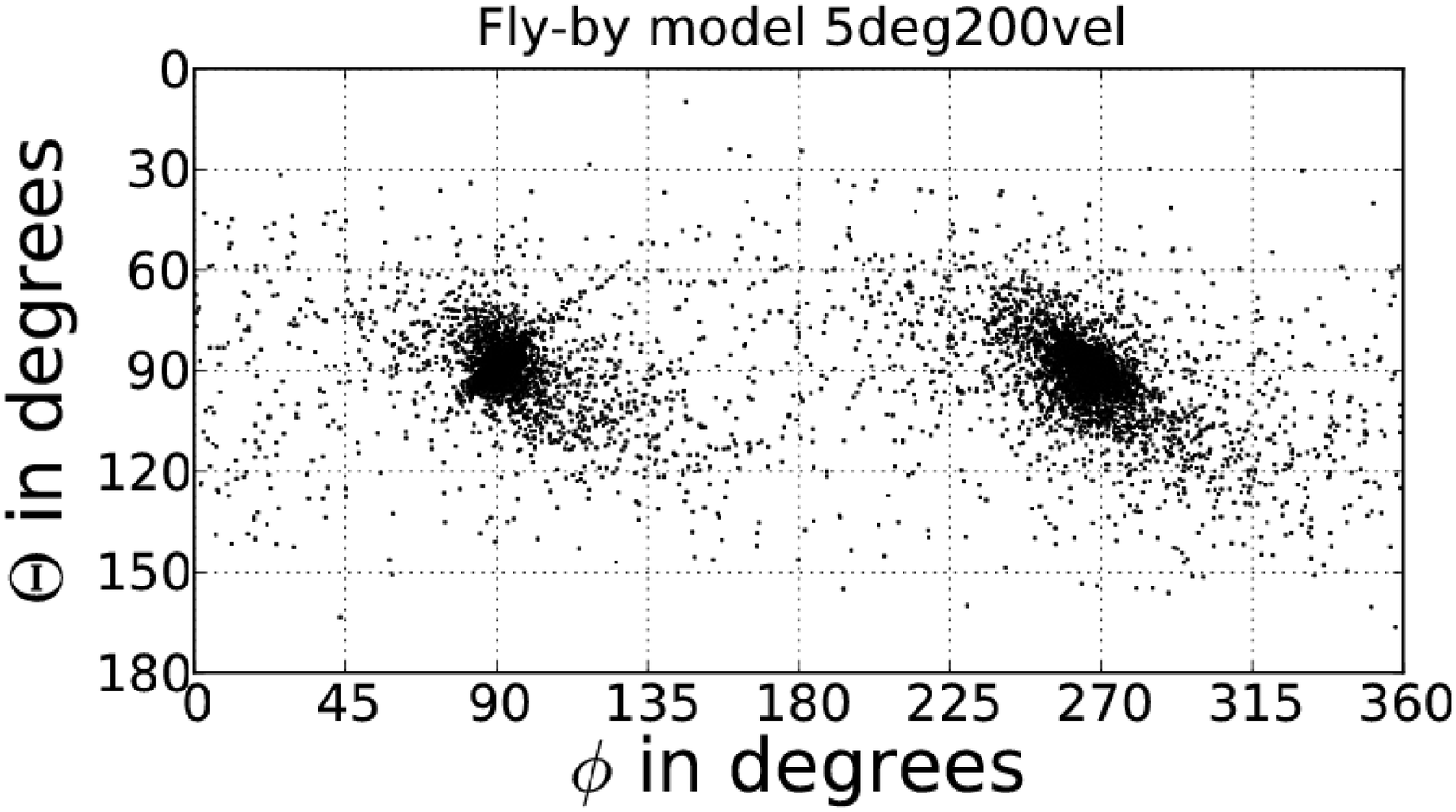}
 \caption{Directions of the orbital angular momenta of particles of the infalling galaxy for merger model 7.5deg100vel (left) and fly-by model 5deg200vel (right) at the final snapshot after 10 Gyr of calculation. Particles which have prograde motions have $0^\circ < \phi < 180^\circ$, those with retrograde motions have $180^\circ < \phi < 360^\circ$. In both cases two distinct and significant populations, one pro- and one retrograde, show up. $\theta = 0^{\circ}$ points towards the target galaxy spin direction (''north'').
 Merger: the 20000 plotted particles were randomly chosen from all $10^5$\ particles with radial distances $0~\rm{kpc} \leq r \leq 170~\rm{kpc}$\ from the centre of the merged galaxies. They show a wide distribution, characteristic for the more chaotic motions after the merger.  It can be seen that the bulk of pro- and retrograde particles have precessed away from the expected positions at $\theta = 90^\circ$, $\phi = 90^\circ$\ or $\phi = 270^\circ$\ respectively, but still lie within the 60--degree circles-of-acceptance (COA) for mergers.
 Fly-by: the plot includes all particles which have radial distances of less than 400 kpc from the target galaxy. Both the particles with prograde and the ones with retrograde motion remain close to the expected orbital angular momenta. They lie well within the 30--degree COAs for fly-bys. Compare to Fig. \ref{lplotMW} which plots the MW satellite galaxy distribution in a similar coordinate system.}
 \label{lplotrun}
\end{figure*}

\begin{figure*}
\centering
 \includegraphics[width=88mm]{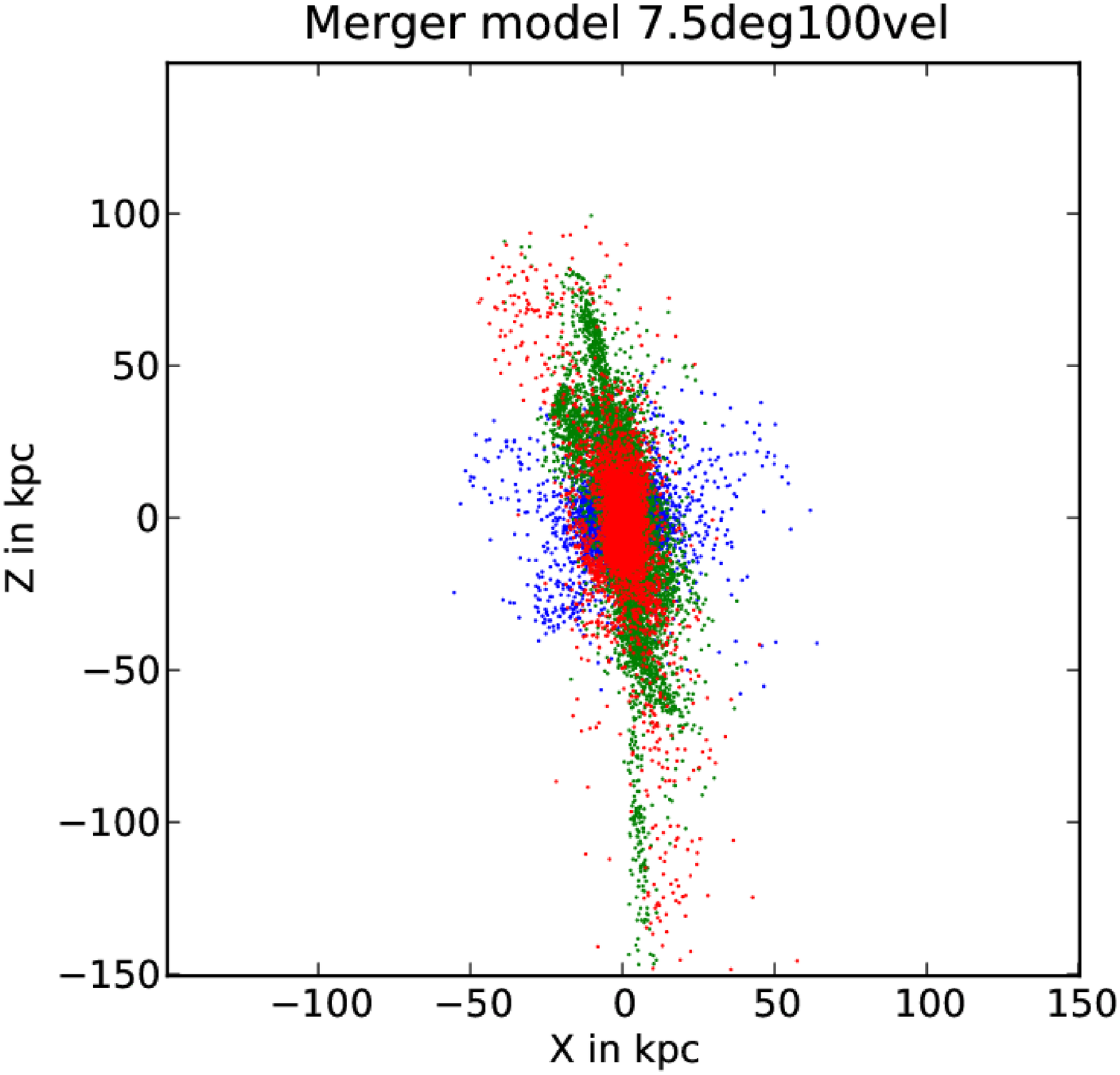}
 \includegraphics[width=88mm]{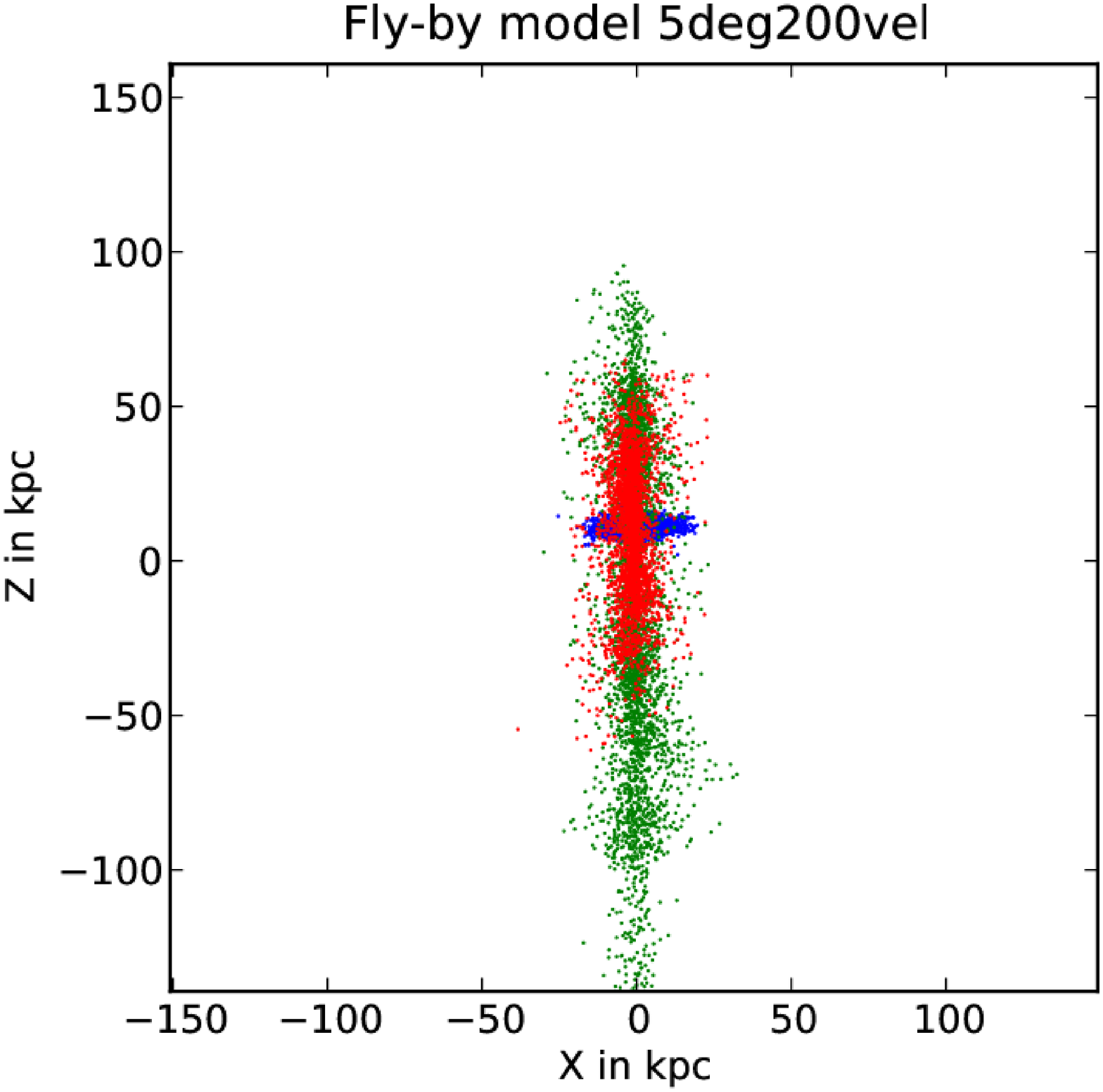}
 \caption{Edge-on views of the tidal debris discs in the merger model 7.5deg100vel (left panel) and the fly-by model 5deg200vel (right panel) at the end of the calculations. Retrograde particles are plotted in red onto prograde particles (green), both are plotted on top of 20000 particles arbitrarily selected from the target galaxy (blue). In the merger case, only the first 20000 particles of the pro- and retrograde population are plotted. The tidal debris form a disc around the central galaxy which is oriented in the plane of the interaction. Note that the thickness of the tidal debris distribution is less than the thickness of the MW DoS \citep{metz07,kroupa10}. In a less-ideal interaction geometry (e.g. when the infalling disc or the galaxy-orbit is not perfectly perpendicular to the target disc) and through subsequent dynamical evolution, the debris-disc can be expected to be thicker. See the on-line edition of the article for a colour version of this figure.}
 \label{edgeonviews}
\end{figure*}

The final distribution of orbital angular momenta distributions for the merger model 7.5deg100vel is plotted in the left panel of Fig. \ref{lplotrun}. For the sake of clarity only a subset of 20000 particles is shown. The right panel plots the angular momenta for all particles within 400 kpc of the target galaxy in the fly-by model 5deg200vel. The angular momenta in the merger model show a wider distribution than in the fly-by, but in both cases two populations near the expected positions of pro- and retrograde material can be clearly seen. Until the end of the calculations, the debris is distributed in a disc. This can be seen in Fig. \ref{edgeonviews}, where the tidal debris is plotted in a projection edge-on to the orbital plane of the galaxies.

To identify the pro- and retrograde particles, a circular area (circle of acceptance, COA) around the expected positions at $(\theta, \phi)_{\rm{pro}} = (90, 90)$ for a prograde and $(\theta, \phi)_{\rm{retro}} = (90, 270)$ for a retrograde orientation is looked at. The circular area has a radius of 30$^{\circ}$\ in the case of a fly-by and of 60$^{\circ}$\ in the case of a merger, to account for the stronger changes in angular momenta directions due to precession and torquing during the merger process in the latter case.

To check whether most particles attracted by the target are either in the pro- or in the retrograde COA, the situation at the end of the calculations is looked at closer. For this, all particles with radial distances $r \leq\ 400\ \rm{kpc}$\ from the target at the end of the calculations (after 10 Gyr) are counted. For the merger model 7.5deg100vel, $2.1 \times 10^5$\ pro- and $1.6 \times 10^5$\  retrograde particles are counted in the respective circles of acceptance. Relative to the total number of $5.0 \times 10^5$\ particles considered, material outside the COAs accounts for merely 26 per cent. For the fly-by model 5deg200vel, 4500 pro- and 3000 retrograde particles are counted from a total of 9100 particles within 400 kpc of the centre of density of the target galaxy. Thus, only 17 per cent are neither in the pro- nor in the retrograde COA.

\begin{figure*}
\centering
 \resizebox{\hsize}{!}{\includegraphics{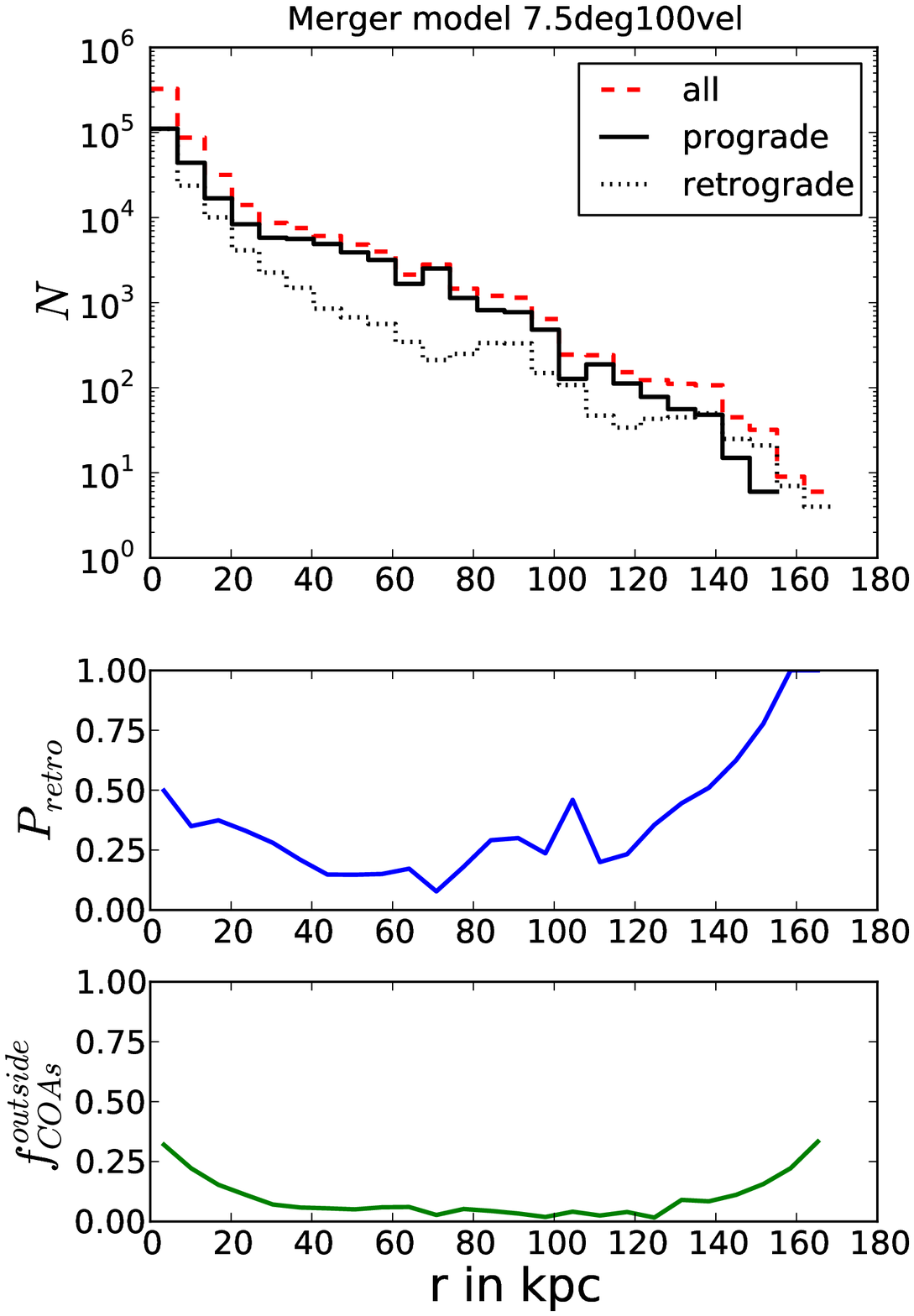} \includegraphics{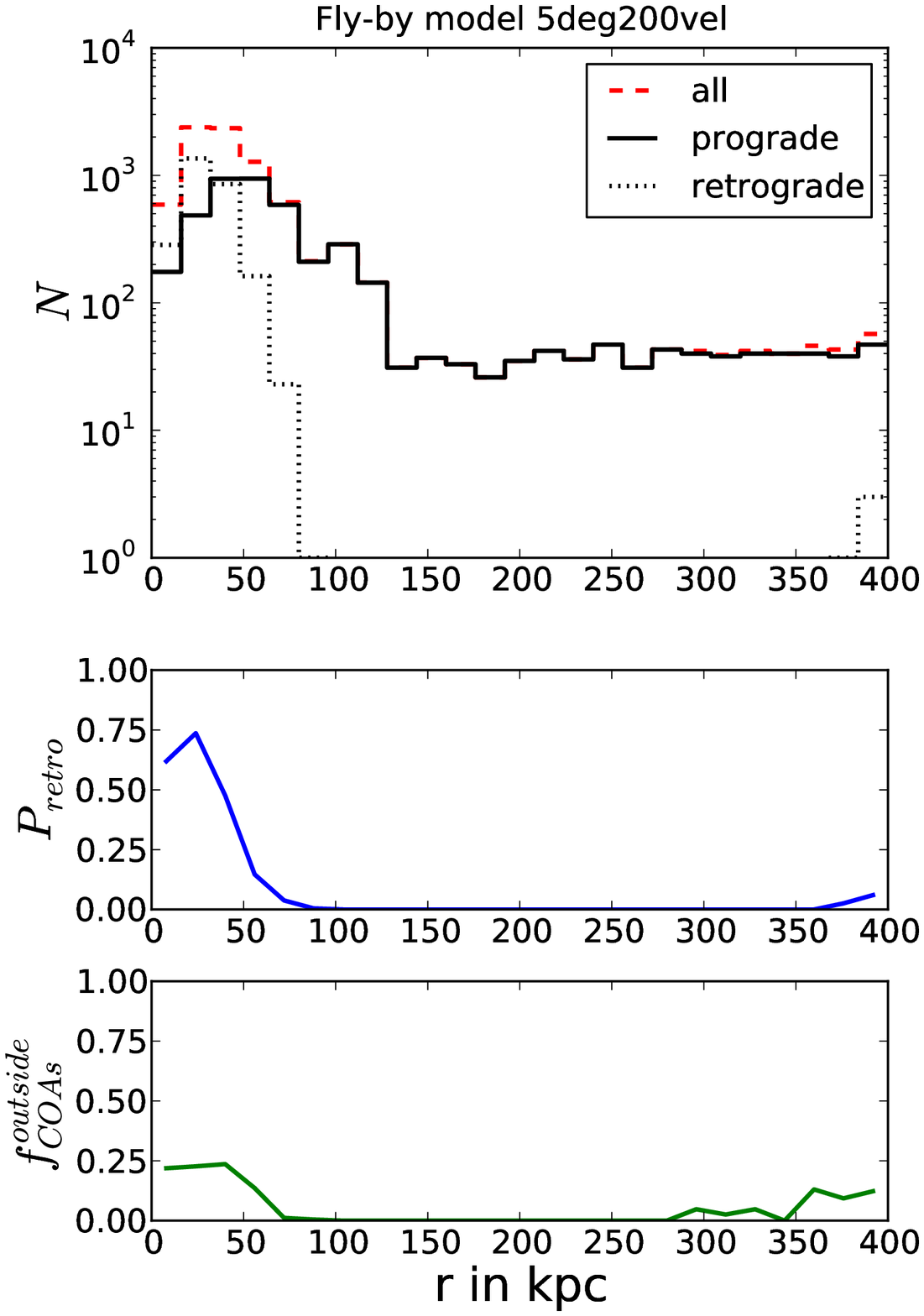}}
 \caption{
 The radial distribution of particles in the merger model 7.5deg100vel (left) and fly-by model 5deg200vel (right) at the final snapshot. The top panels give the number of particles in each bin, the 25 bins are plotted logarithmically to account for the vast range in numbers. They have a width of 6.8 kpc in the merger and 16 kpc in the fly-by case. Particles counted as prograde and retrograde are those within the respective COAs of 60$^{\circ}$\ (merger) and 30$^{\circ}$\ (fly-by) radius. The middle panel shows $P_{\rm{retro}}$, the fraction of retrograde material. The bottom panel plots $f^{\rm{outside}}_{\rm{COAs}}$, the fraction of all particles that are neither in the pro- nor in the retrograde COA.
 Merger: both pro- and retrograde material exists out to a maximum distance of about 160 kpc. $P_{\rm{retro}}$ is highest in the centre and at the maximum distance, varying between approximately 20 and 40 per cent. $f^{\rm{outside}}_{\rm{COAs}}$ is highest in the central region, where the most chaotic motions are expected, but particles outside of the COAs are a minority at all distances.
 Fly-by: while retrograde particles are strongly centred and do not exist beyond about 80 kpc, particles on prograde orbits can be found all the way between the two galaxies. Thus, a fly-by can in principle produce prograde tidal debris at all distances between the interacting galaxies. The plot of $P_{\rm{retro}}$\ shows that only in the inner region retrograde material dominates. As it is formed early, when the two galaxies are still closer to each other, it has lower average apocentres. The lower panel of $f^{\rm{outside}}_{\rm{COAs}}$\ shows that the majority of all material is found within the COAs, as supported by the strong concentration of the particles in Fig. \ref{lplotrun} (right panel). The maximum is found in the inner region, where about 25 per cent of the particles are outside the COAs.
See the on-line edition of the article for a colour version of this figure.
}
 \label{raddistr}
\end{figure*}

Figure \ref{raddistr} shows the number of particles as a function of radial distance from the target galaxy at the end of the integrations for both models.
The relative amount of retrograde particles $P_{\rm{retro}}$\ is expressed as the fraction of the number of particles ${N_{\rm{retro}}}$\ that are found to be retrograde compared to the sum of pro- (${N_{\rm{pro}}}$) and retrograde particles,
\begin{equation}
P_{\rm{retro}} := \frac{N_{\rm{retro}}}{N_{\rm{pro}} + N_{\rm{retro}}} .
\label{PretroDef}
\end{equation}

The fraction of particles that are outside of the COAs is given relative to the total number of particles $N_{\rm{all}}$\ as 
\[
f^{\rm{outside}}_{\rm{COAs}} := \frac{N_{\rm{all}} - \left(N_{\rm{pro}} + N_{\rm{retro}} \right)}{N_{\rm{all}}}.
\]
It is highest near the centre for both models. Low apocentre particles will later be excluded as they suffer most from interactions in the central part of the target galaxy.

Particles with distances up to 160 kpc exist in the merger model. The majority of the infalling galaxy has merged with the target and is thus found at the centre.
$P_{\rm{retro}}$\ varies with radius, starting with an equal distribution at low radial distances, dropping below 0.25 for intermediate distances and rising again to a maximum of 1.0 at the highest distance of 160 kpc. For radial distances larger than about 30 kpc $f^{\rm{outside}}_{\rm{COAs}}$ is less than 10 per cent. It rises again at the outermost bins, where the total particle numbers are of the order of tens only and velocities are slowest.

In contrast, in the fly-by model 5deg200vel, most particles are within 130 kpc, but there is a constant value of approximately 30--40 particles per 16-kpc-bin at larger distances. Particles bridge the whole gap between target and infalling galaxy, which have a separation of 720 kpc by the end of the calculation. The transition from retrograde particles produced in phase one of the encounter to prograde particles from phase two can be clearly seen. Outside of 80 kpc, no retrograde particles are found. $P_{\rm{retro}}$\ consequently drops to zero, but has a maximum of 0.75 in the inner 50 kpc. As in the merger case, the vast majority of particles is inside the pro- or retrograde COA and $f^{\rm{outside}}_{\rm{COAs}}$ is again highest in the central region, but still reasonably low at less than 25 per cent.

To be accepted for the detailed analysis, particles have to approach both an apo- and a pericentre within the analysis-time between 3 and 10 Gyr. Those particles which are found within the respective (pro- and retrograde) COA in at least 12 out of the 15 analysed snapshots are then identified as prograde and retrograde particles\footnote{The direction of $\mathbf{L}$ might change in the course of time, especially particles of highly eccentric trajectories can be strongly perturbed by local gradients in the potential. Furthermore, as the potential is not spherically symmetric, the path of the particles might be changed slightly on their orbit by precession and torques, changing their orbital angular momentum vectors and moving them out of the COA. Thus, demanding the particles to appear inside the COA for all snapshots was deemed to be unrealistic.}. Finally, the apocentre distance has to be at least 30 kpc. Thus the numbers of included particles are lower limits only. The benefits, however, outweight this: all accepted particles are on bound orbits around the target galaxy and can be assigned an orbital eccentricity (see Sect. \ref{eccensect} for details). 

In the case of merger model 7.5deg100vel, there are $36800$ prograde and $6000$\ retrograde particles with apocentre distances of at least 30 kpc. The fraction of retrograde particles with respect to the sum of pro- and retrograde ones is $P_{\rm{retro}} = 0.14$. Particles with apocentres larger than 60 kpc are considered to be `far'. For these, the numbers are: $15000$ prograde and $2250$ retrograde particles, the fraction of retrograde particles with apocentres above 60 kpc is $P_{\rm{retro}}^{\rm{far}} = 0.13$.

For fly-by model 5deg200vel, the numbers are $3450$\ prograde and $2300$\ retrograde particles with apocentre distances $\geq 30$~kpc. In this case $P_{\rm{retro}}$ is $0.40$. The numbers for `far' particles in this case are: $1920$ pro- and $170$ retrograde, leading to a lower retrograde fraction $P_{\rm{retro}}^{\rm{far}} = 0.08$, as expected from the two-phase origin in the fly-by.

In the two models, significant numbers of particles orbiting in both directions are found. This is, as far as the authors of the present contribution know, the first time such an analysis was performed, showing that it is indeed possible to form particles on opposed orbits in both, mergers and close passings of two galaxies. This shows that, as tidal dwarf satellites form from such material and thus share the same phase-space region, co- and counter-rotating orbits of TDGs can in principle be created in a single galactic encounter.

\subsection{Eccentricities and apocentre distances}
\label{eccensect}

\begin{figure*}
\centering
 \includegraphics[width=88mm]{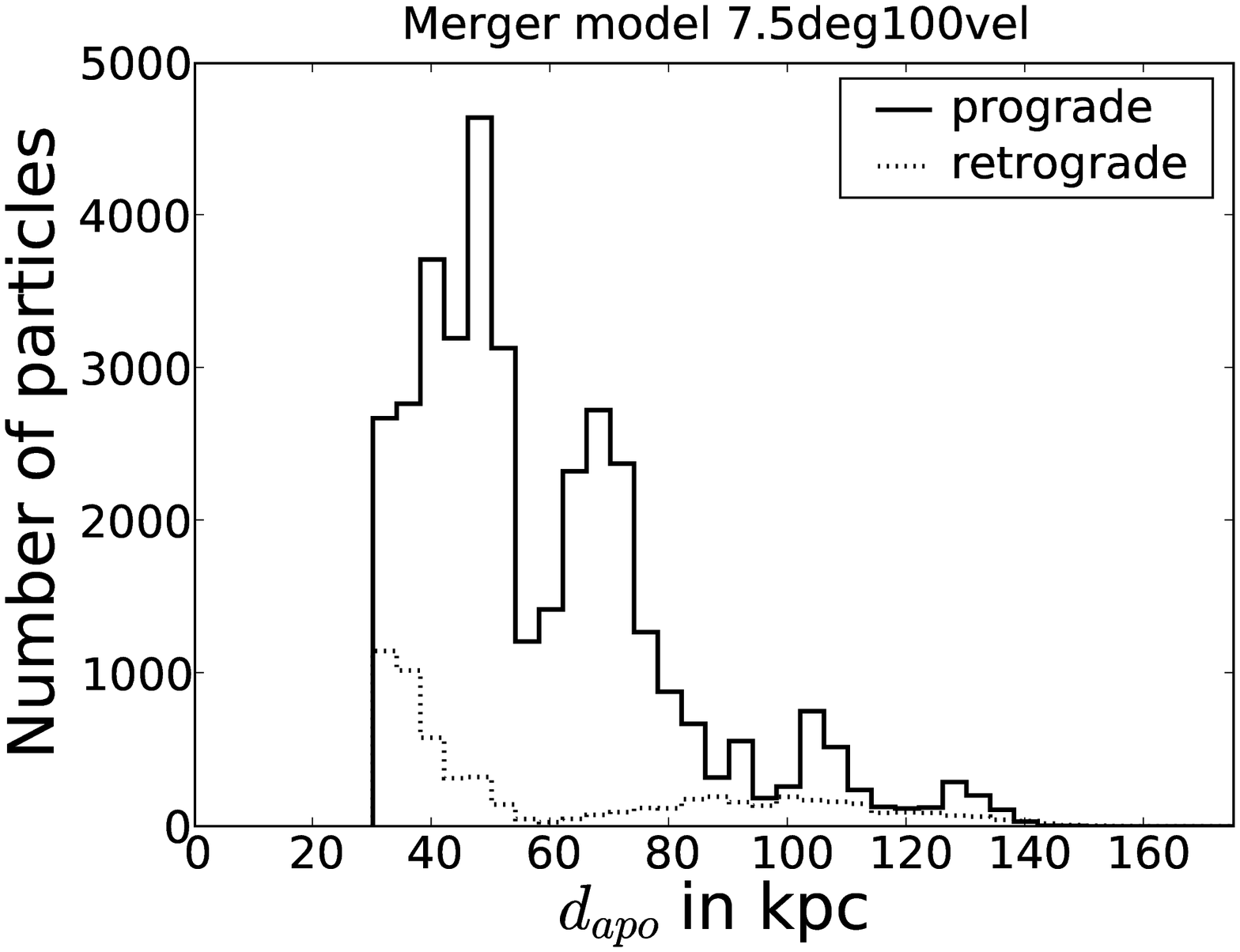}
 \includegraphics[width=88mm]{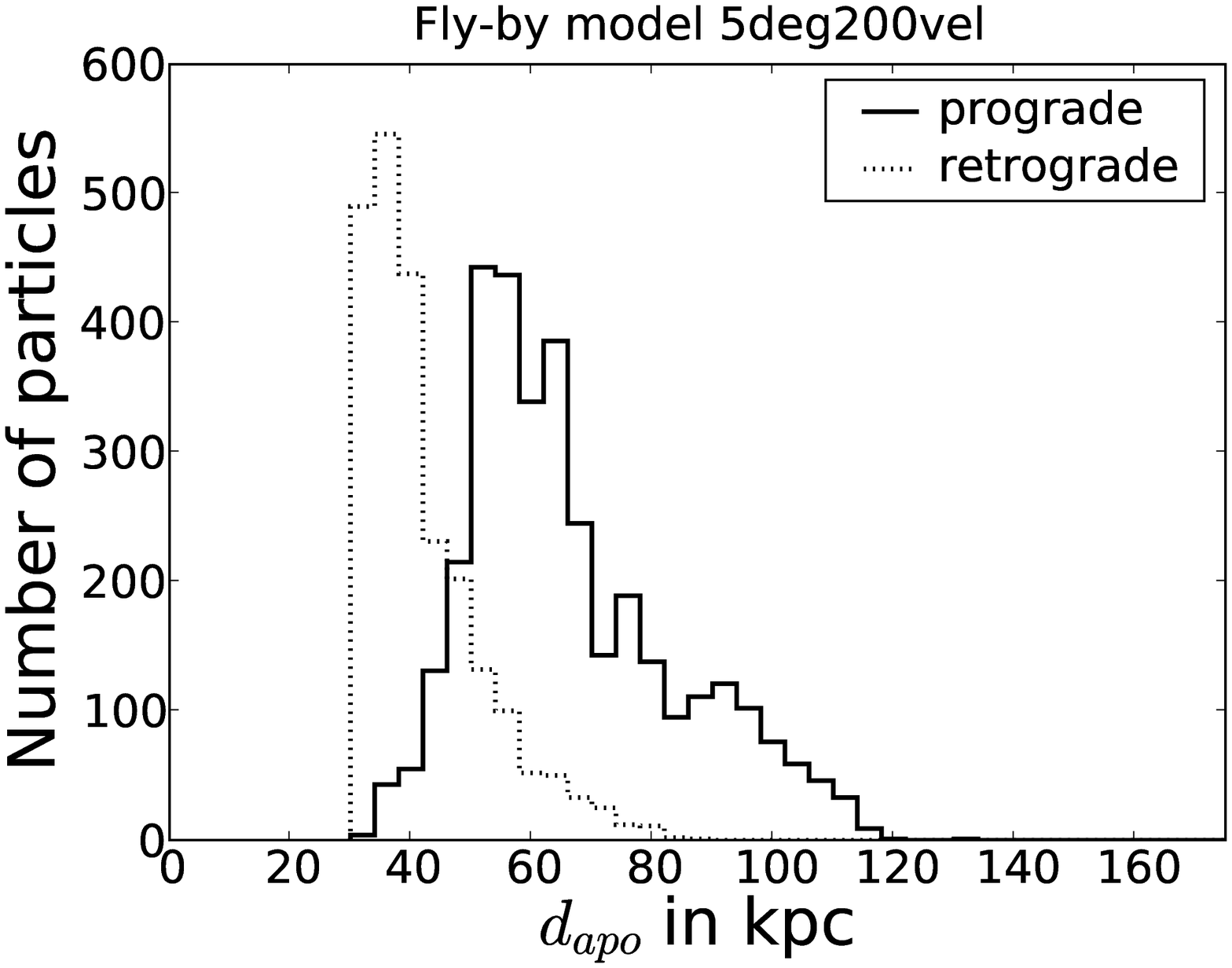}
 \caption{Apocentre distances for merger model 7.5deg100vel (left) and fly-by model 5deg200vel (right). Plotted are the numbers of particles in bins of $d_{\rm{apo}}$\ with width 4.0 kpc, starting at 30 kpc. In case of a merger, particles on pro- and retrograde orbits (within the respective COAs ) can be found at all apocentre distances from 30 up to about 140 kpc. In the fly-by case, in contrast, pro- and retrograde particles have distinct apocentre properties. The retrograde particles have lower apocentres (up to about 80 kpc) while the prograde particles have maximum apocentre distances of 120 kpc, but only few of them are found with $d_{\rm{apo}}$\ below 50 kpc. This hints at the two-phase origin typical for fly-by interactions. Particles with retrograde orbits are formed first, while the passing galaxies are still close to each other, and thus have lower average apocentre distances.}
 \label{Apohistrun}
\end{figure*}

The particles identified a posteriori as pro- and retrograde ones are now analysed in detail. For this analysis, all particles that are found within the afore mentioned 60-- or 30--degree circles of acceptance in at least 12 analysed steps are taken into account. This number of steps results in particle numbers closest to the ones derived from averaging over the results with minimum numbers of 10 to 15 demanded steps.

\label{eccendescription}

To describe the elongation of particle orbits each particle is assigned an orbit 'eccentricity' $e$\ by determining the apocentre-,$d_{\rm{apo}}$, and pericentre distances, $d_{\rm{peri}}$, of its orbit,
 \[
 e = \frac{d_{\rm{apo}} - d_{\rm{peri}}}{d_{\rm{apo}} + d_{\rm{peri}}}.
 \]
To determine $d_{\rm{apo}}$ and $d_{\rm{peri}}$ the particle trajectories, in a coordinate system centred either on the target galaxy's centre of density (for fly-bys) or on the average of the centres of density of target and infalling galaxy (for mergers), is followed starting with the last snapshot and proceeding backwards in time. The pericentre distance $d_{\rm{peri}}$\ is estimated from a circle fitted to the three time-steps ($\pm 25$ Myr) closest to the minimum distance of the particle. This avoids an overestimation of $d_{\rm{peri}}$\ which would result in an underestimation of $e$\ in case of fast particles on highly eccentric orbits which have a coarse trajectory sampling.

In the case of the merger model 7.5deg100vel the orbit eccentricities stretch from low values, 0.5 for pro-, 0.35 for retrograde particles, up to an eccentricity near 1.0, which describes a parabolic orbit. The prograde particles are centred around their mean at $e_{\rm{pro}} = 0.76$. The standard deviation of the distribution has a value of 0.09. The orbit eccentricities of the retrograde particles in contrast are distributed more widely. Their mean value is $e_{\rm{retro}} = 0.76$\ with a standard deviation of 0.18.
Both particle groups have apocentre distributions, plotted in Fig. \ref{Apohistrun}, populated mostly near the minimum accepted distance of 30~kpc, beyond which their numbers drop. Particles of both groups reach out to 140~kpc. There thus is co- and counter-orbiting material at similar radial distances from the target galaxy.
The average eccentricity of the subset of 'far' prograde particles ($d_{\rm{apo}}\ >$\ 60 kpc) is essentially the same as for 'all' prograde particles (0.77), but it is higher (0.94) for 'far' retrograde particles.

The situation is different for the fly-by model 5deg200vel. As expected, the later-infalling prograde particles have high eccentricities, with a minimum value of about 0.65. Their mean value is $e_{\rm{pro}} = 0.82$, around which the distribution is concentrated, the standard deviation is only 0.08. The retrograde particles have a minimum orbit eccentricity of about 0.5, the mean is $e_{\rm{retro}} = 0.80$\ and the standard deviation is 0.11. As in the merger case, with 0.93, the average eccentricity of the subset of 'far' retrograde particles is higher, while there is no significant change in average eccentricity when looking only at 'far' prograde particles.
The apocentre distance distribution of the fly-by model is shown in the right panel of Fig. \ref{Apohistrun}. In this plot, prograde and retrograde particles have clearly distinguishable properties. The prograde orbits have apocentres in the range between the accepted minimum of 30 to approximately 120~kpc, with a maximum near 60~kpc. Note that, as obvious from the right panel of Fig. \ref{raddistr}, prograde particles spread out to over 400 kpc. They are not included in Fig. \ref{Apohistrun} because they are still falling in towards the target galaxy at the end of the calculation, so for them no eccentricity is determined and they thus are excluded from the detailed analysis. In contrast to the prograde particles, the retrograde particles are distributed more closely to the centre of the target galaxy. Retrograde particles of a significant number only appear up to about 80~kpc.

\subsection{Site of origin within the precursor disc}

\begin{figure*}
\centering
 \includegraphics[width=88mm]{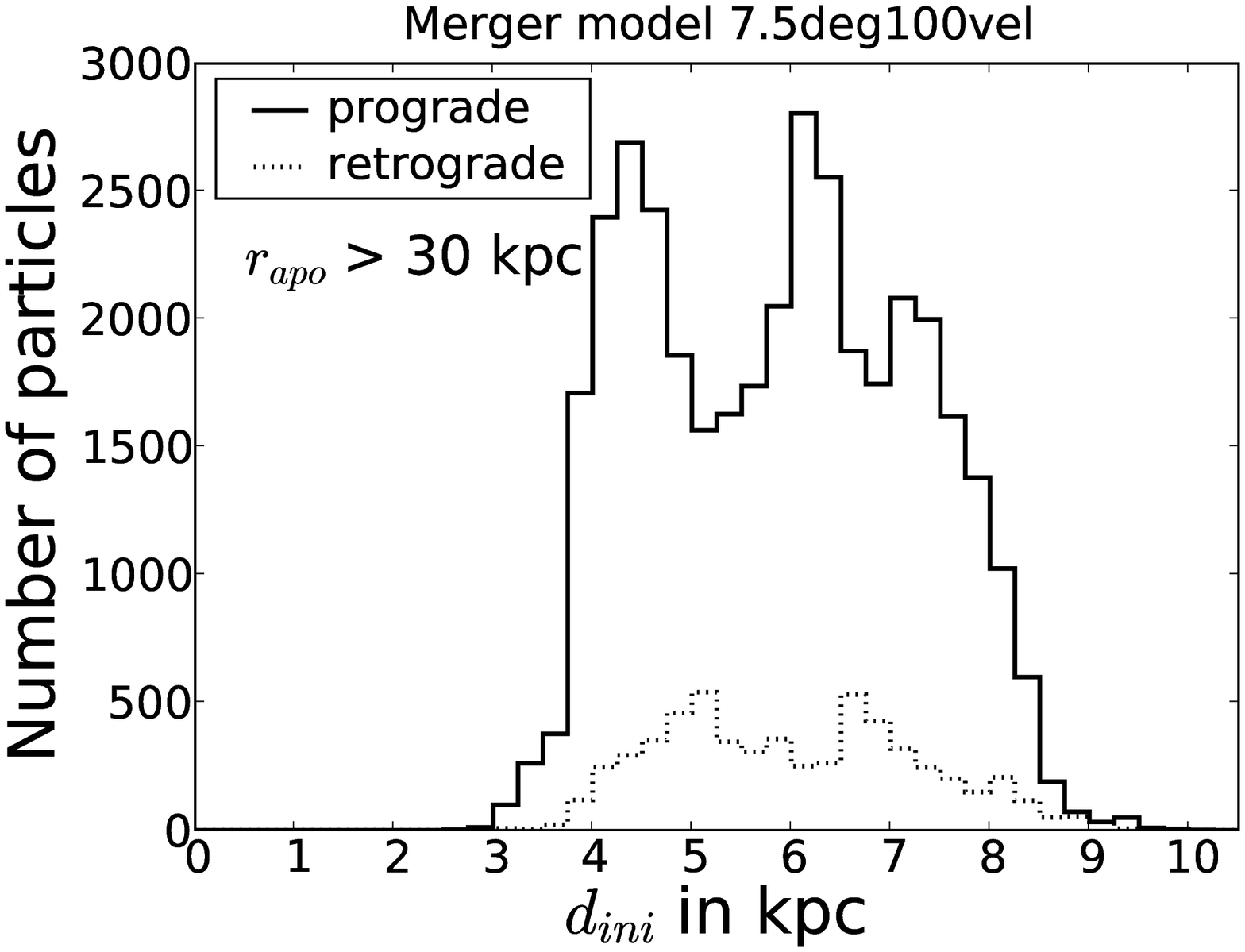}
 \includegraphics[width=88mm]{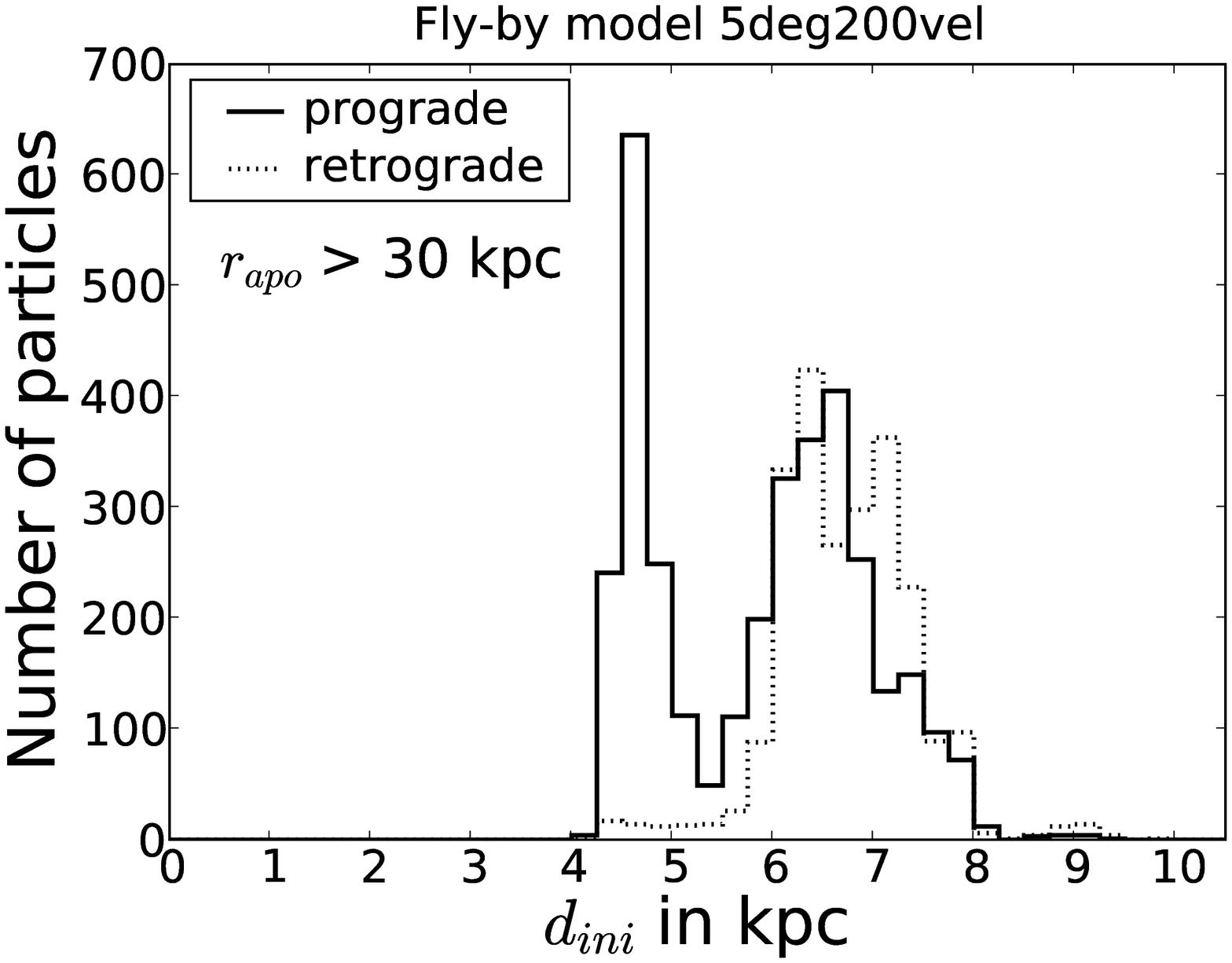}
 \caption{Histogrammes of initial distances $d_{\rm{ini}}$\ within the infalling galaxy for merger model 7.5deg100vel (left) and fly-by model 5deg200vel (right) in bins of 0.25 kpc width. All particles within the respective COAs with apocentres above 30 kpc are included.
 In the merger case, the particles of both populations initially had similar radial distances from the centre of the infalling disc galaxy.
 In the fly-by case, the particles on retrograde orbits had on average higher initial distances from the infalling disc galaxy's centre. Thus, they originate more closer to the rim of the galaxy. There are particles on prograde orbits coming from a similar region, but in additon to those many come from further inside the infalling's disc. Thus, on average the prograde material originates from a more central part of the infalling galaxy.
}
 \label{disthistrun}
\end{figure*}

For the particles found to be pro- and retrograde, the site of origin within the infalling galaxy is determined. This is the average radial position measured from the centre of the infalling galaxy before it interacts with the target galaxy. In Fig. \ref{disthistrun} histogrammes for both models show the distribution of initial distances for the pro- and retrograde particles respectively.

In merger model 7.5deg100vel, particles have initial distances as close as 3 kpc from the centre, the mean values $D^{\rm{all}}$\ for both populations are similar: $D^{\rm{all}}_{\rm{pro}} = 5.9~\rm{kpc}$\ for the prograde particles and $D^{\rm{all}}_{\rm{pro}} = 6.0~\rm{kpc}$\  for the retrogrades. The widths of the distributions are similar, too. This shows that co- and counter-orbiting tidal derbis in this merger come from similar regions in the infalling galaxy.

In fly-by model 5deg200vel the particles originate from a similar region in radial distance, spreading from about 4 kpc to the edge of the initial disc at around 8 kpc\footnote{Some particles have an origin beyond the initial cut-off radius of 8.0 kpc because the disc slightly expands during virialisation.}. However, the distribution within this region is different for pro- and retrograde particles. The former originates from further inside the disc when compared to the retrograde particles, they show a peak at around 4.5~kpc (and a second, lower but wider one at 6.5~kpc) in contrast to the retrograde particles' peak above 6.5 kpc. The mean values of the distributions are $D^{\rm{all}}_{\rm{pro}} = 5.9~\rm{kpc}$\ in case of the prograde particles and $D^{\rm{all}}_{\rm{retro}} = 6.7~\rm{kpc}$\ for the retrograde ones. Thus the mean initial distances differ by about 800~pc.

As visible in Fig. \ref{5deg200vel}, the prograde particles appear later than the retrograde ones. They appear when the tidal tail of the infalling galaxy sweeps over the centre of the target galaxy. The outer part of the tidal tail, near it's tip, consists of particles that are more weakly bound to the infalling galaxy and thus originate from the rim of the disc. This part of the tidal tail forms retrograde particles. Correspondingly the prograde particles, formed further down the tidal tail, originate from closer to the infalling galaxy's disc centre.

\begin{table}
 \caption{Pro-and retrograde particle properties}
 \label{compiletable}
 \begin{center}
 \begin{tabular}{@{}lcccccccccccc}
  \hline
  Symbol & \multicolumn{2}{c}{all ($d_{\rm{apo}}\ >$\ 30 kpc)} & \multicolumn{2}{c}{far ($d_{\rm{apo}}\ >$\ 60 kpc)} \\
     & prograde & retrograde & prograde & retrograde \\
  \hline
  7.5deg100vel &  &  &  & \\
  (Merger) &  &  &  & \\
  $N$ & $36800$ & $6000$ & $15000$ & $2250$\\
  $f_N$~[\%] & 7.36 & 1.20 & 3.0 & 0.45 \\
  $M$~[$10^{8}$~$\rm{M}_{\sun}$] & $5.89$ & $0.96$ & $2.40$ & $0.36$\\
  $P_{\rm{retro}}$ & \multicolumn{2}{c}{$0.14$} & \multicolumn{2}{c}{$0.13$} \\
  $e$ & 0.76 & 0.76 & 0.77 & 0.94\\
  $\Delta e$ & 0.09 & 0.18 & 0.08 & 0.02\\
  $<d_{\rm{ini}}>$~[kpc] & 5.89 & 6.02 & 6.75 & 5.38\\
  $\Delta <d_{\rm{ini}}>$~[kpc] & 1.33 & 1.27 & 0.95 & 0.86\\
  $d_{\rm{ini}}^{\rm{med}}$[kpc] & 5.96 & 5.92 & 6.85 & 5.20\\
  \hline
  5deg200vel &  &  &  & \\
  (Fly-by) &  &  &  & \\
  $N$ & $3450$ & $2300$ & $1920$ & $170$\\
  $f_N$~[\%] & 0.69 & 0.46 & 0.384 & 0.034 \\
  $M$~[$10^{7}$~$\rm{M}_{\sun}$] & $5.52$ & $3.68$ & $3.07$ & $0.27$ \\
  $P_{\rm{retro}}$ & \multicolumn{2}{c}{$0.40$} & \multicolumn{2}{c}{$0.08$} \\
  $e$ & 0.82 & 0.80 & 0.80 & 0.93\\
  $\Delta e$ & 0.08 & 0.11 & 0.08 & 0.03\\
  $<d_{\rm{ini}}>$~[kpc] & 5.89 & 6.73 & 6.07 & 7.23\\
  $\Delta <d_{\rm{ini}}>$~[kpc] & 1.05 & 0.68 & 1.07 & 0.71\\
  $d_{\rm{ini}}^{\rm{med}}$~[kpc] & 6.02 & 6.68 & 6.23 & 7.30\\
  \hline
 \end{tabular}
 \end{center}

 \small \smallskip
 $N$\ is the number of particles with measured apocentre of at least 30 (all) and 60 (far) kpc, corresponding to a total mass $M$ and $f_N$, the fraction of total particle number of the infalling galaxy.
 The mean orbit eccentricity is denoted with $e$, having a standard deviation of $\Delta e$. Similarly, $<d_{\rm{ini}}>$ stands for the mean initial radial distance within the infalling galaxy, given in kpc. $\Delta <d_{\rm{ini}}>$ is the standard deviation and $d_{\rm{ini}}^{\rm{med}}$ the median value. The values were determined by demanding the particles to be inside the circles of acceptance (COA) for at least 12 of the 15 analysed steps.
\end{table}

\section{Exploration of the parameter space}
\label{parascansect}
Varying the initial relative velocities and their directions between the two galaxies produces a wide range in all analysed properties. For fly-by models, 16 calculations with four different initial velocities (1.8, 2.0, 2.2 and 2.4 times the parabolic velocity $v_{\rm{parab}}$) and four different initial velocity directions (5.0, 6.0, 7.0 and 8.0 degrees from the line connecting the two galaxies, which is the z-axis) are performed. Mergers are set up with both pro- and retrograde oriented infalling galaxy discs, in total there are 18 for each. They have initial velocities of 0.5, 1.0 and 1.5 times $v_{\rm{parab}}$ and directions of 0 (head-on collision), 2.5, 5.0, 7.5, 10.0, 15.0 and 20.0 degrees from the z-axis. The fastest models only include directions up to 7.5$^{\circ}$. Furthermore, several models with unequal galaxy masses are calculated. Their mass ratio is 4 : 1 for Target : Infalling. 10 fly-by models and 6 merger models of each orientation are analysed.

The values extracted from the calculations (e.g. particle numbers and retrograde-fraction, average initial distances, orbit eccentricities) are compiled in Tables \ref{flybynumbers} to \ref{4to1mergerproparameters} in the Appendix for each model type (equal mass fly-by, both orientations in mergers and the same for the 4-to-1 mass fraction interactions).

Before discussing the results for the different types of interactions in detail, a general remark on the expected radial distributions of tidal material is asked for. A fly-by can, in principle, produce tidal debris at all distances, see Fig. \ref{raddistr}. The tidal bridge between the two departing galaxies becomes stretched and will thin out, but the tidal debris near its centre has a small velocity dispersion and will feel both the attracting gravitational force of the infalling and of the target galaxy. Thus, it might stay near the centre of gravity of the interactions while the two precursor galaxies recede. 
A merger, in contrast, can only produce tidal debris up to distances determined by the energy of the merger process. 

Faster initial velocities lead to later mergers because the galaxies oscillate more strongly before settling to their common centre of mass. Therefore, the analysis starts later for models with high initial velocities. This delay is reflected in the maximum orbital periods, they are shorter, reducing the maximum apocentre the analysis can find.

Before discussing the different interaction types, some general remarks on the typical remnant galaxy shall be made. Its morphology depends on the type of interaction. In the case of equal-mass mergers the discs get destroyed and a spheroidal, bulge-like remnant forms. In contrast to this, 4-to-1 mergers leave the more massive target galaxy disc identifiable, but it heats up and gets tilted. The inclusion of gas dissipative processed would probably help the formation of a spheroidal remnant. The target disc is preserved in fly-by interactions of both mass-ratios, however it is disturbed. It heats up and a bar instability forms, which might result in a pseudobulge-component as observed in the MW \citep{Babusiaux2010}.

The spin angular momentum vector of the target galaxy disc lies within the plane of the interaction in all models. Due to this interaction geometry, no pronounced tidal tails form out of the target galaxy. Therefore, the analysis concentrates on the particles of the infalling galaxy.

\subsection{Equal-mass fly-bys}
\label{equalmassflybysection}
Particles in both orbital directions are found in all analysed models. The calculated fly-by models all show the same simple two-phase behaviour: once the two galaxies pass, a tidal tail evolves from the infalling galaxy. Falling towards the target galaxy, the tail first forms retrograde particles (phase one). After some time, as the galaxies depart further, the tidal tail sweeps over the centre of the target galaxy. This changes the orbital direction of the particles falling onto the target afterwards, making them prograde (phase two).

The total number of pro- and retrograde particles within the COAs for which apo- and pericentre were determined, $N_{\rm{all}} = N_{\rm{pro}} + N_{\rm{retro}}$ lies in the range of about 800 to some 6000 particles. Up to 2000 particles reach apocentre distances above 60 kpc, the lowest number found there is about 400.

\begin{figure}
\centering
 \includegraphics[width=88mm]{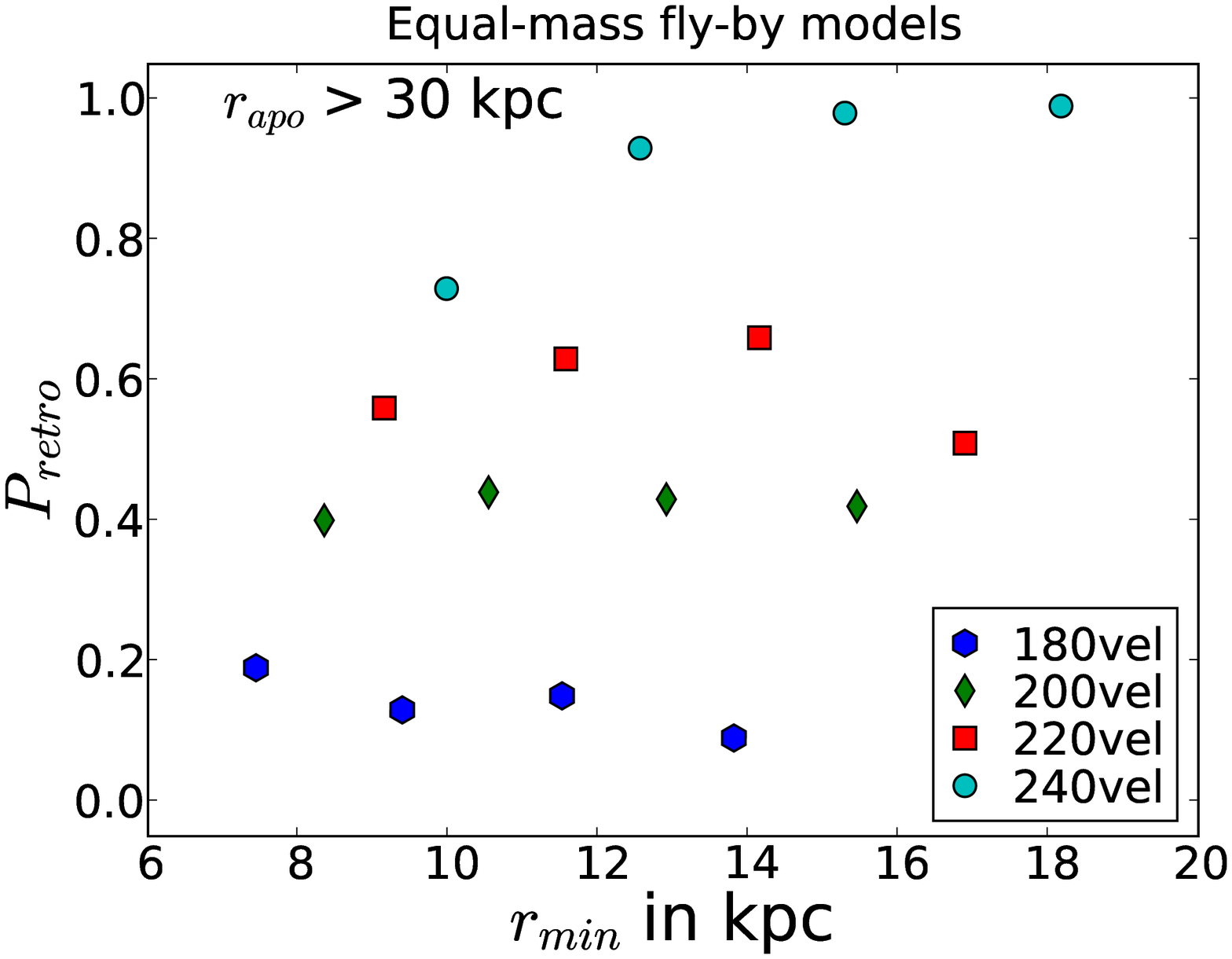}
 \includegraphics[width=88mm]{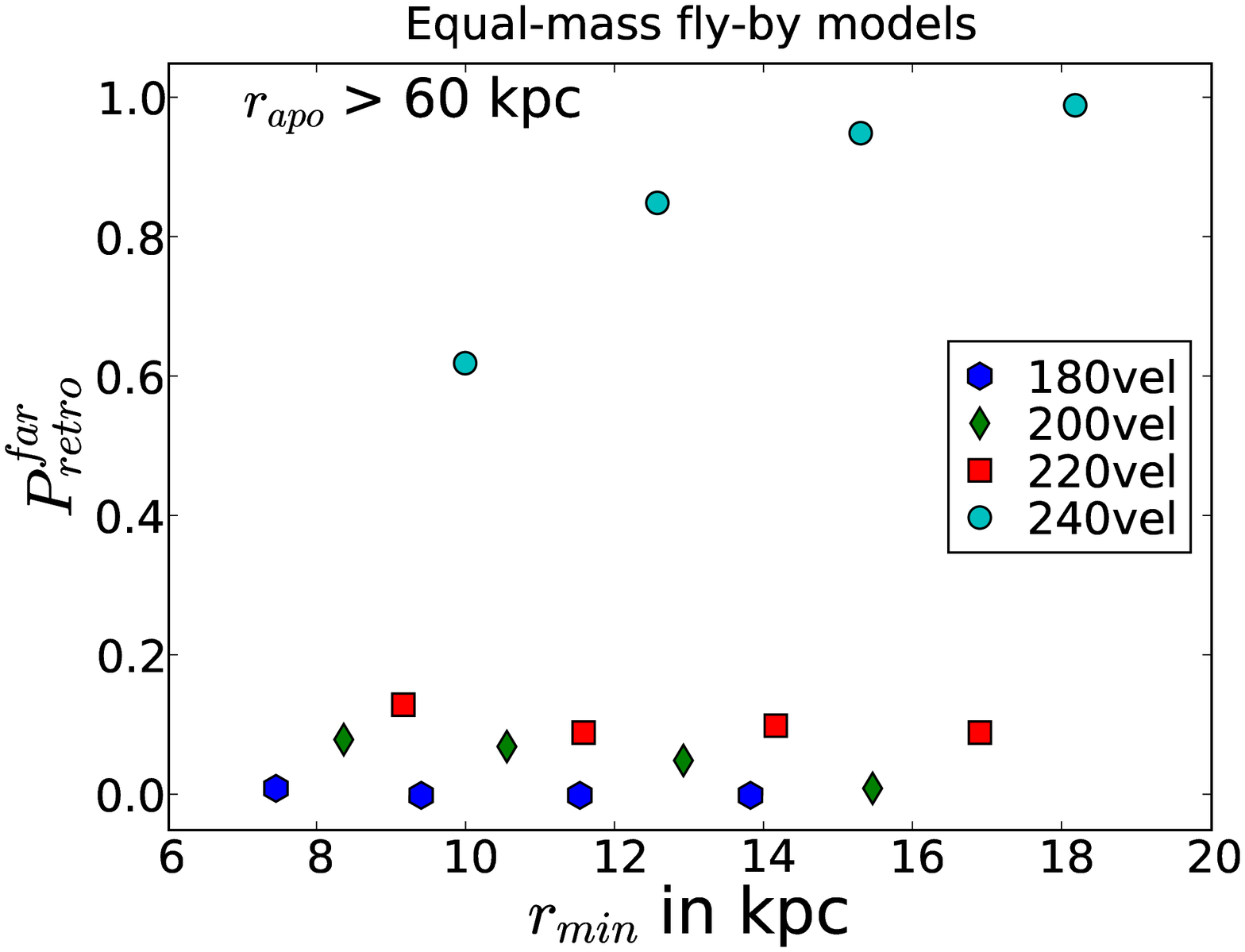}
 \caption{The fraction of particles on retrograde orbits after the fly-by interaction of two equal-mass galaxies. They have apocentres above 30 kpc ($P_{\rm{retro}}$, upper panel), and above 60 kpc ($P_{\rm{retro}}^{\rm{far}}$, lower panel). See Tables \ref{flybynumbers} and \ref{flybyparameters} for the values. The fraction of retrograde material is well defined by the initial velocity only, without much dependence on the minimum distance $r_{\rm{min}}$\ at which the two galaxies pass each other. $P_{\rm{retro}}$ increases with increasing initial velocity, which ranges from 1.8 (`180vel') to 2.4 (`240vel') times the the parabolic velocity. See the on-line edition of the article for a colour version of this figure.}
 \label{flybypretro}
\end{figure}

The fraction of retrograde particles, $P_{\rm{retro}}$ (Eq. \ref{PretroDef}), can be seen in Fig. \ref{flybypretro}. It shows a clear trend: the faster the galaxies pass, the higher is $P_{\rm{retro}}$. In a slow passage, prograde particles dominate, while in the fastest passages calculated, the retrograde particles are the dominating population. This seems to be almost independent of $r_{\rm{min}}$, the minimum distance of the two galaxies during their passage. The $r_{\rm{min}}$-$P_{\rm{retro}}$-plane is quite evenly populated, with the region of small $r_{\rm{min}}$\ and large $P_{\rm{retro}}$\ being empty. $P_{\rm{retro}}$ ranges from 0.09 in model 8deg180vel, the slowest passage with high galaxy-galaxy-distance at pericentre, to 0.99 in model 8deg240vel, the model with the highest initial velocity and farthest pericentre distance between the galaxies.

\begin{figure}
\centering
 \includegraphics[width=88mm]{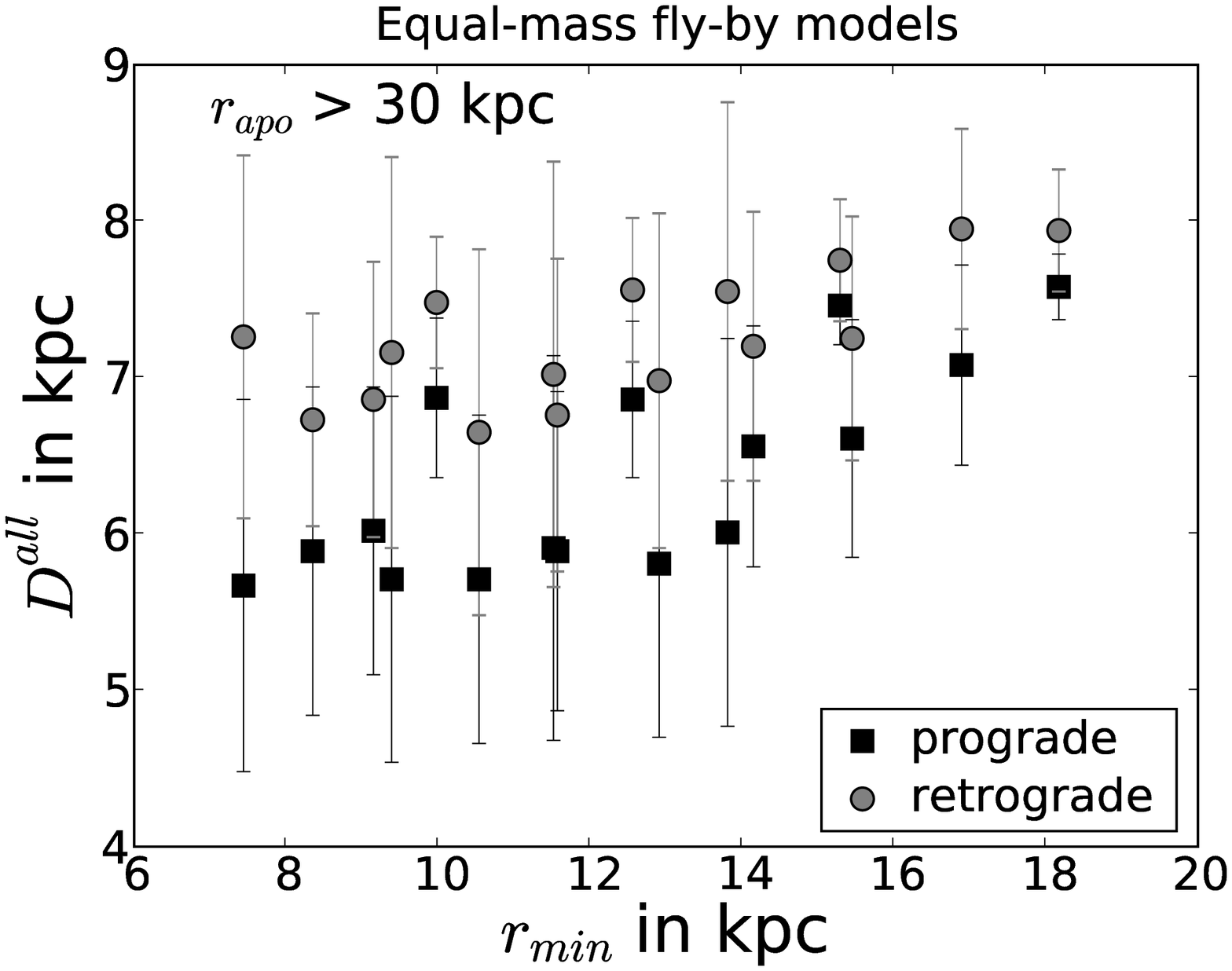}
 \includegraphics[width=88mm]{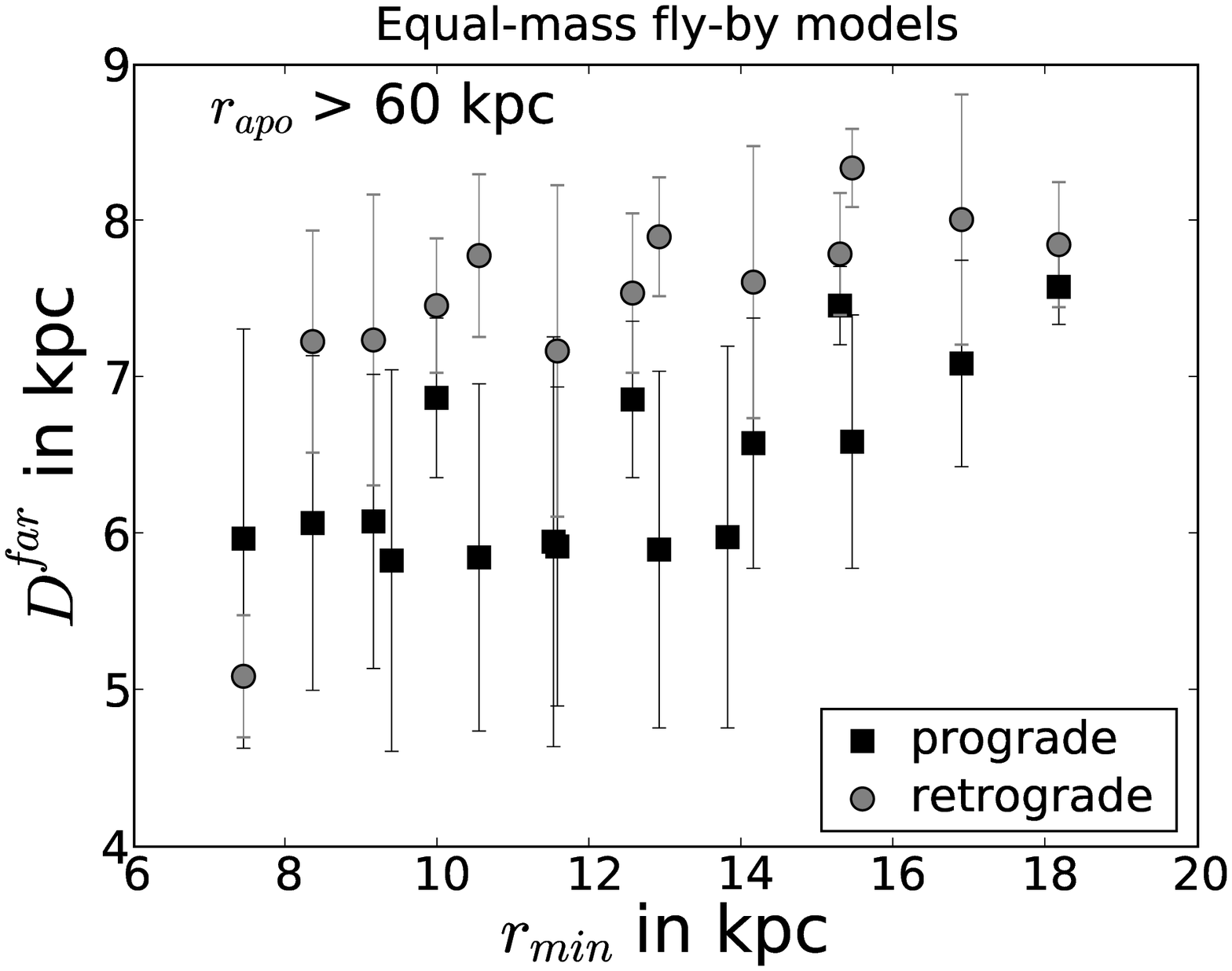}
 \caption{Prograde equal-mass fly-bys: the mean initial distances to the infalling galaxy's centre, $D_{\rm{pro}}^{\rm{all}}$\ and $D_{\rm{retro}}^{\rm{all}}$, for particles on pro- and retrograde orbits after the interaction. Upper panel: all particles with apocentre distances above 30 kpc are taken into account. There is a clear trend that retrograde particles originate from further out. Lower panel: the mean initial distances $D_{\rm{pro}}^{\rm{far}}$\ and $D_{\rm{retro}}^{\rm{far}}$\ show a better distinction, the before mentioned trend is more pronounced when only considering particles with apocentres above 60 kpc. All models of Tables \ref{flybyparameters} were included, except models 6deg180vel, 7deg180vel and 8deg180vel as there are no far retrograde particles found.}
 \label{flybydini}
\end{figure}

Considering only 'far' particles with apocentres above 60 kpc (lower panel in the figure), this trend is still seen, but most fractions drop strongly. For the slowest passages (relative velocity of 1.8 times the parabolic velocity), virtually no retrograde particles with higher apocentre distances are found. For the fastest models, in contrast, $P_{\rm{retro}}^{\rm{far}}$\ only changes slightly. Therefore, $P_{\rm{retro}}^{\rm{far}}$\ still shows a wide spread, from 0.00 to 0.99. However, all but the fastest passage models have $P_{\rm{retro}}^{\rm{far}} < 0.2$. It thus seems that more extreme ratios of co- and counter-orbiting material are preferred in fly-by scenarios. The reason for the drop in $P_{\rm{retro}}$\ when considering only far particles for most models is the sharp transition between the two phases, producing retrograde particles with low apocentre distances first. Many of these do not reach out to over 60 kpc.

It can be concluded that the relative velocity of the two interacting galaxies has a dominating effect on the ratio of prograde to retrograde particles. The perigalactic distance during the passage has a smaller effect. This is of importance for the reconstruction of galaxy-encounter histories. Within the scenario of an early, similar-mass fly-by event, the fraction of retrograde material (e.g. in the form of satellite galaxies) gives direct hints at the relative velocity at which this encounter happened. But one has to keep in mind that there is an ambiguity: pro- and retrograde orbits in the models can not directly be identified with co- or counter-rotating orbits. This ambiguity has to be broken with additional information. For this, the use of the radial (apocentre) distribution of the fraction of counter-orbiting tidal debris is suggested. If it decreases strongly for larger distances from the central galaxy, the debris could be identified with the retrograde material in the models which shows the same behaviour. This information then would not only break the ambiguity, but also allow a reconstruction of the orientation of the orbital angular momentum of the early encounter.

Plotting the average initial distance $D$ versus pericentre passage distance $r_{\rm{min}}$\ between the galaxies in Fig. \ref{flybydini}, it emerges that retrograde particles originate from further out in the infalling disc. Prograde particles have average values of $D$ of about $D_{\rm{pro}}^{\rm{all}} \approx 6\ \rm{kpc}$, retrograde ones of approximately $D_{\rm{retro}}^{\rm{all}} \approx 7\ \rm{kpc}$. For every model, the average value of $D$ is higher for the retro- than the prograde particles. This effect is more obvious when only considering the far particles which have apocentre distances of at least 60 kpc.

The rise in $D$\ with increasing $r_{\rm{min}}$, by about 1 kpc from $r_{\rm{min}} \approx 8$\ to $r_{\rm{min}} \approx 16~\rm{kpc}$ can be explained as follows: when the galaxies pass at greater distances, the tidal interaction is less severe. The number of stripped particles decreases with increasing $r_{\rm{min}}$. As it is easiest to unbind particles from the rim of a galactic disc, these outside particles will still contribute to the pro- and retrograde populations, while particles with lower initial distances remain bound to the infalling galaxy.

The orbital eccentricities for both particle populations rise slightly with increasing $r_{\rm{min}}$. Increasing velocities have a stronger effect on the average orbit eccentricities for prograde particles, going from $e_{\rm{pro}} = 0.74$\ to $e_{\rm{pro}} = 0.97$. The corresponding value for the retrograde population ranges only from $e_{\rm{retro}} = 0.79$\ to $e_{\rm{retro}} = 0.90$. As can be expected, far particles in all models show higher eccentricities on average.

\subsection{Equal-mass mergers}
For merging galaxies, both a pro- and a retrograde orientation of the infalling disc's direction of spin compared to the orbit lead to significant numbers of particles orbiting the merger remnant.

\begin{figure*}
 \centering
 \resizebox{\hsize}{!}{\includegraphics{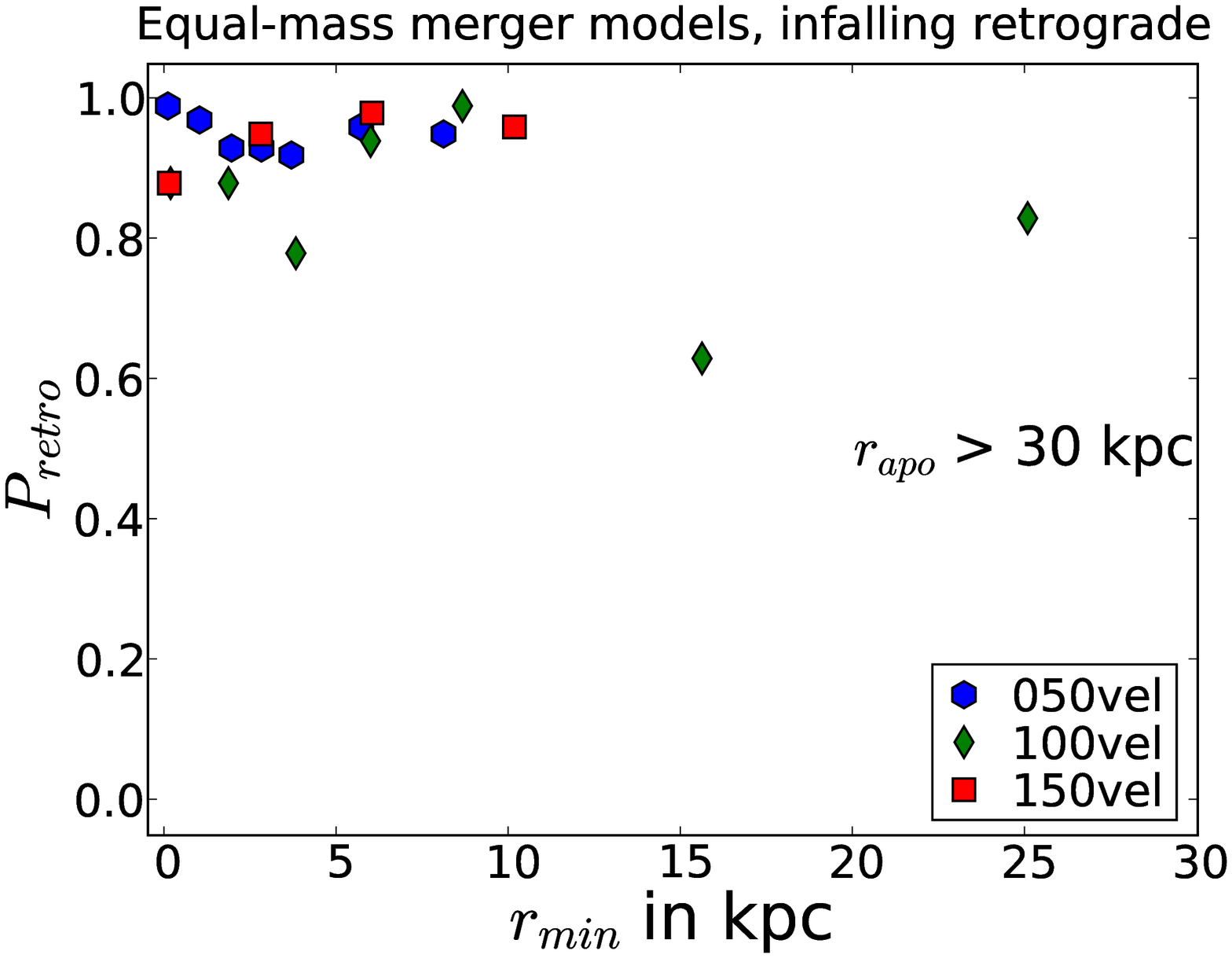} \includegraphics{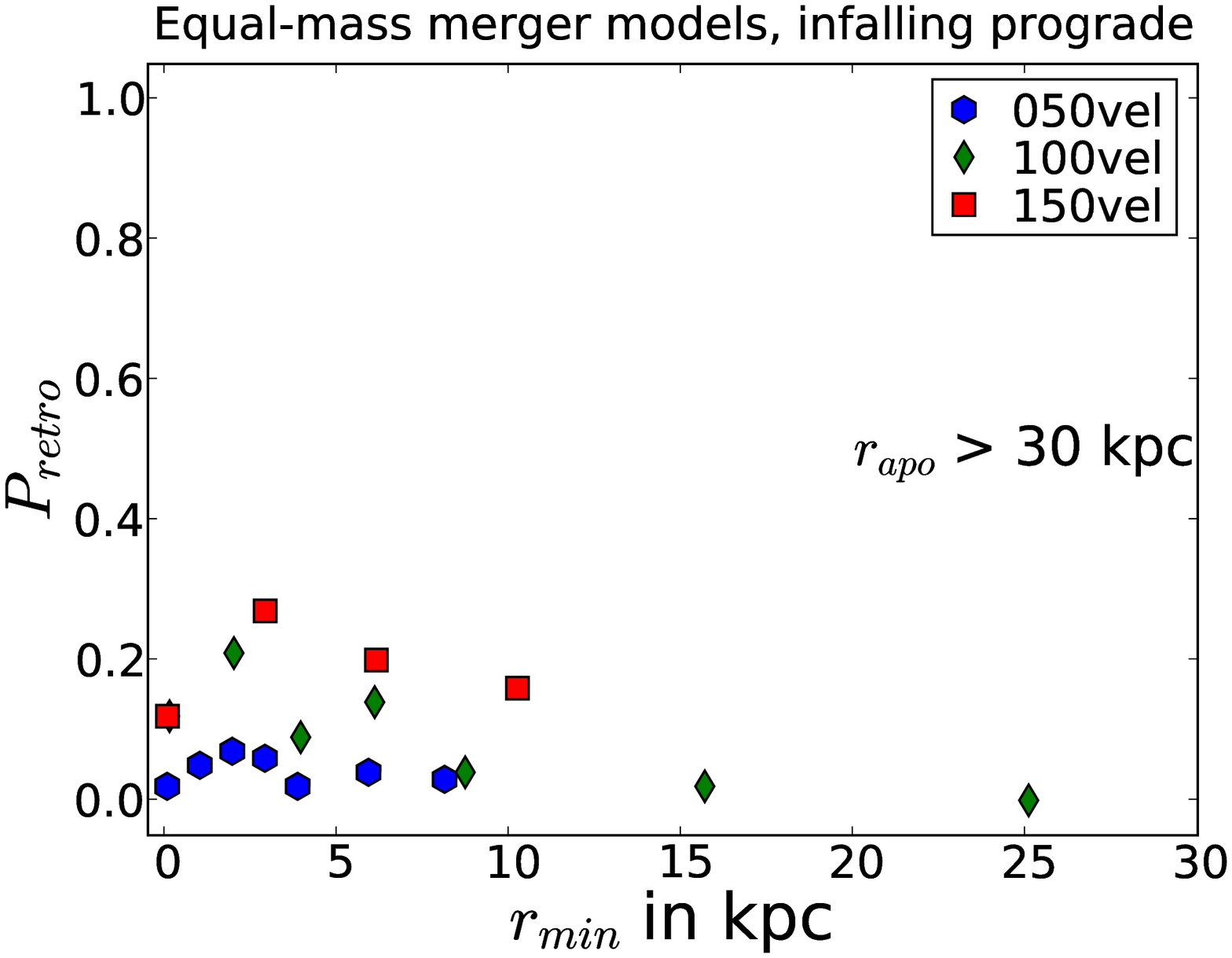}}
  \resizebox{\hsize}{!}{\includegraphics{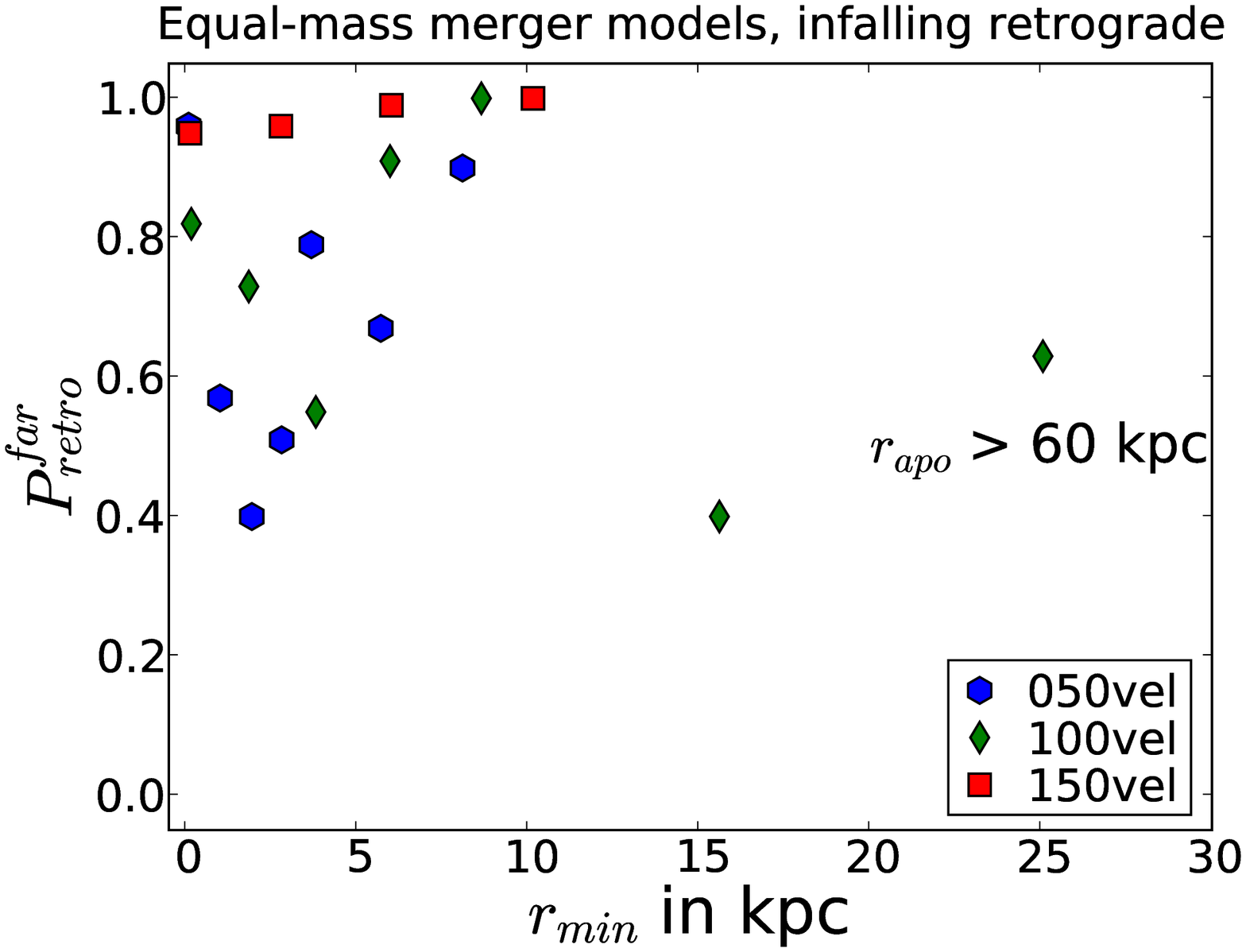} \includegraphics{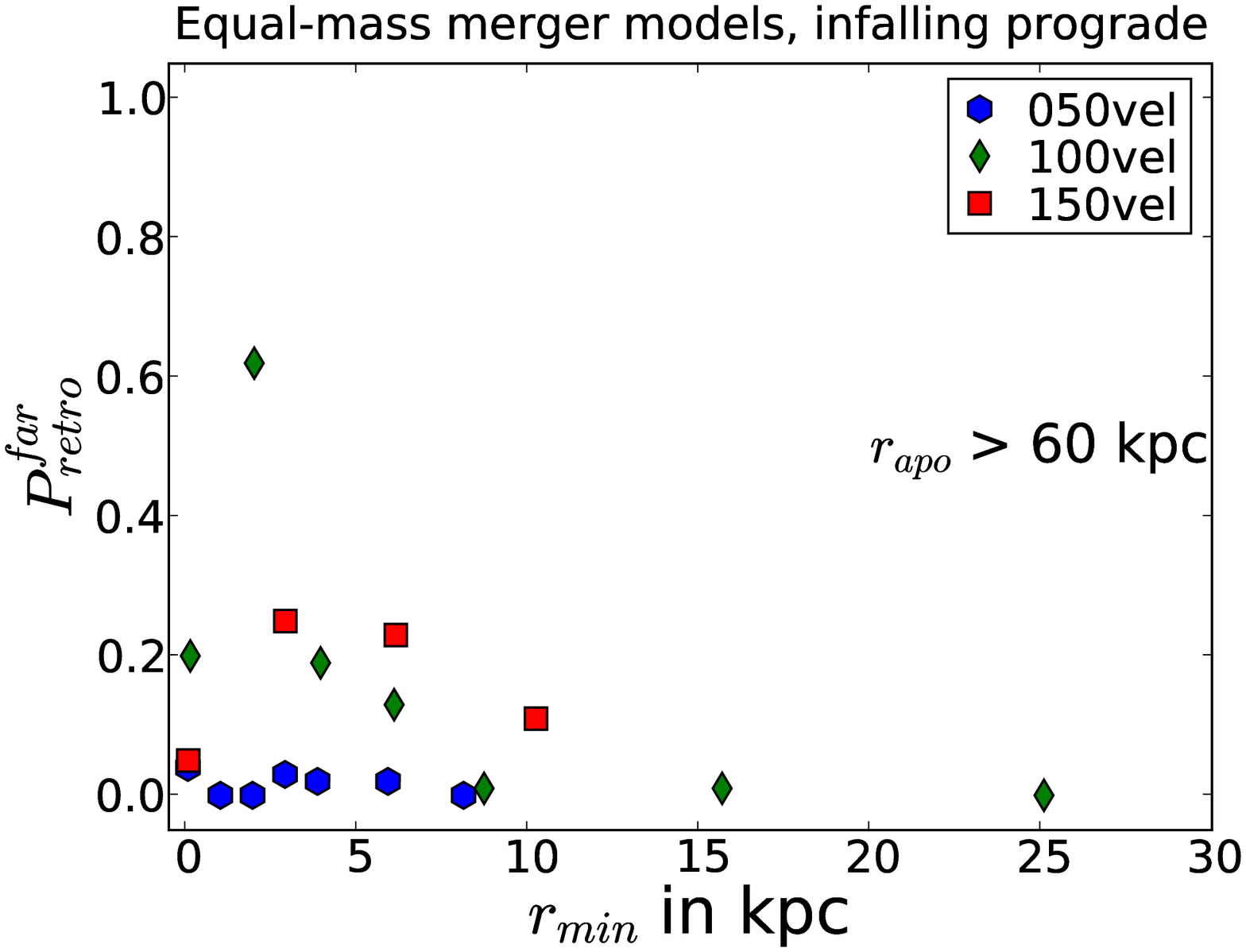}}
 \caption{Equal-mass-mergers: the fraction of particles on retrograde orbits with apocentres above 30 kpc ($P_{\rm{retro}}$, upper panels), and above 60 kpc ($P_{\rm{retro}}^{\rm{far}}$, lower panels) after the merger. The panels on the left show models with retrograde, those on the right prograde spin orientation of the infalling galaxy. See Tables \ref{mergerretronumbers} to \ref{mergerproparameters} for the values. The symbols reflect the initial velocities, with `050vel' standing for 0.5, `100vel' for 1.0 and `150vel' for 1.5 times the parabolic velocity, calculated assuming the initial galaxies are point masses. Mergers where the infalling disc has a retrograde angular momentum (left panels) lead to the highest fractions of retrograde material in the models computed. There is no simple dependence on the initial relative velocity of the galaxies, but for the subset of far particles, faster interactions tend to lead to higher values in $P_{\rm{retro}}$.
Mergers where the infalling disc has a prograde spin angular momentum (right panels) lead to low fractions of retrograde material. Similar to the fly-by interactions, faster initial velocities lead to higher fractions of retrograde material. But in contrast to the fly-by cases, $P_{\rm{retro}}$\ does not drop when considering only the subset of far particles, because of the more similar radial distribution of particles on pro- and retrograde orbits. See the on-line edition of the article for a colour version of this figure.}
 \label{mergerpretro}
\end{figure*}

In the case of merging galaxies, the outcoming distributions of pro- and retrograde particles seem not to follow a distinctive trend, the parameters do not depend in a simple way on the initial velocity or first pericentre between the two galaxies. The prograde orientation tends to give a higher absolute number of pro- and retrograde particles. $N_{\rm{all}}$\ ranges from approximately 5000 to 43000 for prograde infalling galaxies, it is between 6000 and 27000 particles if the infalling galaxy spins retrogradely. For $r_{\rm{min}}$ above about 4 kpc, $N_{\rm{all}}$\ is higher for a prograde than for a retrograde infalling galaxy, below it is smaller on average.
These numbers drop when considering only far particles. The drop-off is most severe for models with low initial velocities (0.5 times the parabolic velocity), low $r_{\rm{min}}$ (< 4 kpc) and a retrograde infalling galaxy. Essentially no far particles are found for these. For faster and more distant encounters, several $10^3$\ to $10^4$ particles with high apocentres are counted.
In principle, the maximum distance of a satellite galaxy born as a TDG associated with a merger can be used to derive a lower boundary for the relative velocity between the precursor infalling galaxy and the central (target) galaxy. \textit{Therefore it can be concluded that mergers of galaxies with low relative velocities do not produce a sufficient number of far particles to resemble the MW satellite system.}

If the infalling galaxy's spin angular momentum is oriented retrogradely to the orbital one, $P_{\rm{retro}}$\ varies between 0.79 to 1.00. When considering the far particles only, this fraction even drops to 0.38 in one model (2.5deg100vel), with an even spread in $P_{\rm{retro}}^{\rm{far}}$. The retrograde fraction is plotted in Fig. \ref{mergerpretro} against the first perigalactic passage distance of the galaxies, $r_{\rm{min}}$. For a retrograde orientation, plotted in the right panels of Fig. \ref{mergerpretro}, fractions of retrograde particles vary between 0.01 and 0.37. Even $P_{\rm{retro}}^{\rm{far}} = 0.60$\ is reached by one model when taking only the far particles into account. However, there is a large gap in the distribution, the values for the other models spread between 0.00 and 0.25.

The initial distances $D$\ within the infalling galaxy of where the tidal material stem from do not distinguish the pro- and retrograde populations as well as in case of fly-bys of galaxies. The average values differ in general, but the distributions overlap strongly. Furthermore, there is no pronounced trend: for a retrograde spinning infalling galaxy, the average initial distance for prograde particles tends to be higher (at about $D_{\rm{pro}}^{\rm{all}} = 7$~kpc) than that of retrograde ones (at about $D_{\rm{retro}}^{\rm{all}} = 6.5$~kpc). This holds true up to a first perigalactic passage distance $r_{\rm{min}}$\ of about 10 kpc. Thereafter, the three models with higher $r_{\rm{min}}$\ show the opposite trend, with $D_{\rm{pro}}^{\rm{all}}$\ of about 5 kpc. These differences are even smaller for far particles and above $r_{\rm{min}} = 10\ \rm{kpc}$\ the far prograde particles even show slightly higher initial distances. For prograde infalling galaxies there is a slight trend to lower $D$\ for prograde than for retrograde particles.

The average orbit eccentricities of the two populations differ in general. For models in which the infalling galaxy has a retrograde spin, prograde particles show a higher average orbit eccentricity in most cases. This is the other way around in models with prograde spinning infalling galaxies, there retrograde particles have higher eccentricities. These trends hold true when only considering far particles. The orbit eccentricities for these subpopulations increase, however, and the differences between pro- and retrograde particles are reduced.

\subsection{4-to-1 mass fly-bys}
The results of the ten fly-by models with a mass ratio of 4:1 for target:infalling galaxy are compiled in Tables \ref{4to1flybynumbers} and \ref{4to1flybyparameters}. The spread in initial parameters here is higher, the initial velocity varies from 1.75 to 2.5 times the parabolic velocity, the perigalactic distance of the two passing galaxies, $r_{\rm{min}}$, now ranges from 3.7 to 16.9 kpc.

As in the equal-mass-situation, most models show a significant number of both pro- and retrograde particles, their sum being in the range between 1300 and 14000. The fraction of retrograde particles $P_{\rm{retro}}^{\rm{all}}$\ covers a wide range, from less than 0.01 in model 8deg175vel up to 0.94 in model 4deg250vel. The numbers of far particles fall in the range between 600 and 6000. As in the equal-mass models, $P_{\rm{retro}}^{\rm{far}}$\ is usually much lower, in the range of several per cent. But for the fastest initial velocities, again, $P_{\rm{retro}}$\ does not change as much. Retrograde material again forms first and the transition of the two phases, producing at first retro- and then prograde particles, also takes place in interactions of non-equal-mass galaxies. \textit{It can therefore be concluded that the two-phase origin is a general behaviour of fly-by interactions, and not a special effect of a certain mass-ratio of the galaxies.}

One important difference to the equal-mass case is that $P_{\rm{retro}}$\ now seems to change dramatically with $r_{\rm{min}}$. For two models of the same initial velocity the retrograde-fraction can differ by a factor of 15 or more. Still, looking at the maximum values for $P_{\rm{retro}}$, the previously found trend of an increasing fraction of retrograde material with increasing velocity is reproduced. Interestingly, these maximum values are all found for $r_{\rm{min}}$\ in the short range of 6.9 to 9.7 kpc, $r_{\rm{min}}$\ for the other models here are at least a factor 1.5 off. This has to be contrasted to the equal-mass case where the relative differences in $r_{\rm{min}}$\ are only 1.2 and the difference between maximum and minimum in $r_{\rm{min}}$\ does not exceed a factor of two.

It can be speculated that, if more calculations with intermediate $r_{\rm{min}}$ would be included, the $P_{\rm{retro}}$-values would again lie on curves for each initial velocity that are not crossing each other, similar to Fig. \ref{flybypretro}. A reconstruction similar to the one described in Sect. \ref{equalmassflybysection} will be more difficult in this case because the fraction of retrograde material is not only determined by the initial velocity but also by the minimum distance to which the passing galaxy approached the target. Nevertheless, as probably every point in the $P_{\rm{retro}}$-$r_{\rm{min}}$-plane can be associated to a given initial velocity, such a reconstruction is deemed to be in principle possible in the case of a 4-to-1 mass ratio.

The average orbital eccentricities of the pro- and retrograde populations are similar to the equal-mass models. This includes the tendency that the retrograde population has a slightly lower orbital eccentricity in principle. The same is true for the site of origin in the precursor disc: again, retrograde particles originally had a larger radial distance $D$\ to the infalling disc's centre. The difference now on average is 0.75 kpc for all ($d_{\rm{apo}}\ >$\ 30 kpc) and 0.5 kpc for the far ($d_{\rm{apo}}\ >$\ 60 kpc) particles. For model 2deg200vel no significant differences are found, while in others the difference can exceed 1.5 kpc. 

\subsection{4-to-1 mass mergers}
The tidal particle numbers in the merger models are again higher than in the fly-by case of this mass-ratio. The sum of pro- and retrograde particles $N_{\rm{all}}$\ reaches up to 60000 particles and more in some models. The number of far particles reaches up to 30000 for both, pro- and retrograde orientations of the infalling galaxy. While the numbers are somewhat higher than in the equal-mass case, one has to keep in mind that one particle now has only half the mass. Nevertheless, the total amount of material on pro- and retrograde orbits is comparable.

In all $2 \times 6$\ merger models, particles on pro- and on retrograde orbits are found. In contrast to the equal-mass merger models, the fraction of retrograde material, $P_{\rm{retro}}$, now is more evenly distributed. While in the equal-mass case the infall of a disc with retrograde spin leads to high values of $P_{\rm{retro}}$, in the 4-to-1-mass case $P_{\rm{retro}}^{\rm{all}}$\ is in the intermediate regime of 0.23 to 0.63. Considering only particles with apocentres above 60 kpc, $P_{\rm{retro}}^{\rm{far}}$\ changes to 0.03 and 0.06 for the two models with the biggest $r_{\rm{min}}$\ (they have the smallest $P_{\rm{retro}}^{\rm{all}}$-values, too: 0.23 and 0.30 respectively). The fraction is not much different in the other four models. A prograde spin of the infalling galaxy results in lower values of $P_{\rm{retro}}$, but not as low as in the equal-mass case: $P_{\rm{retro}}^{\rm{all}}$\ ranges from 0.01 to 0.44, $P_{\rm{retro}}^{\rm{far}}$ even reaches 0.66 in model 2.5deg050vel.
As far as the six models each tell, $P_{\rm{retro}}$\ drops with increasing $r_{\rm{min}}$. The models with lower velocities have higher $P_{\rm{retro}}$, too, but are at the same time the ones with the lowest values of $r_{\rm{min}}$.

The average orbital eccentricities $e$\ are generally on the higher end of the distribution found for equal-mass mergers. They vary between $e = 0.76$ and $e = 0.90$, with the retrograde ones being higher on average. This holds true for both spin orientations of the infalling galaxy but is more pronounced for a prograde infalling galaxy. Considering the far particles only, the average orbital eccentricities are again higher.

The initial distance of the tidal material to the centre of the infalling galaxy, $D^{\rm{all}}$, does not show a conclusive behaviour when looking at all particles. But, considering only the far particles, $D^{\rm{far}}$\ is higher for retro- than for prograde particles ($D_{\rm{retro}}^{\rm{far}} > D_{\rm{pro}}^{\rm{far}}$) in most models (the ones with faster initial velocities). This is the same trend already observed in all fly-by interactions, but interestingly not in the equal-mass mergers where, if at all, an opposite trend was evident.

\section{Discussion}

\label{discusssect}

\subsection{Observational Predictions}
\label{comparesect}

To compare the results with observations, the ratio between pro- and retrograde particles can be expressed with respect to the lower-numbered population. This counter-rotating ratio $R_{\rm{countrot}}$\ is defined as the fraction of particles orbiting in an opposite sense compared to the majority of pro- or retrograde particles. By this definition any knowledge about the orbital direction of the galaxy-interaction is removed from the data derived from the models, just as is the case for the observable situation around galaxies. This implies the following definition:
\[
R_{\rm{countrot}} = \left\{ 
\begin{array}{l l}
  N^{\rm{retro}} : N^{\rm{pro}} & \quad \mbox{if $N^{\rm{retro}} < N^{\rm{pro}}$},\\
  N^{\rm{pro}} : N^{\rm{retro}} & \quad \mbox{if $N^{\rm{retro}} > N^{\rm{pro}}$}.\\
\end{array} \right.
\]

\subsubsection{Numbers, masses and radial distribution}
\label{numberdistrsect}

A range in $R_{\rm{countrot}}$\ of less than $1 : 100$\ up to about $1 : 1$\ can be easily reproduced with fly-bys and mergers of both mass-ratios investigated.
As all stellar particles in a calculation have the same mass, these number-ratios also reflect the mass-ratios. The maximum mass available can be calculated from the total number of orbiting particles. 
The mass of all particles orbiting the target galaxy with an apocenter larger than 30 kpc is up to $6.8  \times 10^{8}~\rm{M}_{\sun}$, 8.6 percent of the infalling galaxy mass in the equal-mass merger models. It is up to $9.2 \times 10^{7}~\rm{M}_{\sun}$\ (1.2 percent) for equal-mass fly-bys, up to $5.6 \times 10^{8}~\rm{M}_{\sun}$\ (14 percent) and $1.1 \times 10^{8}~\rm{M}_{\sun}$\ (2.8 percent) for 4-to-1 mass ratio mergers and fly-bys, respectively. As only particles that approach an apo- and pericentre are counted, these are only lower limits to the total available mass. \textit{On average, a higher fraction of the infalling galaxy's particles is stripped when it is less massive than the target, favouring interactions of non-equal mass galaxies to disrupt the infalling galaxy more effectively.}

The two-phase origin of pro- and retrograde material in fly-by interactions results in a characteristic radial distribution of the orbits of tidal material. Particles on retrograde orbits are distributed up to a maximum apocenter, while particles on prograde orbits can spread out much further. Thus, there is a maximum distance from the target galaxy up to where both co- and counter orbiting material exists. Further out, only one orbital direction, the prograde one, can be found. \textit{This might be used as a hint on the original orbital direction when reconstructing past galaxy-galaxy interactions.}

\subsubsection{Metallicity}
Even though the models do not include chemical evolution, some predictions on the metallicity distributions for pro- and retrograde tidal material can still be deduced. It is a common feature of disc galaxies to show a radial metallicity gradient, with metallicity decreasing with increasing radius $D$\ \citep[see for example the review by][]{henry99}.
The infalling galaxy model has similar parameters compared to those adopted by \citet{magrini} for M33. Therefore, their computed [Fe/H] gradient at more than 8 Gyr before today of $-0.08~\rm{dex~kpc}^{-1}$\ is taken for the model. Note that the following discussion strongly depends on the adopted metallicity gradient for the infalling disc galaxy.

The different mean initial distances of the ultimately pro- and retrograde populations hint at a metallicity difference between those particle groups. In the fly-by models, the mean initial distance for prograde material is on average 0.9 kpc closer to the infalling galaxy's centre than that of retrograde material. This results in an average intrinsic [Fe/H] difference of 0.07 dex. In all fly-by models analysed, retrograde material has a lower metallicity compared to prograde material. This is irrespective of the mass ratio of the two galaxies. The intrinsic metallicity thus might be a useful tool to reconstruct the encounter-histories of stellar streams and maybe TDG systems around galaxies.

The largest difference of average initial distances is 2.0 kpc in case of model 7deg200vel and only considering the 'far' particles. This predicts a mean metallicity of prograde material of about 0.16 dex higher than that of retrograde material, given the adopted metallicity gradient in the infalling disc. The values strongly depend on the assumed metallicity gradient, but generally lies below the uncertainties of most metallicity determinations \citep[e.g.][]{grebel03}. It should become accessible once better statistics are available. However, the spread in initial distance within one population is of the same order as the difference between the two. So, while a general trend can be expected, the metallicity can not be used to determine whether material orbits pro- and retrograde. The results of merger models show that there the distribution in initial distance follows no clear trend.

The scenario of a tidal origin for the MW streams and satellite galaxies of the MW predicts that in fly-by cases the intrinsic metallicity of retrograde material is lower than that of prograde material. With better models including chemical evolution and more precise observational data, it might become possible to test this prediction: while the distributions of intrinsic metallicities of pro- and retrograde material overlap, retrograde material is expected to be slightly less metal rich \textit{on average}.. But caution is advised: the accretion of unprocessed material by newly formed dwarf galaxies and the internal chemical self-evolution \citep{recchi07} will most probably smear out this effect. It will thus be of importance to observe the initial (i.e. pre-encounter) metallicity as present in the oldest stars in stellar streams and dSphs.

\subsection{The MW satellites as TDGs}

The dSph satellite galaxies of the MW show properties that are difficult to understand in terms of them being cosmological sub-structures \citep{McGaugh10}. Instead, they are more readily understood as ancient TDGs \citep{metz07, kroupa10}. Compared with the predictions of the presented models, it is found that the MW satellite system is also in agreement with having formed out of tidal material and a possible scenario for their origin can be distilled from the data.

Is is sometimes observed that star-forming-regions in tidal tails are distributed in a beads-on-a-string morphology \citep{Mullan11}, that is, star formation takes place at spots about equidistantly distributed along the tidal tails. This supports an approximately homogeneous distribution of the formation sites of TDGs along the tails. Further hints for such a distribution come from figure 2 in \citet{bournaud06}, which plots the number of objects formed as a function of their formation radius, normalized by the total radial extend of the tidal tail. For their initial distribution (200 Myr after the interaction), the radial distribution is approximately constant, the bin-heights vary only by a factor of two. Therefore, in the following the most simple assumption on the formation efficiency of TDGs is adopted, their formed number being proportional to the number of particles on  pro- and retrograde orbits.

\subsubsection{Orbital Poles}

\begin{figure*}
\centering
 \includegraphics[width=160mm]{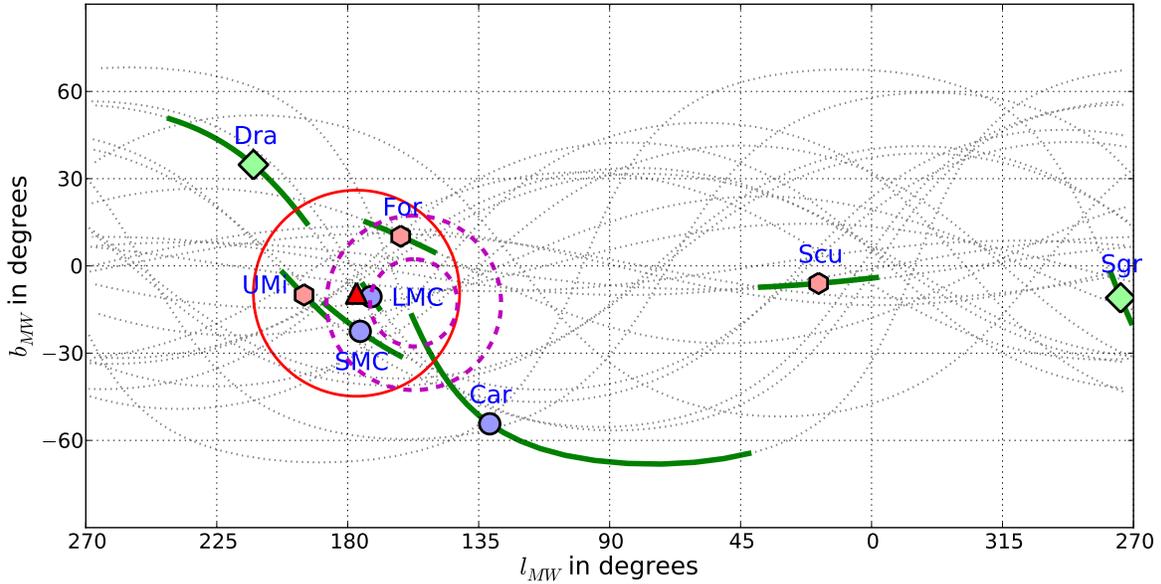}
 \caption{Possible orbital poles (dotted lines) in galactocentric galactic coordinates for the 24 known MW satellite galaxies as defined by their angular positions. For eight MW satellite galaxies proper motions are available, the resulting orbital poles are marked by the symbols. The green lines give the uncertainties in direction of the orbital poles. The red triangle marks the mean direction of the orbital poles of the six satellites close to $l_{\rm{MW}} = 180^\circ$, excluding Sagittarius and Sculptor. The red circle marks the spherical standard deviation from this average direction, the dashed circles are the regions $15^{\circ}$\ and $30^{\circ}$\ degrees from the normal of the DoS fitted to the 11 classical MW satellites. The data for the orbital poles are taken from \citet{metz}. The possible orbital poles (dotted lines) are densest at $l_{\rm{MW}} \approx 0^{\circ}$\ and $180^{\circ}$, where they are expected to lie in the case of a rotationally supported DoS.
 The known satellite orbital poles fall close to these positions with the exception of Sagittarius.
 A comparison of this plot with the results of particle orbital poles from the numerical calculations in Fig. \ref{lplotrun} shows that the orbital poles of the MW satellite galaxies have a similar distribution. There are two distinct populations with an offset of about $180^\circ$. The positions of the Draco and Carina orbital poles even hint at a diagonal distribution of orbital poles (elongated from upper left to lower right), similar to the calculated distributions. This shows that the tidal origin model is in agreement with the observations of the MW satellite system. See the on-line edition of the article for a colour version of this figure.}
 \label{lplotMW}
\end{figure*}

The distribution of orbital poles of particles in models 5deg200vel and 7.5deg100vel as presented in Figure \ref{lplotrun} shows two distinct populations, offset by 180 degrees. This agrees with the positions of the MW satellite orbital poles, which show one concentration, from which Sculptor is offset by about 180 degrees. In addition to that, it is especially interesting to note the shape of the distributions: they are elongated in diagonal directions in both models. This is suggestive for the distribution of the orbital poles of the co-orbiting MW satellites, as found by \citet{metz}. Their figure 1 is reproduced in Fig. \ref{lplotMW} in the same projection as Fig. \ref{lplotrun} here. Here the $l_{\rm{MW}}$-axis is shifted to align with the positions of pro- and retrograde particles. The satellite galaxies Draco and Carina are oriented in a similar diagonal direction compared to the other co-orbiting galaxies.

The resemblance to the result for the merger model in Fig. \ref{lplotrun} is most striking: not only are the most likely positions of the orbital poles for the satellites reproduced by the model, but also the segments of the great circles (green lines), which represent the uncertainties of the orbital poles, align with the distribution of angular momenta. The counter-orbiting population is only made up of one galaxy, Sculptor, thus there is no structure that could be compared to the models.

While the position of Sculptor's orbital pole is in agreement with the results of the tidal formation scenario assumed in this paper, further data on the proper motions of the MW satellite galaxies will allow the assessment of the validity of this scenario. Finding more orbital poles close to the one of Sculptor will allow to compare the angular-momenta substructures to the observed situation in more detail. This will test the validity of the tidal origin scenario and, if the MW satellite galaxies are TDGs, constrain the properties of the early interaction.

\subsubsection{Number- and mass-ratios}

Comparing the mass ratios and number ratios to the eight satellite galaxies with measured proper motions, the number-ratio of counter- to co-orbiting satellites in the known MW dwarfs is found to be $R_{\rm{countrot}} = 1:6$\ within the uncertainties, but $R_{\rm{countrot}} = 1:4$, excluding the Draco and Carina dwarf that show large orbital pole uncertainties, might be possible, too. These numbers lie well between the extremes of the models discussed in Sect. \ref{resultssect}.

The corresponding stellar-mass-ratios of the dwarf satellites are, using values from \citet{Woo}: $R_{\rm{countrot}}^{\rm{mass}} = 1 : 1700$. Considering the mass distributions, the model results (Sec. \ref{numberdistrsect}) have ratios of 1:100 to 1:1. But the observed ratio is strongly dominated by the LMC and SMC, which could be the infalling remnant of an initially passing galaxy (see Sect. \ref{reconstruct} below). In that case, a ratio of about $1 : 17$ excluding LMC and SMC is well reproduced by many of the far subpopulations, especially in the fly-by models with $P_{\rm{retro}}^{\rm{far}}$ in the range between 0.05 and 0.10 and 0.95 to 0.99, but also for a number of merger models.

Concerning the comparison of models with the MW, it is too early for a concluding statement because there are many dwarfs for which no proper motion measurements are available. It is well possible that more dwarf galaxies on counter-rotating orbits will be found. Precise proper-motion data will be available through the GAIA mission \citep{perryman01, gaia08}, another method to determine proper motions was presented by \citet{propermotion}.

Looking at the mass available in tidal material, the satellite dwarf galaxies of the MW have stellar masses in the range between a few $10^4$\ and $10^8~\rm{M}_{\sun}$, excepting LMC and SMC \citep{strigari08}. The maximum masses of pro- and retrograde material in the calculations are sufficient to form several of such satellite galaxies, their number depending on the TDG formation rate in the tidal material. Judging from the two models analysed in detail, the formation of retrograde tidal dwarf satellites of high mass seems to be unlikely. But in both calculations there is plenty of mass on these orbits available to make up several Sculptor-type dwarfs.

\subsubsection{Possible reconstruction of the early MW encounter}
\label{reconstruct}
As an illustrative example, a possible early encounter the MW suffered shall now be reconstructed, under the assumption that the MW satellites formed from tidal material expelled during that interaction. As the results of merger models give no conclusive trends, the discussion in this section concentrates on fly-by models of equal mass for simplicity.

The counter-rotating ratio $R_{\rm{countrot}}$\ of the MW satellites is high, at least 1 : 4 considering satellite numbers only. Therefore in the models one has to look for $P_{\rm{retro}}$-values above 0.8 or below 0.2. As it is known that the satellite galaxies have apocentre distances above 60 kpc, only the results of far particles (lower panel in Fig. \ref{flybypretro}) are considered. Several of the models might fit approximately and none can be excluded right away. Either the fly-by was fast (models with 2.4 times the parabolic velocity), then $P_{\rm{retro}}$ is high, so what is seen today as counter-orbiting satellite Sculptor was a prograde orbit compared to the relative motion of the interacting galaxies. Or $P_{\rm{retro}}$\ is low, then all other models would fit and the counter-orbiting satellite comes from retrograde material. One way to discriminate between high or low $P_{\rm{retro}}$\ would be the initial metallicity.

The recent discovery of the most metal-poor star currently known in a MW satellite galaxy in Sculptor \citep{frebel10} is in agreement with Sculptor having a lower average initial metallicity than other satellites. This might hint at an origin farther away from the centre of an infalling galaxy, which on average is the case for retrograde material. Sculptor is, as a counter-orbiting satellite, in the minority, so $P_{\rm{retro}}$\ would thus have to be small.

Reconstructing the initial encounter with these assumptions leads to the following scenario: about 10 Gyr ago the MW had an off-centre fly-by encounter with a similar disc galaxy which produced tidal dwarf galaxies. The infalling galaxy was on a parabolic to mildly hyperbolic orbit (not too fast as the models with $2.4 \times v_{\rm{parab}}$\ are excluded) with an orbital pole in the same direction as that of the later formed co-orbiting part of the DoS. But there remains a question: where did the infalling galaxy go?\footnote{Assuming a relative velocity of about 250 km/s, after 10 Gyr it will have a distance of 2.5 Mpc, at the fringes of the Local Group.}

Maybe the initial interaction was not of similar-sized galaxies but the infalling one was smaller and can indeed still be seen today: the Large Magellanic Cloud. Maybe it is now on its second approach to the MW. The orbital-pole-direction of the interaction must have been close to the normal vector of the Disc of Satellites, which points to $l_{\rm{DoS}} = 157.3^{\circ}$\ and $b_{\rm{DoS}} = -12.7^{\circ}$ \citep{metz07}. These values are similar to the orbital pole of the LMC, which is $l_{\rm{LMC}} = 172.0^{\circ}$\ and $b_{\rm{LMC}} = -10.6^{\circ}$ \citep{metz}. The LMC spin direction, however, can not be used to constrain this scenario. \citet{Weinberg00} predicts that the LMC disc precesses and nutates. This was observationally supported by \citet{vanderMarel02}, who found that the inclination of the LMC disc currently changes at a rate of about 100 degrees per Gyr. It is thus not possible to deduce the LMC's orientation several Gyr ago. The orbital direction of the LMC agrees with Sculptor being a retrograde satellite, as it is on a counter-rotating orbit, leading to a low $P_{\rm{retro}}$. This is produced by most 1:1 and 4:1 models, especially at larger distances from the target galaxy. If an extrapolation of the results to more extreme mass-ratios is possible, an assumption which is supported by the finding that less massive infalling galaxies are more effectively disrupted, the results are thus compatible with this scenario\footnote{However, a differing accretion history of the MW and the LMC progenitor might have made the MW grow more rapidly, thus the initial masses might have been more similar.}. A low $P_{\rm{retro}}$\ then hints at a not too fast fly-by (the highest-velocity models can be excluded, see Fig. \ref{flybypretro}), in better agreement with an orbit that does not allow the LMC to escape. Furthermore, most models lie in the regime of lower $P_{\rm{retro}}$\ values, \textit{therefore no fine-tuning of the initial parameters is needed to arrive at situations similar to today's MW satellite system}. Additionally, fly-by models with lower initial velocity lead to higher particle numbers stripped from the infalling galaxy, thus more material is available to form several TDGs. The radial distribution of the observed satellites supports this, too: Sculptor is not too far from the MW ($\approx$\ 80 kpc) and has one of the lowest radial velocities, suggesting a smaller apocentre distance than for the other satellites. This is the natural outcome for retrograde particles due to the phase-one-origin in the fly-by models, as can be seen in Fig.~\ref{Apohistrun} for model 5deg200vel. That such models naturally form a disc of tidal debris is demonstrated in Fig. \ref{edgeonviews}.

This work has demonstrated above all else that the orbits of stellar streams and the dSph satellites will be a powerful tool for reproducing the early events that shaped the MW if they are interpreted to be of tidal origin. A more detailed modelling of this scenario is deemed to be of huge interest for the search of the origin of the MW satellite distribution. This would be in beautiful agreement with Lynden-Bell's \citeyearpar{Lynden-Bell76} original suggestion that the LMC could be the major surviving part of a Greater Magellanic Galaxy, parts of which have been torn off by tides forming dSph galaxies and streams.

\section{Conclusions}
\label{concludesect}
In an attempt to approach a solution to the Fritz-Zwicky paradox \citep{kroupa10} and the peculiar spacial arrangement and internal properties of the Milky Way satellite galaxies, the generic properties of tidally expelled material in galaxy-galaxy interactions at an early cosmological epoch are here studied numerically. The galaxy models are set up by scaling the MW properties down to those of an early, bulge-less galaxy at $z \approx 2$, about 10 Gyr ago.

\begin{figure}
 \centering
 \includegraphics[width=75mm]{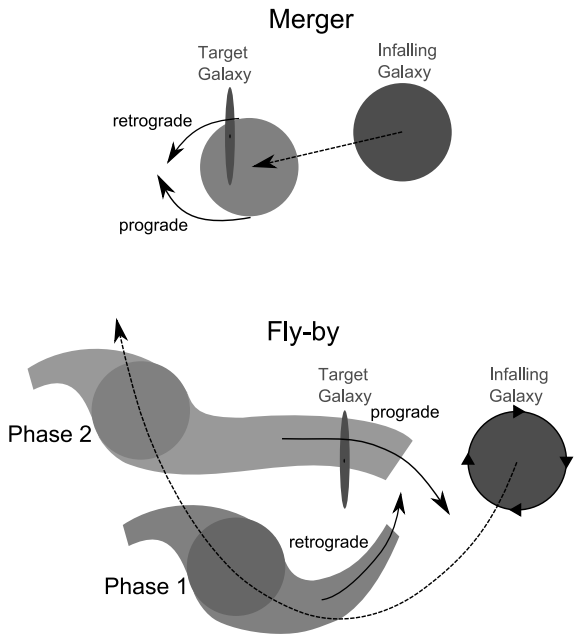}
 \caption{Sketch of the typical origins for pro- and retrograde material in merger (top) and fly-by (bottom) interactions. In both cases, the infalling galaxy starts on the right and follows the path sketched by the dashed line. The direction of spin of its disc is indicated by the arrows along the edge in the fly-by sketch (prograde with respect to the orbit), for mergers both orientations of spin were modelled. The target galaxy is seen edge on.
 In a merger, the collision of the two galaxies involves particles of the infalling galaxy to pass the centre of mass of the target on both sides, resulting in pro- and retrograde orbits. As a merger might involve several passings of the galaxies before the final merging, other sources of pro- and retrograde material, like tidal tails, also play a role.
 In a fly-by, after the galaxies pass, the infalling one developes tidal tails. The tidal material streaming towards the target galaxy first follows back the infalling's path, resulting in retrograde material appearing first (Phase 1). Later, as the galaxies further depart from each other, the tidal tail is swept over the target's centre of mass. This changes the orbital direction of the particles falling towards the target galaxy, making their orbits prograde (Phase 2).}
 \label{finalsketch}
\end{figure}

Computations of interactions between two similar disc galaxies as well as those with mass-ratios of 4-to-1 for infalling to target galaxy have been performed. It has be shown for the first time that it is possible to form particles in tidal streams with both pro- and retrograde orbits in one single galactic encounter. This is true for mergers and, maybe more surprisingly, for fly-bys, of nearly perpendicularly oriented disc galaxies. The typical origins for pro- and retrograde material are sketched in Fig. \ref{finalsketch}. In the merger case, the close passing of the infalling galaxy near the centre of mass of the target results in particles passing the centre on both sides. They therefore end up on opposing orbits around the merged galaxies. In addition, material tidally slung away falls back at the merged galaxies in pro- and retrograde directions. In the case of a fly-by, the galaxies depart and the infalling one evolves a tidal tail, from which particles fall back towards the target galaxy. The direction of the orbital motion for particles falling in is changed from retrograde, with respect to the interaction geometry, to prograde, after the tidal tail sweeps-over the centre of the target galaxy. This marks two phases of the encounter, effectively forming two classes of particles on opposing orbits in a fly-by encounter.

TDGs will form from the tidal material, occupying the same phase-space volume because the effect of dynamical friction on them can be neglected. This scenario explains the otherwise unlikely distribution of satellite positions in a Disc of Satellites and orbital poles and naturally resolves the riddle of the counter-orbiting satellite Sculptor. The number of particles on pro- and retrograde orbits are in agreement with common satellite galaxy masses of some $10^{7}~\rm{M}_{\sun}$, up to 14 per cent of the infalling galaxy's particles finally are on orbits around the target galaxy with apocentre distances beyond 30 kpc.

Moreover, it is possible to extract constraints on the early properties of such a disc of tidal material. While the situation in case of a galaxy merger is complicated and depends strongly on the actual parameters of the interaction (initial relative velocity, perigalactic passing distance), fly-by events show a clear trend: the fraction of retrograde particles varies strongly with the initial relative velocity between the two galaxies, increasing for higher velocities. Thus, the fraction of retrograde material observed constrains the relative velocity at which the encounter happened. Considering particles with an apocentre of 60 kpc or more, the increase is not as strong as retrograde particles have lower apocentres. Prograde particles fall in later when the galaxies are more separated again, giving them higher apocentre distances. They are also found to be on more eccentric orbits initially, but some circularisation of their orbit can be expected \citep[see for example][]{hashimoto03}.

Of observational interest is the site of origin of pro- and retrograde particles in the infalling galaxy's disc in fly-by encounters. Prograde ones originate from a more central position by about 1 kpc. Combined with a metallicity gradient in the infalling disc this results in an average difference in metallicity of pro- and retrograde material, with the latter being less metal rich by about 0.1 dex in [Fe/H]. The effect, however, will have been strongly affected and probably been washed out later because of individual chemical evolution \citep{recchi07} and accretion of unprocessed material from the inter-galactic medium onto condensed objects such as TDGs. The results of merger models are not as conclusive concerning the metallicities of pro- and retrograde material.

Within the scenario of a tidal origin of the MW satellite galaxies, some first constraints on the early, major encounter the MW experienced can be derived from the present results. A merger with a small initial relative velocity of the two galaxies can be excluded as the origin of the MW dwarf satellite galaxies because no particles with high enough apocentre distances are produced. In principle, the maximum distance of a MW satellite born as a TDG associated with a merger can be used to derive a lower boundary for the relative velocity between the merging galaxies. In a fly-by scenario, an early interaction with the LMC progenitor shows an interesting consistency: the initial direction of motion of the infalling galaxy is known, making Sculptor retrograde. This results in a low fraction of retrograde material, in agreement with the results of the majority of models. Furthermore, this fits with Sculptor's comparably lower apocentre and predicts a lower initial metallicity, as discussed above\footnote{Note that the LMC spin is likely to be precessing and nutating, such that its current direction can not be used to constrain this scenario \citep{Weinberg00, vanderMarel02}.}. Such a scenario is in beautiful agreement with Lynden-Bell's \citeyearpar{Lynden-Bell76} original suggestion that a Greater Magellanic Galaxy might be the parent of the then-known dSph satellite galaxies. It might be the most probable origin of the MW satellite distribution, but needs to be addressed in more detail in the future. The implications of a tidal origin of the MW satellite galaxies for cosmological theory are expanded in \cite{kroupa10}.

The models discussed in this contribution did not include gas dynamics, nor was the growth of the Milky Way disc after the interaction included. No direct formation of satellite galaxies was investigated. These points will be addressed and included in future works, but will mainly change the detailed parameters of pro- and retrograde particle populations, not the fact that both orbital directions are produced naturally in interactions of disc galaxies.

\begin{acknowledgements}
M.P. acknowledges support through the Bonn-Cologne Graduate School of Physics and Astronomy and through DFG reseach grant KR 1635/18-1 in the frame of the DFG Priority Programme 1177, “Witnesses of Cosmic History: Formation and evolution of galaxies, black holes, and their environment”. 
We thank Manuel Metz and Ingo Thies for very useful discussions.

\end{acknowledgements}

\bibliographystyle{aa}
\bibliography{15021}

\appendix

\section{Scaling of the Galaxy Models}
\label{scalingappendix}

\subsection{Assumptions}
The model is based on several assumptions used to determine the main structural properties of the disc galaxy model by scaling current MW parameters to a smaller disc mass. A constant and time-independent proportionality between luminosity and mass is assumed, to allow to treat surface brightness and surface density similarly. This allows to apply the Tully-Fisher relation \citep{Tully-Fisher-Relation} to scale the circular velocities. Thus, also the validity of the Tully-Fisher relation independent of age or redshift and size of the disc galaxy is assumed, which seems to be acceptable, at least to a redshift $z \approx 1$, following \citet{vogt1}. \citet{maomo} have found no strong indications for a change in the relation, too. The gaseous galaxy-component is ignored, assuming a progenitor galaxy described by a  pure exponential disc embedded in a spherical Hernquist dark-matter halo, which is used to model the empirically found flat rotation curves in disc galaxies and constitutes, in a non-Newtonian framework, the zeroth-order approximation to non-Newtonian dynamics.

\subsection{Determination of parameters}
To determine the parameters of the model, the stellar mass of the bulge component of the MW as seen today is taken as a starting point. Values range from $1.6 \times 10^{10}~\rm{M}_{\sun}$\ \citep{bulge} to $1.8 \times 10^{10}~\rm{M}_{\sun}$\ \citep[calculated using values from][]{BT} or might be as low as $1.3 \times 10^{10}~\rm{M}_{\sun}$\ \citep{gerhard}. Based on the presupposition that a merger of two identical galaxies created the bulk of the bulge component, the mass of the model galaxy has to be about half of today's bulge mass. Some mass loss might occur due to matter transported outwards along tidal arms. This leads to a stellar mass of $M_{\rm{d}} = 8.0 \times 10^9~\rm{M}_{\sun}$ for the initial small disc galaxies in the case of the equal mass encounters. For the 4-to-1 mass ratio interactions, disc masses of $4.0 \times 10^9~\rm{M}_{\sun}$\ and $M_{\rm{d}} = 1.6 \times 10^{10}~\rm{M}_{\sun}$ are adopted. After formation of the bulge the early Milky Way would still have accreted matter in a disc, forming today's disc component within about $10^{10}$\ years when accreting $\sim 5~\rm{M}_{\sun}~\rm{yr}^{-1}$. The dark matter halo component is chosen to be ten times more massive than the early disc.

Furthermore, the following common values are adopted as those of today's MW and applied as starting points for the down-scaling: the disc scale length is chosen to be $R_{\rm{d}}^{\rm{today}} = 3.5~\rm{kpc}$. Observations reveal similar values like 3.3 kpc from star counts from 2MASS performed by \citet{2mass-scalelength}, but lower values, like 2.1 kpc \citep{gerhard}, have been reported, too. The larger value is chosen here because the thick disc seems to have a larger scale length \citep[3.7 kpc according to][]{ojha01} and, as it is composed of older stars, should trace the original mass distribution better.
It is assumed that the mass of the MW disc is $M_{\rm{d}}^{\rm{today}} = 5.6 \times 10^{10}~\rm{M}_{\sun}$. \citet{gerhard} gives a value for the mass enclosed within 8 kpc of $5.5 \times 10^{10}~\rm{M}_{\sun}$ including the bulge.
Also, the International Astronomical Union (IAU) standard \citep{kerr86} for the circular velocity near the Sun $v_{\rm{max}}^{\rm{today}} = 220~\rm{km}~\rm{s}^{-1}$ and the Galactocentric distance of the Sun $R_{\sun}^{\rm{today}} = 8.5~\rm{kpc}$ is adopted. The adopted parameters are compiled in Table \ref{mwparams}.

\begin{table}
 \caption{Assumed MW parameters.}
 \label{mwparams}
 \begin{center}
 \begin{tabular}{@{}lll}
  \hline
  Symbol & Value & Description\\
  \hline
  $M_{\rm{d}}^{\rm{today}}$ & $5.6 \times 10^{10}~\rm{M}_{\sun}$ & mass of disc \\
  $M_{\rm{bulge}}$ & $1.3~\rm{to}~1.8 \times 10^{10}~\rm{M}_{\sun}$ & mass of bulge \\
  $R_{\rm{d}}^{\rm{today}}$ & 3.5 kpc & disc scale length \\
  $R_{\sun}^{\rm{today}}$ & 8.5 kpc & Solar radius\\
  $v_{\rm{max}}^{\rm{today}}$ & $220~\rm{km}~\rm{s}^{-1}$ & circular velocity near Sun \\
  $\Sigma(r_{\sun}^{\rm{today}})$ & $50~\rm{M}_{\sun}~\rm{pc}^{-2}$ & disc surface density near Sun \\
  \hline
 \end{tabular}
 \end{center}

 \small \smallskip

 Assumed parameters of today's MW disc, applied as starting points for the down-scaling of the disc to its younger version, 10 Gyr ago.
\end{table}

\subsection{Estimating the size of the young MW}
\label{discscaling}
The most important parameters to determine for an exponential disc are the scale length and the rotation curve. Furthermore the disc should be stable, even though small galaxies are often observed to be irregular (e.g. the LMC).

The circular velocity $v_{\rm{rot}}(R)$\ at a radius $R$\ around a mass distribution characterised by an enclosed mass of $M(<R)$\ is given by
\begin{equation}
v_{\rm{rot}}(R) = \sqrt{\frac{G \cdot M(<R)}{R}} ,
\label{vrotequation}
\end{equation}
where $G$\ is the gravitational constant. Using a two-component galaxy, the rotation curve is a result of the interplay between the disc- and the halo-mass-distribution, the respective circular velocities being added in quadrature \citep{hernquist-setup}.

As a first estimate for the disc scale length, the central surface brightness is demanded to be constant for all galaxies and independent of the Hubble parameter \citetext{\citealp{kruit}, see also \citealp{BT}}. With the previously mentioned assumption of a proportionality between luminosity and mass, this translates into a constant central surface density $\Sigma_0$. \citet{naab} applied the same assumption of a constant central surface density when studying the evolution of the MW. For the central surface brightness the following equation holds in the case of a thin exponential disc with a total mass $M_{\rm{d}}$\ and a scale length $R_{\rm{d}}$,
\begin{equation}
\Sigma_0 = \frac{M_{\rm{d}}}{(2 \pi \cdot R_{\rm{d}}^2)} .
\label{surfacedensity}
\end{equation}
This leads to the scaling relation
\[
\frac{R_{\rm{d}}}{R_{\rm{d}}^{\rm{today}}} = \sqrt{\frac{M_{\rm{d}}}{M_{\rm{d}}^{\rm{today}}}} ,
\]
Such that $R_{\rm{d}} = 1.3~\rm{kpc}$.

Another possibility to scale $R_{\rm{d}}$\ avoids inclusion of the MW disc mass but starts from the disc surface density $\Sigma(R_{\sun}^{\rm{today}})$\ in the Solar neighbourhood. Assuming an exponential disc, the central MW surface density $\Sigma_0$\ is calculated via
\[
\Sigma(R_{\sun}^{\rm{today}}) = \Sigma_0 \exp(- R_{\sun}^{\rm{today}} / R_{\rm{d}}^{\rm{today}})
\]
and then Eq. \ref{surfacedensity} is solved for the scale length, yielding $R_{\rm{d}} = 1.5~\rm{kpc}$\ when accepting a value of $\Sigma(R_{\sun}^{\rm{today}}) = 50~\rm{M}_{\sun}~\rm{pc}^{-2}$ as indicated from observations \citep[see][and references therein]{naab}.

An estimate completely independent of today's MW properties can be derived from typical properties of galaxies. \citet{BT} stated that the typical disc scale length for galaxies scales with $R_{\rm{d}} = 3 / h$~kpc, where $h$\ is the dimensionless Hubble-parameter defined by $H_{0} \equiv 100~h~\rm{km~s}^{-1}~\rm{Mpc}^{-1}$. Assuming a closed universe with currently $H_0 = 71~\rm{km}~\rm{s}^{-1}~\rm{Mpc}^{-1}$, mass-density parameter $\Omega_{\rm{M}} = 0.27$\ from \citet{wmap03} and vacuum density $\Omega_{\rm{vac}} = 0.73$, the time about 10 Gyr ago corresponds to a redshift of $z \approx 2$. The Hubble parameter at that redshift is given by
\[
H(z) = H_{\rm{0}} \left[ \Omega_{\rm{vac}} + (1 - \Omega_{\rm{vac}} - \Omega_{\rm{M}}) (1 + z)^2 + \Omega_{\rm{M}} (1 + z) ^3 \right]^{0.5} ,
\]
and was approximately $H(z = 2) \approx 200~\rm{km}~\rm{s}^{-1}~\rm{Mpc}^{-1}$\ at that time, so $h(z = 2) \approx 2.0$. This again leads to $R_{\rm{d}} = 1.5$~kpc which is in good agreement with the first two estimates. Finally, a slightly larger scale length of 1.6 kpc is  choosen for the model, because a bigger value increases the stability of the disc. This is because more halo mass (in relation to disc mass), with particles of random motions, is inside the disc's volume. The disc is truncated at 5 times this scale length, close to the frequently used value of 4.5 times \citep{kruit}. Note that the disc of a gas component today is understood to be more extended than the stellar disc.

The rotation curve is demanded to be flat and the maximum velocity is estimated by applying the Tully-Fisher relation. The maximum rotational velocity to adjust the density profiles to is thus scaled using the mass ratio of today's and the model galaxy's disc mass. This leads to: 
\[
v_{\rm{max}}^{\rm{early}} = 115...135~\rm{km~s^{-1}},
\]
depending on the slope of the Tully-Fisher relation. An intermediate value is adopted, aiming at a rotation velocity of $v_{\rm{circ}}^{\rm{tot}} \approx 125~\rm{km~s^{-1}}$.

As the primary objective is to set up the disc component, whose mass-distribution is already fixed with the scale length, the rotation curve is adjusted with the dark matter halo. The halo properties are thus determined by demanding it to give that circular velocity. 

For the disc, the rotational velocity $v_{\rm{circ}}^{\rm{disc}}$\ at the cut-off radius can be estimated using Equation \ref{vrotequation}, inserting the total disc mass $M_{\rm{d}}$\ and the maximum disc radius $R_{\rm{d}}^{\rm{max}}$. Having $v_{\rm{circ}}^{\rm{tot}}$ and $v_{\rm{circ}}^{\rm{disc}}$, $v_{\rm{circ}}^{\rm{halo}}$ is easily calculated making use of the fact that circular velocities add in quadrature. This results in $v_{\rm{circ}}^{\rm{halo}} = 105\ \rm{km~s^{-1}}$.

In a Hernquist dark-matter halo profile \citep{hernquistprofile}, the density $\rho_{\rm{h}}$\ at a radial distance $R$ is given by
\[
\rho_{\rm{h}} = \frac{M_{\rm{H}}}{2 \pi} \frac{a}{R (R + a)^3} ,
\]
with the core radius $a$ and the total halo mass $M_{\rm{H}}$. The mass inside a radius $R$ is
\begin{equation}
\label{hernquistmass}
M(<R) = M_{\rm{H}} \cdot \left(1 - \frac{2}{\frac{R}{a} + 1} + \frac{1}{\left(\frac{R}{a} + 1\right)^2}\right) .
\end{equation}
Again making use of Equation \ref{vrotequation}, the mass within $R = R_{\rm{d}}^{\rm{max}}$, the maximum disc particle radius, is determined. With the use of Eq. \ref{hernquistmass} the halo core radius can be estimated.

The model's halo needs to have $1.9 \times 10^{10}\ \rm{M_{\sun}}$ within a radius of $R_{\rm{d}}^{\rm{max}} = 8.0$~kpc. In the case of a finite Hernquist profile with maximum radius $R_{\rm{H}}^{\rm{max}} = 100~\rm{kpc}$ this results in a core radius of $a = 10.0$~kpc. After virialisation, the rotation curve of the disc component has a maximum at $\approx 122~\rm{km~s}^{-1}$.

It should be emphasized that this `dark-matter' halo has not been constructed for consistency with the CDM hypothesis, but merely to model the actual gravitational behaviour of the empirical flat rotation curves of the young galaxy. What the true explanation for this gravitational behaviour might be is the topic of other studies \citep[see e.g.][]{kroupa10}.

For a quicker reference, the model parameters are compiled in Table \ref{modelparams}. Interestingly, the resulting values are quite similar to those found for M33, which is a bulge-less galaxy, too \citep[see for example][]{magrini}, justifying the choice of parameters as physically reasonable.

Furthermore, several models with a mass ratio of 4:1 between the target and the infalling galaxy are calculated. Their parameters are determined in the same manner and can also be found in Table \ref{modelparams}, while their results are discussed in Sect. \ref{parascansect}.

\subsection{Model setup}
\label{modelsetupsec}
The model galaxy is created using the code \textsc{MaGalie} \citep{magalie}. It is based on a method by \citet{hernquist-setup} to generate multi-component galaxies. The code is highly efficient for generating multi-component disc galaxies with many particles as the computational time scales only linearly with the particle number.

The disc components are set up with $5 \times 10^5$\ particles, the halo components are made up of $10^6$\ particles. These numbers represent a compromise between resolution (demanding high particle numbers) and calculation time. They result in disc particle masses of $1.6 \times 10^{3}\ \rm{M_{\sun}}$ for the equal-mass model or $0.8 \times 10^{3}\ \rm{M_{\sun}}$ and $3.2 \times 10^{3}\ \rm{M_{\sun}}$ for the infalling and target galaxy in the 4-to-1 mass ratio models.

As the integration code \sbpp\ demands all particles in one galaxy to be of equal mass, the following standard work-around is used to allow the modelling of the stellar disc with a suitable number of particles without increasing the total particle number to an unfeasible value. The idea is to split up the galaxy into a disc and a halo component and to treat them as separate galaxies whose centres of mass have the same position. That way, the mass per halo particle can be higher, resulting in fewer halo particles and thus speeding up the computations. Furthermore, the splitting in two models allows to tailor the grid sizes to the respective components' size and therefore gives a higher resolution to the important disc components. The total number of grids is doubled by this splitting, leading to an increase in computational time by about a factor two. But the reduced computation time through the smaller number of halo particles, scaling about linearly, outweights this.

The number of grid cells per dimension is 64.
The gridsizes (radii) are 10, 30, and 400 kpc for the disc component galaxy model and 40, 120 and 400 kpc for the dark matter halo component. This gives grid cells with side lengths of 0.33, 1.0 and 13.33 kpc for the disc component and 1.33, 4.0 and 13.33 kpc for the halo component.

The newly set up model still suffers from minor numerical set-up mismatches because velocities and positions are stored as discrete values. Consequently the galaxy has to virialise to arrive at dynamical equilibrium. For this the model, consisting of disc and halo component, is calculated in isolation with \sbpp\ for 3000 steps with a time-step size of 0.5 Myr, corresponding to about 7 orbiting times at a radius of 4 kpc. The galaxy is in dynamical equilibrium thereafter. The total energy error during virialisation is less than 0.2 per cent, the Lagrange-Radii expand slightly as the sharp edge at the maximum particle distance cut-off is smoothed out.

\section{Tables of parameter scans}
\label{tablesect}

\begin{table*}
\begin{minipage}{180mm}
 \caption{Equal-mass fly-by: particle numbers}
 \label{flybynumbers}
 \begin{center}
 \begin{tabular}{@{}lccccccccc}
  \hline
  Model & $v_{\rm{ini}}$ [$v_{\rm{parab}}$] & $r_{\rm{min}}$ [kpc] & Steps analysed & $N^{\rm{all}}$ & $N^{\rm{far}}$ & $P^{\rm{all}}_{\rm{retro}}$ & $P^{\rm{far}}_{\rm{retro}}$ & $N^{\rm{60-250 kpc}}$ & $P^{\rm{60-250 kpc}}_{\rm{retro}}$\\
  \hline
5deg180vel & 1.8 & 7.44 & 15 & 4898 & 805 & 0.19 & 0.01 & 1641 & 0.43 \\
6deg180vel & 1.8 & 9.39 & 15 & 2874 & 474 & 0.13 & 0.00 & 281 & 0.17 \\
7deg180vel & 1.8 & 11.52 & 15 & 2011 & 387 & 0.15 & 0.00 & 322 & 0.07 \\
8deg180vel & 1.8 & 13.81 & 15 & 1476 & 428 & 0.09 & 0.00 & 362 & 0.00 \\
  \hline
5deg200vel & 2.0 & 8.35 & 15 & 5747 & 2090 & 0.40 & 0.08 & 1741 & 0.02 \\
6deg200vel & 2.0 & 10.54 & 15 & 4644 & 1316 & 0.44 & 0.07 & 1119 & 0.04 \\
7deg200vel & 2.0 & 12.91 & 15 & 2848 & 902 & 0.43 & 0.05 & 663 & 0.02 \\
8deg200vel & 2.0 & 15.45 & 15 & 1636 & 655 & 0.42 & 0.01 & 583 & 0.01 \\
  \hline
5deg220vel & 2.2 & 9.15 & 15 & 3233 & 1372 & 0.56 & 0.13 & 1292 & 0.06 \\
6deg220vel & 2.2 & 11.57 & 15 & 3836 & 1195 & 0.63 & 0.09 & 1046 & 0.03 \\
7deg220vel & 2.2 & 14.15 & 15 & 2177 & 650 & 0.66 & 0.10 & 616 & 0.06 \\
8deg220vel & 2.2 & 16.89 & 15 & 849 & 364 & 0.51 & 0.09 & 393 & 0.04 \\
  \hline
5deg240vel & 2.4 & 9.98 & 15 & 576 & 405 & 0.73 & 0.62 & 694 & 0.28 \\
6deg240vel & 2.4 & 12.56 & 15 & 2425 & 1173 & 0.93 & 0.85 & 1171 & 0.55 \\
7deg240vel & 2.4 & 15.29 & 15 & 1873 & 693 & 0.98 & 0.95 & 607 & 0.63 \\
8deg240vel & 2.4 & 18.17 & 15 & 862 & 479 & 0.99 & 0.99 & 363 & 0.80 \\
  \hline
 \end{tabular}
 \end{center}

 \small \smallskip

$v_{\rm{ini}}$: initial relative velocity of the two galaxies in units of the point-mass parabolic velocity.

$r_{\rm{min}}$: pericentre distance between the two galaxy centres of density for the first passage.

Steps analysed: number of snapshot files (in steps of 0.5 Gyr) analysed for determining pro- and retrograde motions. Particles had to appear within the circles of acceptance in 12 out of these snapshots. In those cases where the galaxies merged late, fewer than 12 snapshots have been analysed. In these cases, the particles had to appear within the circles in all snapshots.

$N$: sum of pro- and retrograde particles.

$P_{\rm{retro}}$: fraction of retrograde particles with respect to the sum of pro- and retrograde ones.

$N^{\rm{60-250 kpc}}$: sum of pro- and retrograde particles with galactic distances in the range of 60 to 250 kpc at the final snapshot.

$P^{\rm{60-250 kpc}}_{\rm{retro}}$: fraction of retrograde particles in $N^{\rm{60-250 kpc}}$.
\end{minipage}
\end{table*}

\begin{table*}
\begin{minipage}{180mm}
 \caption{Equal-mass fly-by: properties of the two populations}
 \label{flybyparameters}
 \begin{center}
 \begin{tabular}{@{}lcccccccc}
  \hline
  Model & $ e^{\rm{all}}_{\rm{pro}} $ &  $ e^{\rm{all}}_{\rm{retro}} $ &  $ e^{\rm{far}}_{\rm{pro}} $ &  $ e^{\rm{far}}_{\rm{retro}} $ & $ D^{\rm{all}}_{\rm{pro}} $ [kpc] & $ D^{\rm{all}}_{\rm{retro}} $ [kpc] & $ D^{\rm{far}}_{\rm{pro}} $ [kpc] & $ D^{\rm{far}}_{\rm{retro}} $ [kpc]\\
  \hline
5deg180vel &  0.74  $\pm$  0.11 & 0.79  $\pm$  0.12 & 0.75  $\pm$  0.10 & 0.85  $\pm$  0.05  &  5.67  $\pm$  1.19 & 7.26  $\pm$  1.16 & 5.97  $\pm$  1.34 & 5.09  $\pm$  0.39 \\
6deg180vel &  0.78  $\pm$  0.09 & 0.81  $\pm$  0.11 & 0.82  $\pm$  0.07 & --  &  5.71  $\pm$  1.17 & 7.16  $\pm$  1.25 & 5.83  $\pm$  1.22 & -- \\
7deg180vel &  0.82  $\pm$  0.08 & 0.82  $\pm$  0.08 & 0.81  $\pm$  0.06 & --  &  5.91  $\pm$  1.23 & 7.02  $\pm$  1.36 & 5.95  $\pm$  1.31 & -- \\
8deg180vel &  0.84  $\pm$  0.06 & 0.84  $\pm$  0.07 & 0.85  $\pm$  0.04 & --  &  6.01  $\pm$  1.24 & 7.55  $\pm$  1.21 & 5.98  $\pm$  1.22 & -- \\
  \hline
5deg200vel &  0.82  $\pm$  0.08 & 0.80  $\pm$  0.11 & 0.80  $\pm$  0.08 & 0.93  $\pm$  0.03  &  5.89  $\pm$  1.05 & 6.73  $\pm$  0.68 & 6.07  $\pm$  1.07 & 7.23  $\pm$  0.71 \\
6deg200vel &  0.83  $\pm$  0.06 & 0.79  $\pm$  0.12 & 0.84  $\pm$  0.04 & 0.90  $\pm$  0.04  &  5.71  $\pm$  1.05 & 6.65  $\pm$  1.17 & 5.85  $\pm$  1.11 & 7.78  $\pm$  0.52 \\
7deg200vel &  0.88  $\pm$  0.04 & 0.81  $\pm$  0.08 & 0.88  $\pm$  0.04 & 0.91  $\pm$  0.03  &  5.81  $\pm$  1.11 & 6.98  $\pm$  1.07 & 5.90  $\pm$  1.14 & 7.90  $\pm$  0.38 \\
8deg200vel &  0.91  $\pm$  0.03 & 0.85  $\pm$  0.06 & 0.90  $\pm$  0.03 & 0.93  $\pm$  0.03  &  6.61  $\pm$  0.76 & 7.25  $\pm$  0.78 & 6.59  $\pm$  0.81 & 8.34  $\pm$  0.25 \\
  \hline
5deg220vel &  0.89  $\pm$  0.04 & 0.81  $\pm$  0.11 & 0.89  $\pm$  0.04 & 0.93  $\pm$  0.03  &  6.02  $\pm$  0.92 & 6.86  $\pm$  0.88 & 6.08  $\pm$  0.94 & 7.24  $\pm$  0.93 \\
6deg220vel &  0.91  $\pm$  0.03 & 0.81  $\pm$  0.08 & 0.91  $\pm$  0.03 & 0.93  $\pm$  0.02  &  5.89  $\pm$  1.02 & 6.76  $\pm$  1.00 & 5.92  $\pm$  1.02 & 7.17  $\pm$  1.06 \\
7deg220vel &  0.93  $\pm$  0.02 & 0.85  $\pm$  0.06 & 0.93  $\pm$  0.02 & 0.92  $\pm$  0.04  &  6.56  $\pm$  0.77 & 7.20  $\pm$  0.86 & 6.58  $\pm$  0.80 & 7.61  $\pm$  0.87 \\
8deg220vel &  0.94  $\pm$  0.02 & 0.88  $\pm$  0.05 & 0.94  $\pm$  0.01 & 0.93  $\pm$  0.02  &  7.08  $\pm$  0.64 & 7.95  $\pm$  0.64 & 7.09  $\pm$  0.66 & 8.01  $\pm$  0.80 \\
  \hline
5deg240vel &  0.96  $\pm$  0.02 & 0.88  $\pm$  0.09 & 0.96  $\pm$  0.02 & 0.92  $\pm$  0.04  &  6.87  $\pm$  0.51 & 7.48  $\pm$  0.42 & 6.87  $\pm$  0.51 & 7.46  $\pm$  0.43 \\
6deg240vel &  0.97  $\pm$  0.02 & 0.84  $\pm$  0.07 & 0.97  $\pm$  0.02 & 0.89  $\pm$  0.06  &  6.86  $\pm$  0.50 & 7.56  $\pm$  0.46 & 6.86  $\pm$  0.50 & 7.54  $\pm$  0.51 \\
7deg240vel &  0.98  $\pm$  0.01 & 0.85  $\pm$  0.06 & 0.98  $\pm$  0.01 & 0.91  $\pm$  0.04  &  7.46  $\pm$  0.25 & 7.75  $\pm$  0.39 & 7.46  $\pm$  0.25 & 7.79  $\pm$  0.39 \\
8deg240vel &  0.97  $\pm$  0.02 & 0.90  $\pm$  0.04 & 0.98  $\pm$  0.01 & 0.92  $\pm$  0.03  &  7.58  $\pm$  0.21 & 7.94  $\pm$  0.39 & 7.58  $\pm$  0.24 & 7.85  $\pm$  0.40 \\
  \hline
 \end{tabular}
 \end{center}

 \small \smallskip

$e$: mean orbit eccentricity, uncertainties express the standard deviations of the distributions.

$D$: mean initial distances in kpc from the infalling galaxy's centre of density, uncertainties as before.
\end{minipage}
\end{table*}

\begin{table*}
\begin{minipage}{180mm}
 \caption{Equal-mass merger, infalling galaxy retrograde: particle numbers. Parameters defined as with Table \ref{flybynumbers}.}
 \label{mergerretronumbers}
 \begin{center}
 \begin{tabular}{@{}lccccccccc}
  \hline
  Model & $v_{\rm{ini}}$ [$v_{\rm{parab}}$] & $r_{\rm{min}}$ [kpc] & Steps analysed & $N^{\rm{all}}$ & $N^{\rm{far}}$ & $P^{\rm{all}}_{\rm{retro}}$ & $P^{\rm{far}}_{\rm{retro}}$ & $N^{\rm{60-250 kpc}}$ & $P^{\rm{60-250 kpc}}_{\rm{retro}}$\\
  \hline
0deg050vel & 0.5 & 0.08 & 15 & 10355 & 47 & 0.99 & 0.96 & 61 & 0.48 \\
2.5deg050vel & 0.5 & 1.00 & 15 & 16314 & 7 & 0.97 & 0.57 & 99 & 0.68 \\
5deg050vel & 0.5 & 1.93 & 14 & 19276 & 10 & 0.93 & 0.40 & 178 & 0.60 \\
7.5deg050vel & 0.5 & 2.80 & 15 & 26740 & 41 & 0.93 & 0.51 & 123 & 0.56 \\
10deg050vel & 0.5 & 3.67 & 15 & 25168 & 326 & 0.92 & 0.79 & 195 & 0.49 \\
15deg050vel & 0.5 & 5.70 & 15 & 17146 & 1094 & 0.96 & 0.67 & 545 & 0.61 \\
20deg050vel & 0.5 & 8.09 & 14 & 15572 & 3842 & 0.95 & 0.90 & 2959 & 0.89 \\
  \hline
0deg100vel & 1.0 & 0.16 & 15 & 10705 & 770 & 0.88 & 0.82 & 1973 & 0.57 \\
2.5deg100vel & 1.0 & 1.84 & 15 & 13672 & 2657 & 0.88 & 0.73 & 3079 & 0.81 \\
5deg100vel & 1.0 & 3.80 & 15 & 12980 & 4313 & 0.78 & 0.55 & 4271 & 0.54 \\
7.5deg100vel & 1.0 & 5.97 & 15 & 15574 & 6050 & 0.94 & 0.91 & 4322 & 0.76 \\
10deg100vel & 1.0 & 8.64 & 15 & 12442 & 3780 & 0.99 & 1.00 & 1704 & 0.92 \\
15deg100vel & 1.0 & 15.60 & 13 & 5848 & 188 & 0.63 & 0.40 & 516 & 0.23 \\
20deg100vel & 1.0 & 25.06 & 8 out of 8 & 13687 & 579 & 0.83 & 0.63 & 267 & 0.43 \\
  \hline
0deg150vel & 1.5 & 0.12 & 14 & 11907 & 3244 & 0.88 & 0.95 & 2668 & 0.79 \\
2.5deg150vel & 1.5 & 2.78 & 14 & 17636 & 4061 & 0.95 & 0.96 & 2991 & 0.86 \\
5deg150vel & 1.5 & 6.01 & 13 & 20405 & 2932 & 0.98 & 0.99 & 2214 & 0.88 \\
7.5deg150vel & 1.5 & 10.15 & 9 out of 9 & 9008 & 1346 & 0.96 & 1.00 & 2299 & 0.72 \\
  \hline
 \end{tabular}
 \end{center}

\end{minipage}
\end{table*}

\begin{table*}
\begin{minipage}{180mm}
 \caption{Equal-mass merger, infalling galaxy retrograde: properties of the two populations. Parameters defined as with Table \ref{flybyparameters}.}
 \label{mergerretroparameters}
 \begin{center}
 \begin{tabular}{@{}lcccccccc}
  \hline
  Model & $ e^{\rm{all}}_{\rm{pro}} $ &  $ e^{\rm{all}}_{\rm{retro}} $ &  $ e^{\rm{far}}_{\rm{pro}} $ &  $ e^{\rm{far}}_{\rm{retro}} $ & $ D^{\rm{all}}_{\rm{pro}} $ [kpc] & $ D^{\rm{all}}_{\rm{retro}} $ [kpc] & $ D^{\rm{far}}_{\rm{pro}} $ [kpc] & $ D^{\rm{far}}_{\rm{retro}} $ [kpc]\\
  \hline
0deg050vel &  0.78  $\pm$  0.08 & 0.46  $\pm$  0.09 & 0.81  $\pm$  0.04 & 0.84  $\pm$  0.05  &  7.36  $\pm$  0.75 & 5.87  $\pm$  0.97 & 8.67  $\pm$  0.01 & 6.65  $\pm$  0.56 \\
2.5deg050vel &  0.76  $\pm$  0.08 & 0.54  $\pm$  0.14 & 0.86  $\pm$  0.04 & 0.79  $\pm$  0.08  &  7.07  $\pm$  0.89 & 5.78  $\pm$  1.03 & 7.17  $\pm$  2.12 & 6.85  $\pm$  1.75 \\
5deg050vel &  0.77  $\pm$  0.08 & 0.56  $\pm$  0.13 & 0.88  $\pm$  0.04 & 0.80  $\pm$  0.12  &  7.18  $\pm$  0.72 & 5.82  $\pm$  1.08 & 7.55  $\pm$  1.94 & 7.09  $\pm$  2.00 \\
7.5deg050vel &  0.79  $\pm$  0.09 & 0.64  $\pm$  0.14 & 0.87  $\pm$  0.04 & 0.83  $\pm$  0.04  &  7.25  $\pm$  0.86 & 6.05  $\pm$  1.18 & 7.29  $\pm$  1.10 & 7.24  $\pm$  0.91 \\
10deg050vel &  0.78  $\pm$  0.09 & 0.67  $\pm$  0.13 & 0.85  $\pm$  0.04 & 0.85  $\pm$  0.05  &  7.23  $\pm$  0.74 & 6.46  $\pm$  1.17 & 7.13  $\pm$  0.81 & 6.99  $\pm$  0.86 \\
15deg050vel &  0.83  $\pm$  0.13 & 0.70  $\pm$  0.10 & 0.92  $\pm$  0.05 & 0.83  $\pm$  0.04  &  6.78  $\pm$  0.93 & 6.83  $\pm$  1.08 & 6.60  $\pm$  0.77 & 7.18  $\pm$  0.57 \\
20deg050vel &  0.88  $\pm$  0.14 & 0.86  $\pm$  0.07 & 0.95  $\pm$  0.06 & 0.92  $\pm$  0.04  &  6.81  $\pm$  1.06 & 6.33  $\pm$  1.07 & 6.78  $\pm$  0.98 & 6.79  $\pm$  0.73 \\
  \hline
0deg100vel &  0.74  $\pm$  0.11 & 0.66  $\pm$  0.15 & 0.88  $\pm$  0.04 & 0.79  $\pm$  0.07  &  6.80  $\pm$  0.86 & 6.74  $\pm$  1.20 & 7.17  $\pm$  0.97 & 6.56  $\pm$  1.19 \\
2.5deg100vel &  0.78  $\pm$  0.08 & 0.65  $\pm$  0.13 & 0.83  $\pm$  0.05 & 0.82  $\pm$  0.04  &  6.97  $\pm$  0.73 & 6.59  $\pm$  1.14 & 7.05  $\pm$  0.71 & 5.98  $\pm$  1.00 \\
5deg100vel &  0.81  $\pm$  0.12 & 0.71  $\pm$  0.09 & 0.86  $\pm$  0.06 & 0.82  $\pm$  0.04  &  6.94  $\pm$  0.89 & 6.43  $\pm$  1.04 & 6.97  $\pm$  0.87 & 6.42  $\pm$  0.87 \\
7.5deg100vel &  0.80  $\pm$  0.16 & 0.78  $\pm$  0.09 & 0.90  $\pm$  0.08 & 0.84  $\pm$  0.05  &  7.17  $\pm$  0.65 & 5.97  $\pm$  1.09 & 7.07  $\pm$  0.67 & 6.33  $\pm$  1.05 \\
10deg100vel &  0.86  $\pm$  0.06 & 0.80  $\pm$  0.09 & 0.89  $\pm$  0.02 & 0.87  $\pm$  0.04  &  7.28  $\pm$  1.12 & 6.42  $\pm$  1.03 & 5.43  $\pm$  0.23 & 6.49  $\pm$  1.05 \\
15deg100vel &  0.82  $\pm$  0.08 & 0.84  $\pm$  0.10 & 0.95  $\pm$  0.04 & 0.96  $\pm$  0.02  &  5.03  $\pm$  0.97 & 6.35  $\pm$  1.26 & 6.55  $\pm$  1.18 & 5.96  $\pm$  1.14 \\
20deg100vel &  0.92  $\pm$  0.04 & 0.82  $\pm$  0.09 & 0.93  $\pm$  0.04 & 0.96  $\pm$  0.03  &  5.66  $\pm$  1.14 & 6.84  $\pm$  0.98 & 6.25  $\pm$  1.17 & 5.98  $\pm$  1.06 \\
  \hline
0deg150vel &  0.62  $\pm$  0.22 & 0.73  $\pm$  0.12 & 0.84  $\pm$  0.06 & 0.76  $\pm$  0.09  &  7.59  $\pm$  0.80 & 6.30  $\pm$  1.16 & 7.69  $\pm$  1.11 & 6.63  $\pm$  1.18 \\
2.5deg150vel &  0.67  $\pm$  0.20 & 0.73  $\pm$  0.11 & 0.83  $\pm$  0.07 & 0.78  $\pm$  0.06  &  7.55  $\pm$  0.88 & 6.41  $\pm$  1.10 & 7.73  $\pm$  1.10 & 6.64  $\pm$  0.93 \\
5deg150vel &  0.76  $\pm$  0.16 & 0.70  $\pm$  0.14 & 0.88  $\pm$  0.06 & 0.75  $\pm$  0.10  &  7.49  $\pm$  0.81 & 6.04  $\pm$  1.20 & 8.16  $\pm$  0.46 & 6.51  $\pm$  1.25 \\
7.5deg150vel &  0.84  $\pm$  0.09 & 0.77  $\pm$  0.10 & 0.93  $\pm$  0.00 & 0.88  $\pm$  0.07  &  5.00  $\pm$  1.12 & 6.26  $\pm$  1.14 & 6.46  $\pm$  1.19 & 6.23  $\pm$  1.23 \\
  \hline
 \end{tabular}
 \end{center}

\end{minipage}
\end{table*}

\begin{table*}
\begin{minipage}{180mm}
 \caption{Equal-mass merger, infalling galaxy prograde: particle numbers. Parameters defined as with Table \ref{flybynumbers}.}
 \label{mergerpronumbers}
 \begin{center}
 \begin{tabular}{@{}lccccccccc}
  \hline
  Model & $v_{\rm{ini}}$ [$v_{\rm{parab}}$] & $r_{\rm{min}}$ [kpc] & Steps analysed & $N^{\rm{all}}$ & $N^{\rm{far}}$ & $P^{\rm{all}}_{\rm{retro}}$ & $P^{\rm{far}}_{\rm{retro}}$ & $N^{\rm{60-250 kpc}}$ & $P^{\rm{60-250 kpc}}_{\rm{retro}}$\\
  \hline
0deg050vel & 0.5 & 0.06 & 15 & 12092 & 47 & 0.02 & 0.04 & 67 & 0.33 \\
2.5deg050vel & 0.5 & 1.01 & 15 & 10660 & 689 & 0.05 & 0.00 & 222 & 0.04 \\
5deg050vel & 0.5 & 1.95 & 14 & 14060 & 2144 & 0.07 & 0.00 & 1030 & 0.01 \\
7.5deg050vel & 0.5 & 2.90 & 15 & 18794 & 3854 & 0.06 & 0.03 & 2063 & 0.01 \\
10deg050vel & 0.5 & 3.85 & 15 & 22421 & 4936 & 0.02 & 0.02 & 3329 & 0.01 \\
15deg050vel & 0.5 & 5.91 & 15 & 37049 & 6678 & 0.04 & 0.02 & 4832 & 0.05 \\
20deg050vel & 0.5 & 8.12 & 14 & 38768 & 10862 & 0.03 & 0.00 & 5499 & 0.01 \\
  \hline
0deg100vel & 1.0 & 0.13 & 15 & 10642 & 690 & 0.12 & 0.20 & 1906 & 0.44 \\
2.5deg100vel & 1.0 & 2.00 & 15 & 4644 & 925 & 0.21 & 0.62 & 1231 & 0.57 \\
5deg100vel & 1.0 & 3.94 & 15 & 26051 & 9347 & 0.09 & 0.19 & 4440 & 0.28 \\
7.5deg100vel & 1.0 & 6.09 & 15 & 42760 & 17322 & 0.14 & 0.13 & 10387 & 0.20 \\
10deg100vel & 1.0 & 8.72 & 15 & 33369 & 18475 & 0.04 & 0.01 & 12315 & 0.06 \\
15deg100vel & 1.0 & 15.68 & 13 & 36641 & 6143 & 0.02 & 0.01 & 4415 & 0.00 \\
20deg100vel & 1.0 & 25.09 & 8 out of 8 & 35454 & 4620 & 0.00 & 0.00 & 9607 & 0.00 \\
  \hline
0deg150vel & 1.5 & 0.07 & 14 & 11498 & 3363 & 0.12 & 0.05 & 2819 & 0.21 \\
2.5deg150vel & 1.5 & 2.91 & 14 & 12372 & 2246 & 0.27 & 0.25 & 3466 & 0.42 \\
5deg150vel & 1.5 & 6.14 & 13 & 26998 & 7902 & 0.20 & 0.23 & 8727 & 0.39 \\
7.5deg150vel & 1.5 & 10.24 & 9 out of 9 & 25122 & 6423 & 0.16 & 0.11 & 11465 & 0.15 \\
  \hline
 \end{tabular}
 \end{center}

\end{minipage}
\end{table*}

\begin{table*}
\begin{minipage}{180mm}
 \caption{Equal-mass merger, infalling galaxy prograde: properties of the two populations. Parameters defined as with Table \ref{flybyparameters}.}
 \label{mergerproparameters}
 \begin{center}
 \begin{tabular}{@{}lcccccccc}
  \hline
  Model & $ e^{\rm{all}}_{\rm{pro}} $ &  $ e^{\rm{all}}_{\rm{retro}} $ &  $ e^{\rm{far}}_{\rm{pro}} $ &  $ e^{\rm{far}}_{\rm{retro}} $ & $ D^{\rm{all}}_{\rm{pro}} $ [kpc] & $ D^{\rm{all}}_{\rm{retro}} $ [kpc] & $ D^{\rm{far}}_{\rm{pro}} $ [kpc] & $ D^{\rm{far}}_{\rm{retro}} $ [kpc]\\
  \hline
0deg050vel &  0.48  $\pm$  0.09 & 0.79  $\pm$  0.08 & 0.83  $\pm$  0.04 & 0.82  $\pm$  0.04  &  5.77  $\pm$  0.98 & 7.26  $\pm$  0.71 & 6.54  $\pm$  0.52 & 8.67  $\pm$  0.01 \\
2.5deg050vel &  0.52  $\pm$  0.14 & 0.80  $\pm$  0.07 & 0.82  $\pm$  0.03 & --  &  5.79  $\pm$  1.01 & 6.97  $\pm$  0.94 & 6.76  $\pm$  0.68 & -- \\
5deg050vel &  0.66  $\pm$  0.17 & 0.84  $\pm$  0.07 & 0.80  $\pm$  0.04 & 0.85  $\pm$  0.04  &  6.34  $\pm$  1.13 & 5.91  $\pm$  1.12 & 6.96  $\pm$  0.78 & 6.16  $\pm$  0.44 \\
7.5deg050vel &  0.73  $\pm$  0.13 & 0.85  $\pm$  0.06 & 0.76  $\pm$  0.05 & 0.88  $\pm$  0.05  &  6.71  $\pm$  1.00 & 5.59  $\pm$  1.04 & 7.07  $\pm$  0.81 & 5.66  $\pm$  1.10 \\
10deg050vel &  0.74  $\pm$  0.10 & 0.87  $\pm$  0.07 & 0.71  $\pm$  0.05 & 0.91  $\pm$  0.05  &  6.66  $\pm$  1.02 & 5.65  $\pm$  1.07 & 7.08  $\pm$  0.81 & 5.76  $\pm$  1.08 \\
15deg050vel &  0.73  $\pm$  0.11 & 0.79  $\pm$  0.10 & 0.65  $\pm$  0.07 & 0.95  $\pm$  0.02  &  5.89  $\pm$  1.32 & 6.59  $\pm$  1.11 & 7.09  $\pm$  0.75 & 5.19  $\pm$  0.78 \\
20deg050vel &  0.64  $\pm$  0.16 & 0.79  $\pm$  0.11 & 0.67  $\pm$  0.16 & 0.96  $\pm$  0.00  &  5.86  $\pm$  1.21 & 6.96  $\pm$  0.65 & 6.18  $\pm$  1.08 & 7.81  $\pm$  0.00 \\
  \hline
0deg100vel &  0.66  $\pm$  0.15 & 0.74  $\pm$  0.11 & 0.78  $\pm$  0.06 & 0.88  $\pm$  0.04  &  6.74  $\pm$  1.19 & 6.79  $\pm$  0.86 & 6.50  $\pm$  1.22 & 7.21  $\pm$  1.00 \\
2.5deg100vel &  0.66  $\pm$  0.16 & 0.82  $\pm$  0.10 & 0.86  $\pm$  0.05 & 0.88  $\pm$  0.04  &  6.92  $\pm$  1.28 & 6.57  $\pm$  0.79 & 7.36  $\pm$  0.93 & 6.58  $\pm$  0.74 \\
5deg100vel &  0.79  $\pm$  0.11 & 0.89  $\pm$  0.06 & 0.83  $\pm$  0.05 & 0.91  $\pm$  0.04  &  5.98  $\pm$  1.25 & 5.99  $\pm$  1.09 & 6.88  $\pm$  0.71 & 6.00  $\pm$  1.10 \\
7.5deg100vel &  0.76  $\pm$  0.09 & 0.76  $\pm$  0.18 & 0.77  $\pm$  0.08 & 0.94  $\pm$  0.02  &  5.89  $\pm$  1.33 & 6.02  $\pm$  1.27 & 6.75  $\pm$  0.95 & 5.38  $\pm$  0.86 \\
10deg100vel &  0.74  $\pm$  0.10 & 0.85  $\pm$  0.10 & 0.74  $\pm$  0.11 & 0.95  $\pm$  0.05  &  5.98  $\pm$  1.34 & 5.87  $\pm$  1.36 & 6.52  $\pm$  1.23 & 5.48  $\pm$  1.45 \\
15deg100vel &  0.54  $\pm$  0.14 & 0.84  $\pm$  0.08 & 0.67  $\pm$  0.11 & 0.91  $\pm$  0.03  &  5.76  $\pm$  1.22 & 7.51  $\pm$  0.60 & 6.44  $\pm$  1.16 & 7.93  $\pm$  0.31 \\
20deg100vel &  0.51  $\pm$  0.17 & 0.80  $\pm$  0.19 & 0.56  $\pm$  0.13 & --  &  5.62  $\pm$  1.32 & 7.01  $\pm$  0.70 & 6.21  $\pm$  1.22 & -- \\
  \hline
0deg150vel &  0.73  $\pm$  0.12 & 0.63  $\pm$  0.22 & 0.75  $\pm$  0.09 & 0.84  $\pm$  0.06  &  6.31  $\pm$  1.15 & 7.60  $\pm$  0.77 & 6.61  $\pm$  1.18 & 7.88  $\pm$  0.88 \\
2.5deg150vel &  0.77  $\pm$  0.12 & 0.67  $\pm$  0.18 & 0.85  $\pm$  0.06 & 0.85  $\pm$  0.07  &  6.05  $\pm$  1.26 & 7.20  $\pm$  0.85 & 6.44  $\pm$  1.20 & 7.38  $\pm$  0.64 \\
5deg150vel &  0.74  $\pm$  0.15 & 0.68  $\pm$  0.19 & 0.86  $\pm$  0.09 & 0.79  $\pm$  0.13  &  5.61  $\pm$  1.34 & 6.66  $\pm$  1.22 & 6.62  $\pm$  1.29 & 6.90  $\pm$  1.08 \\
7.5deg150vel &  0.67  $\pm$  0.17 & 0.68  $\pm$  0.17 & 0.75  $\pm$  0.12 & 0.80  $\pm$  0.11  &  5.68  $\pm$  1.22 & 6.75  $\pm$  1.35 & 5.82  $\pm$  1.11 & 7.11  $\pm$  1.19 \\
  \hline
 \end{tabular}
 \end{center}

\end{minipage}
\end{table*}

\begin{table*}
\begin{minipage}{180mm}
 \caption{4-to-1-mass fly-by: particle numbers. Parameters defined as with Table \ref{flybynumbers}.}
 \label{4to1flybynumbers}
 \begin{center}
 \begin{tabular}{@{}lccccccccc}
  \hline
  Model & $v_{\rm{ini}}$ [$v_{\rm{parab}}$] & $r_{\rm{min}}$ [kpc] & Steps analysed & $N^{\rm{all}}$ & $N^{\rm{far}}$ & $P^{\rm{all}}_{\rm{retro}}$ & $P^{\rm{far}}_{\rm{retro}}$ & $N^{\rm{60-250 kpc}}$ & $P^{\rm{60-250 kpc}}_{\rm{retro}}$\\
  \hline
4deg175vel & 1.75 & 6.89 & 16 & 11767 & 3813 & 0.33 & 0.00 & 2216 & 0.02 \\
6deg175vel & 1.75 & 11.35 & 16 & 4116 & 2026 & 0.02 & 0.00 & 1581 & 0.04 \\
8deg175vel & 1.75 & 16.86 & 16 & 3951 & 1746 & 0.00 & 0.00 & 1179 & 0.00 \\
\hline
2deg200vel & 2.00 & 3.67 & 16 & 1263 & 628 & 0.03 & 0.05 & 458 & 0.09 \\
4deg200vel & 2.00 & 7.84 & 15 & 12689 & 4745 & 0.50 & 0.01 & 3203 & 0.01 \\
6deg200vel & 2.00 & 13.09 & 16 & 3295 & 2279 & 0.08 & 0.00 & 1660 & 0.01 \\
\hline
4deg225vel & 2.25 & 8.84 & 16 & 13942 & 6095 & 0.64 & 0.21 & 4677 & 0.08 \\
6deg225vel & 2.25 & 14.76 & 16 & 3684 & 2919 & 0.04 & 0.01 & 2303 & 0.01 \\
\hline
4deg250vel & 2.50 & 9.73 & 16 & 9266 & 5837 & 0.94 & 0.90 & 5049 & 0.56 \\
6deg250vel & 2.50 & 16.21 & 15 (steps 4k to 18k) & 1265 & 1138 & 0.17 & 0.19 & 1674 & 0.10 \\
  \hline
 \end{tabular}
 \end{center}

\end{minipage}
\end{table*}

\begin{table*}
\begin{minipage}{180mm}
 \caption{4-to-1-mass fly-by: properties of the two populations. Parameters defined as with Table \ref{flybyparameters}.}
 \label{4to1flybyparameters}
 \begin{center}
 \begin{tabular}{@{}lcccccccc}
  \hline
  Model & $ e^{\rm{all}}_{\rm{pro}} $ &  $ e^{\rm{all}}_{\rm{retro}} $ &  $ e^{\rm{far}}_{\rm{pro}} $ &  $ e^{\rm{far}}_{\rm{retro}} $ & $ D^{\rm{all}}_{\rm{pro}} $ [kpc] & $ D^{\rm{all}}_{\rm{retro}} $ [kpc] & $ D^{\rm{far}}_{\rm{pro}} $ [kpc] & $ D^{\rm{far}}_{\rm{retro}} $ [kpc]\\
  \hline
4deg175vel &  0.83  $\pm$  0.06 & 0.81  $\pm$  0.06 & 0.83  $\pm$  0.06 & 0.88  $\pm$  0.04  &  4.06  $\pm$  0.85 & 4.68  $\pm$  0.82 & 3.98  $\pm$  0.86 & 5.90  $\pm$  0.48 \\
6deg175vel &  0.88  $\pm$  0.05 & 0.84  $\pm$  0.06 & 0.88  $\pm$  0.04 & 0.98  $\pm$  0.00  &  4.07  $\pm$  0.81 & 5.43  $\pm$  0.95 & 4.04  $\pm$  0.82 & 6.17  $\pm$  0.00 \\
8deg175vel &  0.88  $\pm$  0.05 & 0.86  $\pm$  0.05 & 0.91  $\pm$  0.03 & --  &  4.59  $\pm$  0.77 & 5.11  $\pm$  0.97 & 4.41  $\pm$  0.74 & -- \\
\hline
2deg200vel &  0.83  $\pm$  0.07 & 0.95  $\pm$  0.02 & 0.87  $\pm$  0.05 & 0.95  $\pm$  0.02  &  5.68  $\pm$  0.47 & 5.69  $\pm$  0.57 & 5.69  $\pm$  0.57 & 5.67  $\pm$  0.57 \\
4deg200vel &  0.88  $\pm$  0.04 & 0.82  $\pm$  0.06 & 0.88  $\pm$  0.04 & 0.90  $\pm$  0.04  &  4.09  $\pm$  0.77 & 4.88  $\pm$  0.89 & 4.07  $\pm$  0.78 & 5.69  $\pm$  0.90 \\
6deg200vel &  0.93  $\pm$  0.04 & 0.86  $\pm$  0.06 & 0.94  $\pm$  0.02 & 0.94  $\pm$  0.02  &  4.43  $\pm$  0.70 & 5.32  $\pm$  0.68 & 4.40  $\pm$  0.69 & 5.71  $\pm$  0.84 \\
\hline
4deg225vel &  0.94  $\pm$  0.02 & 0.85  $\pm$  0.06 & 0.94  $\pm$  0.02 & 0.91  $\pm$  0.04  &  4.26  $\pm$  0.68 & 4.64  $\pm$  0.74 & 4.26  $\pm$  0.67 & 4.63  $\pm$  0.81 \\
6deg225vel &  0.94  $\pm$  0.03 & 0.89  $\pm$  0.04 & 0.95  $\pm$  0.02 & 0.92  $\pm$  0.03  &  4.48  $\pm$  0.74 & 6.26  $\pm$  0.40 & 4.39  $\pm$  0.75 & 5.96  $\pm$  0.58 \\
\hline
4deg250vel &  0.97  $\pm$  0.01 & 0.87  $\pm$  0.06 & 0.97  $\pm$  0.01 & 0.91  $\pm$  0.04  &  4.40  $\pm$  0.67 & 4.98  $\pm$  0.67 & 4.40  $\pm$  0.67 & 4.84  $\pm$  0.74 \\
6deg250vel &  0.95  $\pm$  0.02 & 0.94  $\pm$  0.03 & 0.96  $\pm$  0.01 & 0.95  $\pm$  0.03  &  5.44  $\pm$  0.30 & 5.87  $\pm$  0.38 & 5.42  $\pm$  0.31 & 5.86  $\pm$  0.38 \\
  \hline
 \end{tabular}
 \end{center}

\end{minipage}
\end{table*}

\begin{table*}
\begin{minipage}{180mm}
 \caption{4-to-1-mass merger, retrograde infalling galaxy: particle numbers. Parameters defined as with Table \ref{flybynumbers}.}
 \label{4to1mergerretronumbers}
 \begin{center}
 \begin{tabular}{@{}lccccccccc}
  \hline
  Model & $v_{\rm{ini}}$ [$v_{\rm{parab}}$] & $r_{\rm{min}}$ [kpc] & Steps analysed & $N^{\rm{all}}$ & $N^{\rm{far}}$ & $P^{\rm{all}}_{\rm{retro}}$ & $P^{\rm{far}}_{\rm{retro}}$ & $N^{\rm{60-250 kpc}}$ & $P^{\rm{60-250 kpc}}_{\rm{retro}}$\\
  \hline
2.5deg050vel & 0.5 & 1.16 & 13 & 62165 & 19714 & 0.63 & 0.51 & 17625 & 0.37 \\
5deg050vel & 0.5 & 2.33 & 13 & 49654 & 23292 & 0.62 & 0.62 & 19567 & 0.45 \\
\hline
2.5deg100vel & 1.0 & 2.28 & 13 & 62118 & 27878 & 0.45 & 0.58 & 28341 & 0.53 \\
5deg100vel & 1.0 & 4.76 & 13 & 28922 & 18078 & 0.44 & 0.48 & 20009 & 0.38 \\
7.5deg100vel & 1.0 & 7.50 & 13 & 13263 & 8192 & 0.23 & 0.03 & 13521 & 0.14 \\
10deg100vel & 1.0 & 11.02 & 11 out of 11 & 15762 & 2067 & 0.30 & 0.06 & 5071 & 0.48 \\
  \hline
 \end{tabular}
 \end{center}

\end{minipage}
\end{table*}

\begin{table*}
\begin{minipage}{180mm}
 \caption{4-to-1-mass merger, retrograde infalling galaxy: properties of the two populations. Parameters defined as with Table \ref{flybyparameters}.}
 \label{4to1mergerretroparameters}
 \begin{center}
 \begin{tabular}{@{}lcccccccc}
  \hline
  Model & $ e^{\rm{all}}_{\rm{pro}} $ &  $ e^{\rm{all}}_{\rm{retro}} $ &  $ e^{\rm{far}}_{\rm{pro}} $ &  $ e^{\rm{far}}_{\rm{retro}} $ & $ D^{\rm{all}}_{\rm{pro}} $ [kpc] & $ D^{\rm{all}}_{\rm{retro}} $ [kpc] & $ D^{\rm{far}}_{\rm{pro}} $ [kpc] & $ D^{\rm{far}}_{\rm{retro}} $ [kpc]\\
  \hline
2.5deg050vel &  0.86  $\pm$  0.07 & 0.86  $\pm$  0.06 & 0.89  $\pm$  0.06 & 0.85  $\pm$  0.05  &  4.08  $\pm$  1.03 & 3.91  $\pm$  0.93 & 4.53  $\pm$  0.89 & 4.34  $\pm$  0.87 \\
5deg050vel &  0.83  $\pm$  0.10 & 0.87  $\pm$  0.06 & 0.88  $\pm$  0.07 & 0.87  $\pm$  0.06  &  4.35  $\pm$  1.02 & 4.16  $\pm$  0.93 & 4.63  $\pm$  0.92 & 4.36  $\pm$  0.83 \\
\hline
2.5deg100vel &  0.85  $\pm$  0.08 & 0.87  $\pm$  0.05 & 0.83  $\pm$  0.10 & 0.88  $\pm$  0.05  &  3.46  $\pm$  1.15 & 4.20  $\pm$  1.00 & 4.40  $\pm$  0.96 & 4.55  $\pm$  0.91 \\
5deg100vel &  0.86  $\pm$  0.09 & 0.88  $\pm$  0.05 & 0.89  $\pm$  0.08 & 0.90  $\pm$  0.04  &  4.20  $\pm$  1.01 & 4.62  $\pm$  0.94 & 4.53  $\pm$  0.93 & 4.96  $\pm$  0.73 \\
7.5deg100vel &  0.87  $\pm$  0.12 & 0.86  $\pm$  0.05 & 0.92  $\pm$  0.05 & 0.92  $\pm$  0.03  &  4.35  $\pm$  1.05 & 4.26  $\pm$  0.86 & 4.42  $\pm$  1.01 & 5.55  $\pm$  0.50 \\
10deg100vel &  0.76  $\pm$  0.11 & 0.84  $\pm$  0.04 & 0.90  $\pm$  0.04 & 0.90  $\pm$  0.03  &  4.25  $\pm$  1.04 & 4.26  $\pm$  0.86 & 4.53  $\pm$  0.88 & 5.10  $\pm$  0.99 \\
  \hline
 \end{tabular}
 \end{center}

\end{minipage}
\end{table*}

\begin{table*}
\begin{minipage}{180mm}
 \caption{4-to-1-mass merger, prograde infalling galaxy: particle numbers. Parameters defined as with Table \ref{flybynumbers}.}
 \label{4to1mergerpronumbers}
 \begin{center}
 \begin{tabular}{@{}lccccccccc}
  \hline
  Model & $v_{\rm{ini}}$ [$v_{\rm{parab}}$] & $r_{\rm{min}}$ [kpc] & Steps analysed & $N^{\rm{all}}$ & $N^{\rm{far}}$ & $P^{\rm{all}}_{\rm{retro}}$ & $P^{\rm{far}}_{\rm{retro}}$ & $N^{\rm{60-250 kpc}}$ & $P^{\rm{60-250 kpc}}_{\rm{retro}}$\\
  \hline
2.5deg050vel & 0.5 & 1.29 & 13 & 42268 & 10307 & 0.44 & 0.66 & 14741 & 0.66 \\
5deg050vel & 0.5 & 2.44 & 13 & 47474 & 12688 & 0.26 & 0.39 & 16696 & 0.50 \\
\hline
2.5deg100vel & 1.0 & 2.38 & 13 & 70820 & 26464 & 0.28 & 0.53 & 27727 & 0.61 \\
5deg100vel & 1.0 & 4.79 & 13 & 62362 & 31521 & 0.06 & 0.09 & 29706 & 0.10 \\
7.5deg100vel & 1.0 & 7.49 & 12 & 46993 & 29979 & 0.01 & 0.00 & 28675 & 0.04 \\
10deg100vel & 1.0 & 11.09 & 11 out of 11 & 33946 & 19700 & 0.05 & 0.01 & 23483 & 0.05 \\
  \hline
 \end{tabular}
 \end{center}

\end{minipage}
\end{table*}

\begin{table*}
\begin{minipage}{180mm}
 \caption{4-to-1-mass merger, prograde infalling galaxy: properties of the two populations. Parameters defined as with Table \ref{flybyparameters}.}
 \label{4to1mergerproparameters}
 \begin{center}
 \begin{tabular}{@{}lcccccccc}
  \hline
  Model & $ e^{\rm{all}}_{\rm{pro}} $ &  $ e^{\rm{all}}_{\rm{retro}} $ &  $ e^{\rm{far}}_{\rm{pro}} $ &  $ e^{\rm{far}}_{\rm{retro}} $ & $ D^{\rm{all}}_{\rm{pro}} $ [kpc] & $ D^{\rm{all}}_{\rm{retro}} $ [kpc] & $ D^{\rm{far}}_{\rm{pro}} $ [kpc] & $ D^{\rm{far}}_{\rm{retro}} $ [kpc]\\
  \hline
2.5deg050vel &  0.82  $\pm$  0.11 & 0.89  $\pm$  0.07 & 0.86  $\pm$  0.09 & 0.94  $\pm$  0.04  &  3.61  $\pm$  1.05 & 3.92  $\pm$  0.83 & 4.52  $\pm$  1.05 & 4.18  $\pm$  0.77 \\
5deg050vel &  0.86  $\pm$  0.08 & 0.90  $\pm$  0.06 & 0.90  $\pm$  0.06 & 0.93  $\pm$  0.04  &  3.46  $\pm$  0.99 & 3.96  $\pm$  0.74 & 3.92  $\pm$  1.16 & 4.12  $\pm$  0.75 \\
\hline
2.5deg100vel &  0.84  $\pm$  0.07 & 0.90  $\pm$  0.07 & 0.90  $\pm$  0.06 & 0.93  $\pm$  0.06  &  3.36  $\pm$  1.04 & 4.05  $\pm$  0.85 & 3.86  $\pm$  1.08 & 4.04  $\pm$  0.81 \\
5deg100vel &  0.85  $\pm$  0.09 & 0.89  $\pm$  0.09 & 0.91  $\pm$  0.05 & 0.92  $\pm$  0.04  &  3.35  $\pm$  0.92 & 4.10  $\pm$  0.98 & 3.64  $\pm$  0.89 & 3.98  $\pm$  0.89 \\
7.5deg100vel &  0.84  $\pm$  0.07 & 0.88  $\pm$  0.09 & 0.87  $\pm$  0.06 & 0.95  $\pm$  0.03  &  3.22  $\pm$  0.89 & 3.85  $\pm$  1.03 & 3.37  $\pm$  0.89 & 4.01  $\pm$  1.04 \\
10deg100vel &  0.82  $\pm$  0.08 & 0.84  $\pm$  0.09 & 0.84  $\pm$  0.06 & 0.89  $\pm$  0.08  &  3.41  $\pm$  0.93 & 3.19  $\pm$  1.05 & 3.43  $\pm$  0.86 & 3.68  $\pm$  0.80 \\
\hline
 \end{tabular}
 \end{center}

\end{minipage}
\end{table*}

\label{lastpage}

\end{document}